\documentclass{aa}  

\usepackage{graphicx}
\usepackage{xcolor}

\usepackage{txfonts}
\usepackage[]{hyperref}

\begin{document} 

\title{EGS-z11-R0: a red, dust-rich galaxy at Cosmic Dawn}

\author{Giulia Rodighiero$^{1,2}$\thanks{Corresponding author, \email{giulia.rodighiero@unipd.it}}
 \and Andrea Ferrara $^{3}$     
\and Michele Catone $^{1,2}$
\and Lorenzo Napolitano $^{4}$     
\and Paolo Cassata $^{1,2}$     
\and Giovanni Gandolfi $^{4}$     
\and Emiliano Merlin $^{2}$     
\and Andrea Grazian $^{2}$     
\and Alvio Renzini $^{2}$     
\and Laura Bisigello $^{2}$     
\and Marco Castellano $^{4}$     
\and Pablo G. P\'{e}rez-Gonz\'{a}lez$^{5}$
\and Borja P\'{e}rez-D\'{\i}az $^{4}$     
\and Edoardo Iani $^{6}$     
\and Carlotta Gruppioni $^{7}$
\and Steven L. Finkelstein $^{8}$
\and Anton M. Koekemoer  $^{9}$
\and Alessandro Bianchetti $^{2}$
\and Francesco Sinigaglia $^{10,11,12,13}$
\and Paola Santini $^{4}$   
\and Giorgia Girardi $^{1,2}$
\and Amelia Vietri $^{1,2}$
\and Roberta Tripodi $^{4,14}$ 
\and Benedetta Vulcani $^{2}$
\and Polet Roman Quesada $^{15}$
\and Gaia E. Esposito $^{2,1}$
}
\institute{
$^{1}$ Dipartimento di Fisica e Astronomia "G. Galilei", Universit\`a di Padova, Vicolo dell'Osservatorio 3, 35131 Padova, Italy \\
$^{2}$ INAF, Osservatorio Astronomico di Padova, Vicolo dell'Osservatorio 5, 35122 Padova, Italy \\
$^{3}$ Scuola Normale Superiore, Piazza dei Cavalieri 7, I-56126 Pisa, Italy\\
$^{4}$ INAF, Osservatorio Astronomico di Roma, Via Frascati 33, 00078 Monteporzio Catone, Roma, Italy \\
$^{5}$ Centro de Astrobiolog\'ia (CAB), CSIC-INTA, Ctra. de Ajalvir km 4, Torrej\'on de Ardoz, E-28850, Madrid, Spain \\
$^{6}$ Institute of Science and Technology Austria (ISTA), Am Campus 1, 3400 Klosterneuburg, Austria \\
$^{7}$ INAF-Osservatorio di Astrofisica e Scienza dello Spazio, via Gobetti 93/3, I-4019, Bologna, Italy\\
$^{8}$ Department of Astronomy, The University of Texas at Austin, Austin, TX, USA \\
$^{9}$ Space Telescope Science Institute, 3700 San Martin Drive,
Baltimore, MD 21218, USA\\
$^{10}$ Institute for Fundamental Physics of the Universe, Via Beirut 2, I-34151 Trieste, Italy \\
$^{11}$ SISSA - International School for Advanced Studies, Via Bonomea 265, 34136 Trieste, Italy \\
$^{12}$ INAF - Osservatorio Astronomico di Trieste, Via G. B. Tiepolo 11, I-34131 Trieste, Italy \\
$^{13}$ INFN – National Institute for Nuclear Physics, Via Valerio 2, I-34127 Trieste, Italy \\
$^{14}$ IFPU- Institute for Fundamental Physics of the Universe, via Beirut 2, I-34151 Trieste, Italy\\
$^{15}$ Department of Physics, New York University (CAS 2027)\\
}

   \date{Received ; accepted}

\abstract{
\textit{Context.} Galaxies discovered  by JWST at $z>10$ are predominantly characterized by extremely blue rest-frame UV slopes, consistent with dust-poor stellar populations. Conversely, the existence of dust-reddened systems at such early epochs has remained largely unconfirmed spectroscopically.

\textit{Aims.} We present the spectroscopic confirmation of EGS-z11-R0 at $z=11.45$, the most distant red galaxy identified and confirmed to date. Discovered serendipitously through inspection of publicly available JWST/NIRSpec data, EGS-z11-R0 offers a rare window into early dust and metal enrichment at cosmic dawn.

\textit{Methods.}We analyze \textit{JWST}/NIRSpec PRISM spectroscopy together with multiwavelength \textit{HST}, NIRCam, and MIRI photometry. 
The spectroscopic redshift is derived from rest-frame UV emission lines detected in the NIRSpec PRISM spectrum. 
In particular, we identify significant detections of the \ion{C}{iv} $\lambda\lambda1548,1551$ and \ion{C}{iii}] $\lambda1908$ transitions, whose line centroids yield a weighted-average redshift of $z_{\rm spec}=11.452 \pm 0.021$. 
We measure rest-frame UV emission-line fluxes and equivalent widths and use these diagnostics to constrain the nature of the ionizing radiation field. 
Finally, we perform spectral energy distribution (SED) modeling with \texttt{CIGALE}, combining the spectroscopy with broadband photometry and including stellar, nebular, dust, and active galactic nuclei (AGN) components.

\textit{Results.} EGS-z11-R0 exhibits a red UV continuum slope ($\beta_{\rm UV} \sim -1.0$), placing it well above the canonical $M_{\rm UV}$-$\beta_{\rm UV}$ relation at $z\sim10$--12 and making it the highest-redshift spectroscopically confirmed member of the emerging “red monster” population. Emission-line diagnostics reveal a hard ionizing spectrum consistent with extreme star formation and compatible with a composite stellar+AGN scenario. The best-fit SED solution favors a stellar mass of $\log(M_\star/M_\odot)\sim9.2$-9.6, a star-formation rate of $\sim10$--40~$M_\odot\,\mathrm{yr^{-1}}$, and substantial dust attenuation ($A_V\sim1.2$~mag), significantly higher than typical values inferred for the bulk of $z>10$ galaxies. 

\textit{Conclusions.} The confirmation of EGS-z11-R0 establishes that chemically evolved, dust-enriched galaxies were already in place at $z\sim11.5$. Its red UV slope demonstrates that rapid metal and dust production occurred within the first few hundred million years of cosmic history. While a moderate AGN contribution cannot be excluded, the data robustly indicate a dust-enriched system observed during an early, possibly obscured phase of galaxy assembly, challenging the prevailing view of an exclusively blue and dust-poor Cosmic Dawn.
}


\maketitle

%-------------------------------------------------------------------

\section{Introduction}
The first few hundred million years after the Big Bang represent a
critical phase in galaxy formation, when the earliest stellar systems
assembled, enriched their interstellar media with heavy elements,
and contributed to cosmic reionization. Since its commissioning,
the \textit{James Webb Space Telescope} (JWST) has revolutionized
our view of this epoch, revealing a surprisingly abundant population
of luminous galaxies at $z>10$, corresponding to cosmic ages
$\lesssim 500$ Myr \citep[e.g.][]{Naidu2022,Harikane2022,Austin,Castellano2022, Castellano2023,Finkelstein2,2023ApJ...951L...1P,PPG,Gandolfi25}.

A defining characteristic of the majority of
these systems is their extremely blue rest-frame ultraviolet (UV)
continuum slopes, often reaching $\beta_{\rm UV} \lesssim -2.5$ \citep{Morales2024,Austin2025,Cullen2024,Chavez2026}. Such steep
slopes are typically interpreted as evidence for young, metal-poor
stellar populations with minimal dust attenuation.

However, recent studies have uncovered a small but intriguing
population of candidate high-redshift galaxies exhibiting significantly redder
UV continua \citep[i.e. $\beta_{\rm UV}>-1.5$, ][]{Rodighiero2023,Mitsuhashi2025,Donnan25,Tang25,Kokorev2025,Napolitano2025b,Taylor2025}. These objects deviate markedly from the canonical
$M_{\rm UV}$-$\beta_{\rm UV}$ relation established at $z\sim6$-9 and
extended to $z\sim10$-12 with JWST. 
The physical origin of
such red slopes at $z>10$ remains uncertain and  given the short age
of the Universe at these epochs ($\lesssim400$ Myr), substantial
reddening is difficult to reconcile with simple models of chemically primitive galaxies.

However, dust attenuation provides the
most direct interpretation, requiring visual extinctions of
$A_V \sim 0.5$--1 mag to reproduce $\beta_{\rm UV} \gtrsim -1.5$.
Such values imply rapid metal production and efficient grain
growth within the first few hundred million years. Alternatively,
strong nebular continuum emission from dense, highly ionized
gas can redden the UV spectrum without invoking large dust
masses, although reproducing the most extreme slopes requires
very high ionization parameters and stellar effective temperatures \citep{Katz2025}.
A further possibility is the presence of a hard ionizing source,
such as an accreting black hole, contributing to a composite
stellar+AGN spectrum \citep[e.g.][]{Hainline2011,Hainline2012,Napolitano2025b}.

Recent theoretical models provide an essential context for these scenarios. Cosmological zoom-in simulations including on-the-fly dust evolution and radiative transfer \citep[e.g.][]{Narayanan2025} demonstrate that bursty star-formation histories alone can generate a wide range of intrinsic UV slopes ($\beta_0 \sim -3$ to $-2.2$).
However, slopes redder than $\beta_{\rm UV} \sim -1.5$ generally require significant dust attenuation once nebular continuum effects are included
\citep[e.g.][]{Katz2025}. 

Other models reinforce this picture. \cite{Ferrara2024} proposes that galaxies at $z>10$ initially undergo radiative feedback-regulated phases leading to a bursty star formation mode. When dust shielding becomes important the galaxy switches to a smooth, substantially obscured star formation regime (which we define here as a "Red Monster" phase). During such an obscured phase that lasts $\simeq 20$\% of the galaxy lifetime, $\simeq 70$\% of the observed stars are formed \citep[at least for JADES-GS-z14-0,][]{Carniani2025}.  As the galaxy becomes super-Eddington, a powerful radiation-driven outflow clears most of the dust that the galaxy transitions into a “blue monster” dominating the bright end of the UV luminosity function. 

Other theoretical frameworks have also explored the rapid buildup of galaxies at cosmic dawn. In particular, \cite{Dekel2023} proposed a feedback-free model in which the early growth of galaxies is driven by intense cold gas inflows and sustained star formation, with feedback playing a limited role during the first stages of galaxy assembly. In this scenario, the efficient accretion of baryons allows galaxies to rapidly build up their stellar mass already within the first few hundred million years after the Big Bang. On the other hand, \cite{Somerville2025} investigated the early evolution of galaxies within cosmological semi-analytic models that implement a density-regulated star formation framework, in which the efficiency of star formation and feedback processes is governed by the gas density and the physical conditions of the interstellar medium.

Furthermore, observational evidence increasingly points toward rapid early stellar assembly of galaxies at these epochs. \cite{Santini2026} show that several $z>10$ systems exhibit unexpectedly high stellar masses, mass-to-light ratios, indicating that substantial stellar assembly has already
occurred within the first few hundred million years. The apparent lack of strong sSFR evolution  at $z>9$ suggests efficient and sustained star formation, possibly occurring during dust-obscured phases predicted by theoretical models. Together, these results imply that red, dust-reddened sources may represent a fundamental and common stage in early galaxy growth rather than rare anomalies.

Despite these theoretical and observational indications, robust
spectroscopic confirmation of strongly dust-reddened galaxies
at $z>10$ is still lacking. Most of the red candidates identified so far
rely on broadband photometry, where the degeneracies between dust,
age, nebular emission, and AGN activity complicate the
interpretation. Secure spectroscopy is therefore essential to
establish whether chemically evolved dust-enriched galaxies
were already in place within the first $\sim400$ Myr of cosmic
history.

In this work, we present the spectroscopic confirmation of
EGS-z11-R0 at $z=11.45$, currently the most distant red galaxy
identified and confirmed to date. The source was discovered
serendipitously through visual inspection of publicly available
JWST/NIRSpec data in the CEERS field. EGS-z11-R0 exhibits a
remarkably flat UV continuum slope ($\beta_{\rm UV}\sim -1.0$),
placing it well above the canonical relation $M_{\rm UV}$ - $\beta_{\rm UV}$
 at $z\sim10$-12. We detect some high-ionization
UV emission lines.

We investigate whether the red UV slope of EGS-z11-R0 is primarily
driven by dust attenuation associated with rapid chemical
enrichment, extreme nebular emission powered by dense star
formation, or a composite stellar+AGN ionizing spectrum.
Combining NIRSpec spectroscopy with multiwavelength HST,
NIRCam, and MIRI photometry, we perform SED modeling
, including stellar, nebular, dust, and AGN components.
Our goal is to assess whether EGS-z11-R0 represents an early,
chemically evolved system observed during a short-lived,
obscured phase of galaxy assembly at cosmic dawn.

This paper is structured as follows. Section~2 describes 
target selection and spectroscopic data. Section~3 presents
the UV slope measurements and their placement in the
$M_{\rm UV}$-$\beta_{\rm UV}$ plane. Section~4 investigates emission-line
diagnostics to constrain the nature of the ionizing source.
Section~5 discusses the results of the SED modeling and the role of dust
and AGN contributions. We conclude with implications for
early metal enrichment and the emerging population of
``red monsters'' at Cosmic Dawn.

Throughout this work, we adopt the cosmological parameters yielded by the latest Planck collaboration release \citep{2020A&A...641A...6P} and a  \cite{Chabrier} initial mass function (IMF).

   \begin{figure*}
   \centering
\includegraphics[width=1\linewidth]{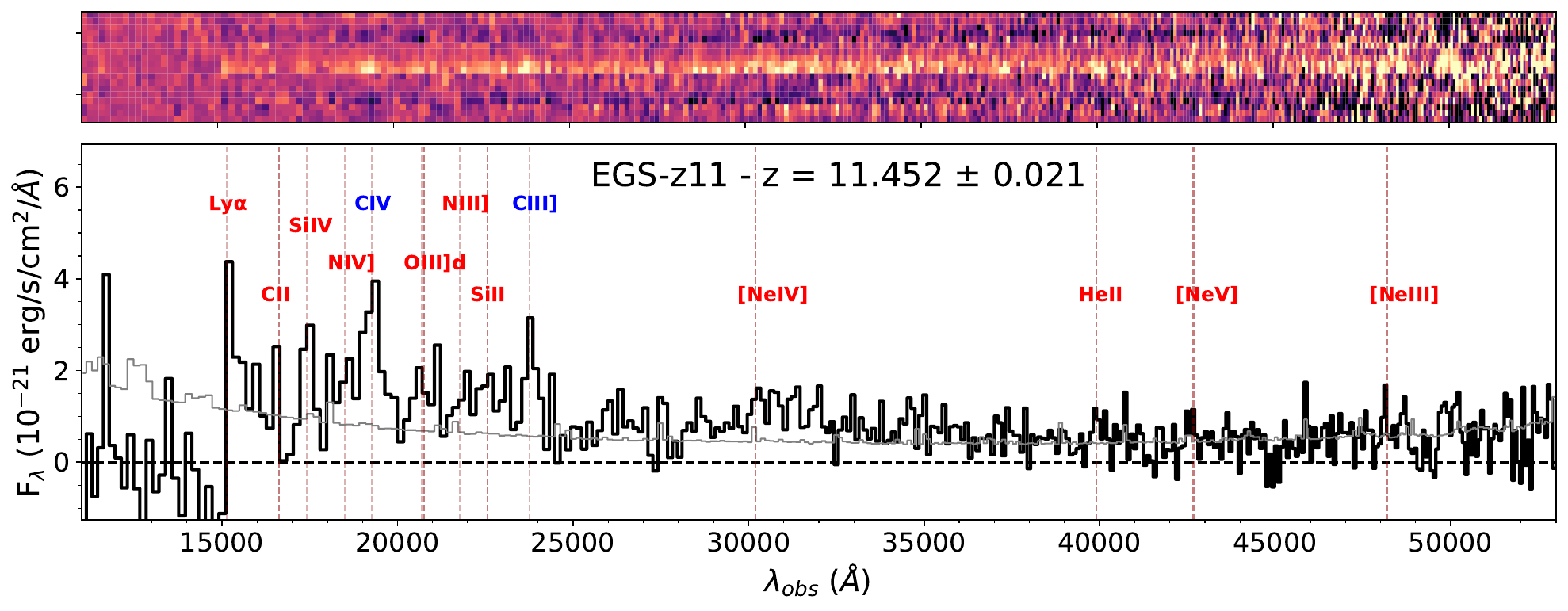}
   \caption{Observed 2D spectrum (top panel) and extracted 1D spectrum (bottom panel) of EGS-z11-R0. 
The pipeline-provided error spectrum is shown in light gray. 
Emission lines with integrated S/N $> 3$ are highlighted in blue, while the locations of lines for which only a $3\sigma$ upper limit is obtained are marked in red.
}
              \label{Fig:2dspec}%
    \end{figure*}

\section{Target selection}
The source analyzed in this work  was initially identified by visual inspection of publicly available JWST spectroscopic products in the \textsc{Dawn JWST Archive} \citep[\textsc{DJA\footnote{\url{https://dawn-cph.github.io/dja/}}},][]{Heintz2024dja, deGraaff2025}. During a systematic search for extreme $z > 10$ candidates, we specifically selected objects displaying unusually red rest-frame UV continua compared to the bulk of blue, dust-poor galaxies commonly found at similar epochs.
EGS-z11-R0 stood out for its markedly red UV slope, inferred from the NIRSpec continuum, and exhibits a clear dropout in the bluest region of the spectrum, consistent with a Lyman break at $z > 11$. 
It is located in the CEERS field \citep{Finkelstein2025}, at 
$\rm RA = 14^h:19^m:26.^s81$, $\rm DEC = 52^\circ:51':42.''66$ with a reported spectroscopic redshift of $z_{\rm spec} = $ 11.4395 (JWST GO program 4106, P.I E. Nelson, DJA srcid = 60147).  The target has been acquired in two MSA/NIRSpec configurations: 1) a low resolution PRISM (R$\sim$100) observation (exposure time 13130 sec); 2) a medium resolution G395M\_F290LP (R$\sim$1000, exposure time 3130 sec) spectrum covering the range of 2.87-5.14 $\mu$m.
We adopt the spectra reduced with the public DJA pipeline (v4)\footnote{\url{https://dawn-cph.github.io/dja/spectroscopy/nirspec/}}.

\subsection{NIRSpec spectral analysis}
\label{sec:nirspec}
We visually inspected the NIRSpec PRISM spectrum (Fig.~\ref{Fig:2dspec}) at the initial redshift solution $z_{\rm spec} = $ 11.4395 suggested by DJA and searched for the Ly$\alpha$-break and emission lines features to measure the spectroscopic redshift and emission lines. We followed the same approach described in \cite{Napolitano2025a} and considered significant only integrated line emissions with a S/N > 3, accounting for errors in both the integrated line flux and the extrapolated continuum at the line position. In particular, the continuum was estimated through a linear interpolation of regions closest to each line that are free of potential features, using the \textsc{emcee} \citep[][]{Foreman_Mackey2013} to perform a Monte Carlo Markov chain analysis (MCMC).

To determine a first estimate of line fluxes, we directly integrated the continuum-subtracted spectrum within a window centered at the expected observed wavelength and spanning 4 $\times$ $\sigma_R(\lambda)$, where $\sigma_R(\lambda)$ represents the expected Gaussian standard deviation for a line observed at resolution R($\lambda$). The resolution is provided by the JWST documentation\footnote{\url{https://jwst-docs.stsci.edu/jwst-near-infrared-spectrograph/nirspec-instrumentation/nirspec-dispersers-and-filters}} with the assumption of a source that illuminates the slit uniformly. Only the UV carbon transitions C \, IV $\lambda\lambda1548,51$ and C \, III] $\lambda1908$ in the PRISM spectrum were found significant with S/N = 3.9 and 3.2, respectively. We recomputed the spectroscopic redshift by taking a weighted average of the line centroids from the carbon emission lines only and found $z_{\rm spec} = $11.452 $\pm$ 0.021, adopted as reference hereafter. We note that this value matches the DJA redshift estimate within the uncertainty of 1$\sigma$.\\
As an additional robustness test of the local continuum interpolation adopted for the line measurements, we also fitted the underlying continuum after masking the spectral regions potentially affected by the C\,{\sc iv}, He\,{\sc ii}, and C\,{\sc iii}] emission lines.
The resulting fit provides an adequate statistical description of the data ($\chi^2_\nu = 0.85$, $p = 0.87$), indicating that there is no statistical evidence that requires a more complex continuum model, such as a power-law continuum. We therefore conclude that the adopted continuum estimation does not significantly affect the UV emission-line measurements presented in this work.

\begin{figure}
   \centering
\includegraphics[width=0.6\linewidth]{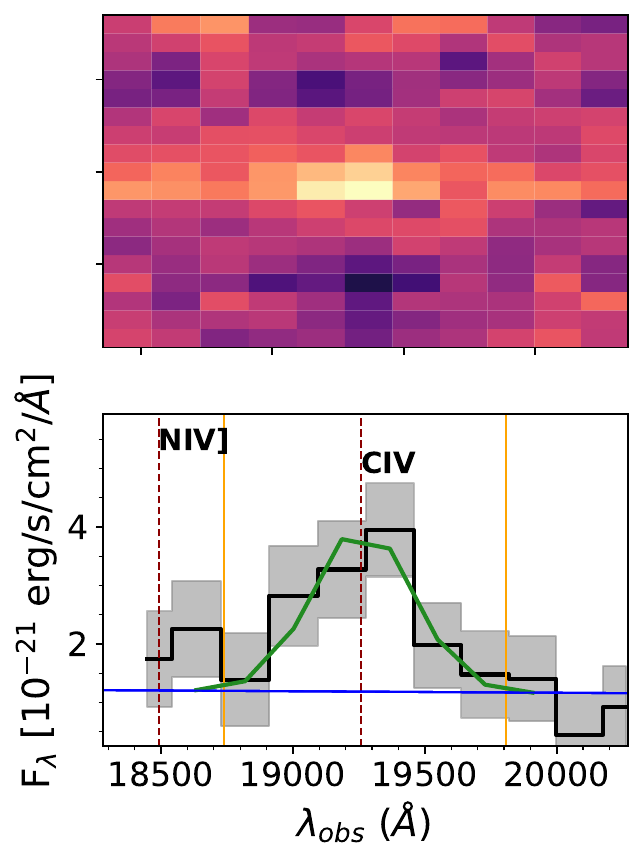}
\includegraphics[width=0.6\linewidth]{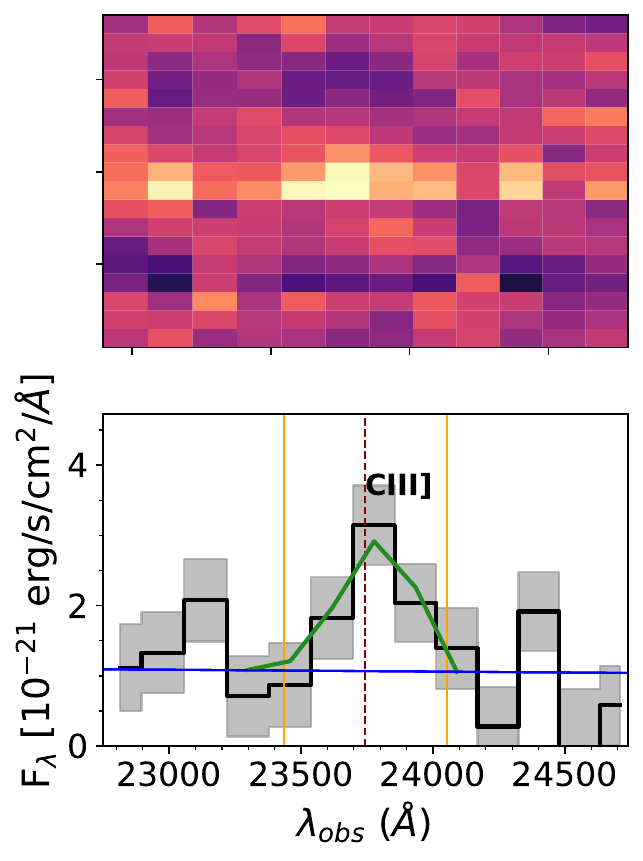}
   \caption{
Top and bottom panels:
Rest-frame UV spectral regions of EGS-z11-R0 extracted from the NIRSpec prism data. 
 Zoom-in around the \ion{C}{iv} emission feature and the \ion{C}{iii}] doublet, respectively.
The gray curves show the observed spectrum, while the shaded regions represent the 1$\sigma$ uncertainties. 
}
    \label{Fig:lines}%
    \end{figure}

As detailed in \cite{Castellano2024}, for the two significant emission features, we then performed a Gaussian fit on the continuum-subtracted flux using the \textsc{specutils} package of \textsc{astropy} \citep{Astropy2013} to obtain a flux and rest-frame equivalent widths (EW) measurements. The Gaussian mean was allowed to vary with $\Delta$z = 0.04, and the Gaussian standard deviation was within 5\% of the nominal $\sigma_R(\lambda)$, to account for the uncertainties of the red shift. We further refined the Gaussian fit employing an MCMC analysis with \textsc{emcee}. The amplitude, standard deviation, and their uncertainty obtained from the best Gaussian model outputted by the \textsc{specutils} package were used to initialize 100 walkers with 100,000 iterations. The best model parameters and the integrated flux were determined by taking the median of the posterior distributions resulting from the MCMC fitting routine. Uncertainties were calculated on the basis of the highest posterior density intervals of the 68-th percentile. EW and their uncertainties were computed based on the integrated flux, the continuum flux determined at the line's position, and the spectroscopic redshift. We present the best fit models in Fig.~\ref{Fig:lines} (top and middle panels) and report line measurements in Table \ref{tab:emission_lines}. We did not significantly detect any other UV emission lines from direct integration. In Table \ref{tab:emission_lines} we report a 3$\sigma$ upper limit for He \, II $\lambda1640$ that is relevant for the analysis in Sect.~\ref{sec:agnVSsfg}.\\
At the spectral resolution of the NIRSpec/PRISM observations, He\,{\sc ii} $\lambda1640$ is unresolved from the adjacent O\,{\sc iii}] $\lambda\lambda1661,1666$ doublet. Therefore, the measured $3\sigma$ upper limit formally applies to the unresolved He\,{\sc ii}+O\,{\sc iii}] blend rather than to He\,{\sc ii} alone. The relative contribution of the two species is expected to depend on the hardness of the ionizing radiation field and on the physical conditions of the gas. In star-forming galaxies, He\,{\sc ii} is generally weak except under extreme ionization conditions and very low metallicities \citep[e.g.][]{Berg2018,Nanayakkara2019,Weida2024}, whereas AGN narrow-line regions can exhibit comparatively stronger He\,{\sc ii} emission \citep[e.g.][]{Feltre2012,PerezMontero2023}.

We also searched for Ly$\alpha$ emission using the dedicated fitting
procedure described in \cite{Napolitano2024}, which models
simultaneously the continuum and the asymmetric Ly$\alpha$ line profile.
A tentative signal ($\sim2\sigma$) Ly$\alpha$ is recovered, with
$EW_0=41.7\pm22.2$ \AA\ and
$F_{\rm Ly\alpha}=(8.7\pm4.4)\times10^{-19}\,
{\rm erg\,cm^{-2}\,s^{-1}}$.
Given its low significance, we do not include Ly$\alpha$ in the
subsequent analysis.

Following \cite{Napolitano2026}, we estimate the UV slope $\beta_{\rm UV}$ by fitting a power-law model ($f_{\lambda} \propto \lambda^\beta$) to the continuum flux in the rest-frame range 1350--2600 $~\AA$ of the NIRSpec PRISM spectrum. To avoid line contamination, we mask the features of CIV$\lambda\lambda$ 1548,51 and CIII] $\lambda$ 1909 using line widths broadened according to instrumental resolution. The fitting is performed using the \textsc{emcee} with 30 walkers and 500,000 steps and adopting a flat prior on $\beta_{\rm UV}$ between -3.5 and 0. We found $\beta_{\rm UV}=-0.68 \pm $0.30, where the best-fit value and uncertainty are derived from the posterior median and standard deviation, respectively. 
We note that the UV continuum slope was measured directly from the reduced DJA NIRSpec spectrum, using the wavelength-dependent path-loss correction provided with the extracted 1D spectrum. No additional rescaling to match the NIRCam photometry was applied prior to the fit. We verified that the path-loss correction varies by less than 0.6\% over the wavelength range used for the $\beta$ measurement (1350--2600\,\AA\ in the rest frame), and therefore has a negligible impact on the derived UV continuum slope.

A full characterization of the various lines and their significance will be discussed in Sect. \ref{sec:agnVSsfg}.
\begin{table}
\caption{Emission-line fluxes.}
\label{tab:emission_lines}
\centering
\begin{tabular}{lcc}
\hline\hline
Line &
Flux &
EW$_0$ \\
&
($10^{-21}$ erg s$^{-1}$ cm$^{-2}$) &
(\AA) \\
\hline
Ly$\alpha$ $\lambda1216$      & $870.0 \pm 440.0$     & $41.7 \pm 22.2$ \\
C\,IV $\lambda\lambda1548,51$ & $1399.0 \pm 359.1$ & $95.1 \pm 26.0$ \\
He\,II +O III] $\lambda1640$          & $<537$             & $<39$ \\
C\,III] $\lambda1908$         & $623.3 \pm 193.1$ & $46.9 \pm 14.1$ \\
\hline
\end{tabular}
\tablefoot{Emission-line measurements for EGS-z11-R0. We note that Ly$\alpha$ line is detected at only the $\sim2\sigma$ significance level and is therefore not considered a secure detection in our analysis.}
\end{table}

\subsection{Photometry}

EGS-z11-R0 has been identified in v1.0 of the public Cosmic Evolution Early Release Science (CEERS; \citealt{Bagley23}) survey's photometric catalog \citep{Cox+25} as source ID 19266, as well as in v0.5 of the UNICORN catalog \citep[][; ID 88738]{Finkelstein2025}, and in the ASTRODEEP-JWST catalog \citep{Merlin+24} as source ID CEERS-77458. We exploit JWST/NIRCam measurements from the v1.0 CEERS catalog and combine these with NIRCam F090W and other photometry from the v0.5 UNICORN catalog (Finkelstein et al. in preparation, which includes HST/ACS and HST/WFC3 imaging from CANDELS \citep{Grogin2011,Koekemoer2011}. We thus assemble the most complete near-infrared photometry compilation available for EGS-z11-R0.

For the JWST/MIRI photometry of EGS-z11-R0, we used observations from the Cycle 2 GO 3794 \textit{MIRI EGS Galaxies Active Galactic Nucleus} survey (MEGA, PI: A. Kirkpatrick; \citealt{Backhaus2025}), which provides imaging data in four broad-band filters from 7 to 21~$\mu\text{m}$ (F770W, F1000W, F1500W, F2100W). The mosaics have been reduced and calibrated with the \texttt{Rainbow} pipeline (Pablo G. Pérez-González, private communication). We then performed template fitting photometry using \textsc{t-phot} v2.0 \citep{Merlin16} and synthetic NIRCam and MIRI PSFs generated with STPSF \citep{Perrin14}. In addition, to make the flux uncertainties consistent with the dispersion of fluxes, we selected different magnitude bins and injected 500 artificial sources of such flux, whose photometry was computed using \textsc{t-phot}. We computed the ratio between the standard deviation of the measured fluxes and the median uncertainty for each bin and extrapolated the rescaling factor as a function of the injected magnitude with a second-degree polynomial. We finally rescaled EGS-z11-R0 flux uncertainties by a multiplicative factor derived from the aforementioned function based on its photometry. The overall MIRI catalog will be presented in Catone et al. (in prep.).

\begin{figure*}
   \centering
\includegraphics[width=.75\linewidth]{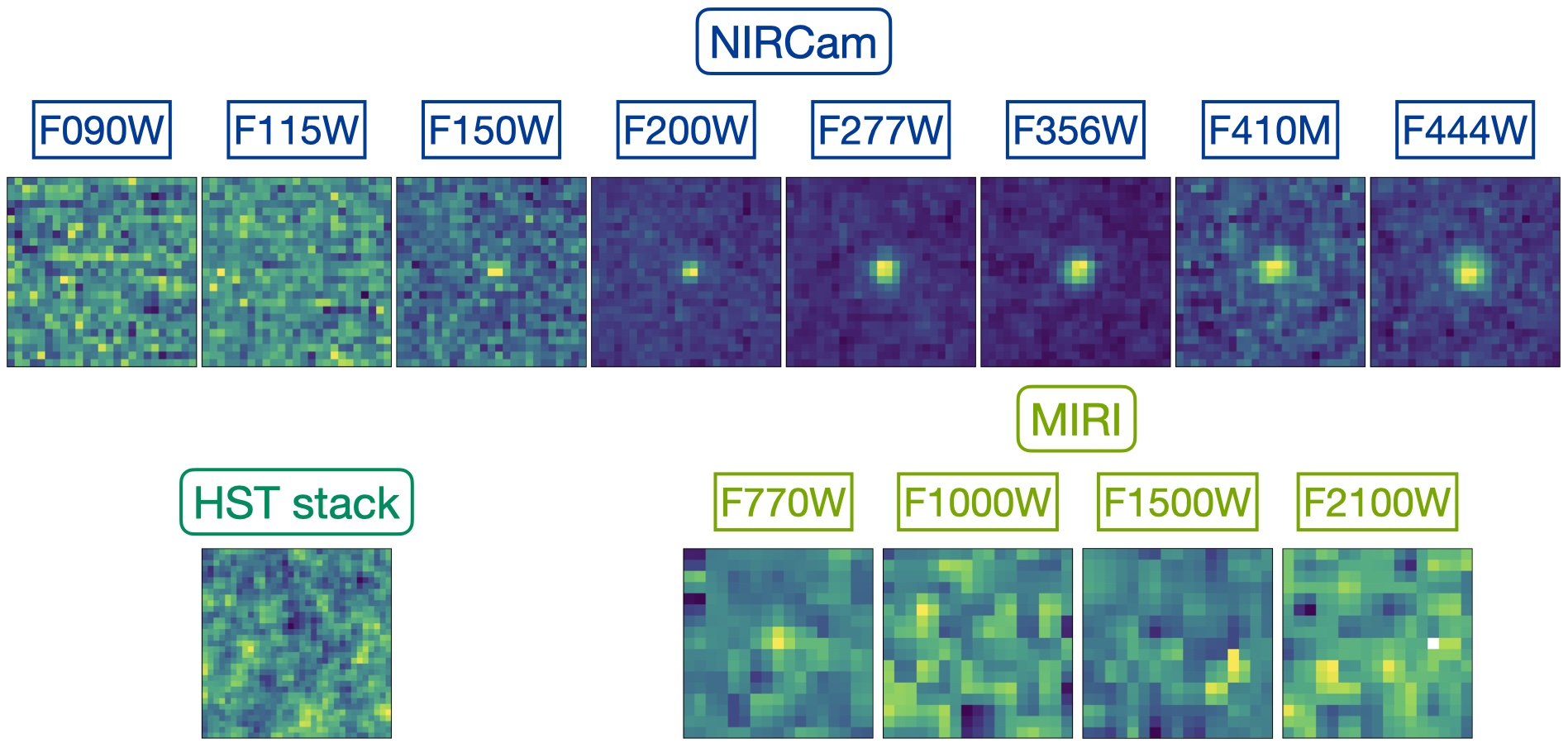}
   \caption{Multiwavelength photometric cutouts of EGS-z11-R0:  NIRCam bands (upper row, from DJA mosaics) and MIRI (bottom right row).  For HST, we report the stacked frame of all the available bands (bottom left), confirming the dropout nature of the source. Each image has a size of 1"$\times$1".}
              \label{Fig:cutouts}%
    \end{figure*}

\begin{figure}
   \centering
   \includegraphics[width=0.5\linewidth]{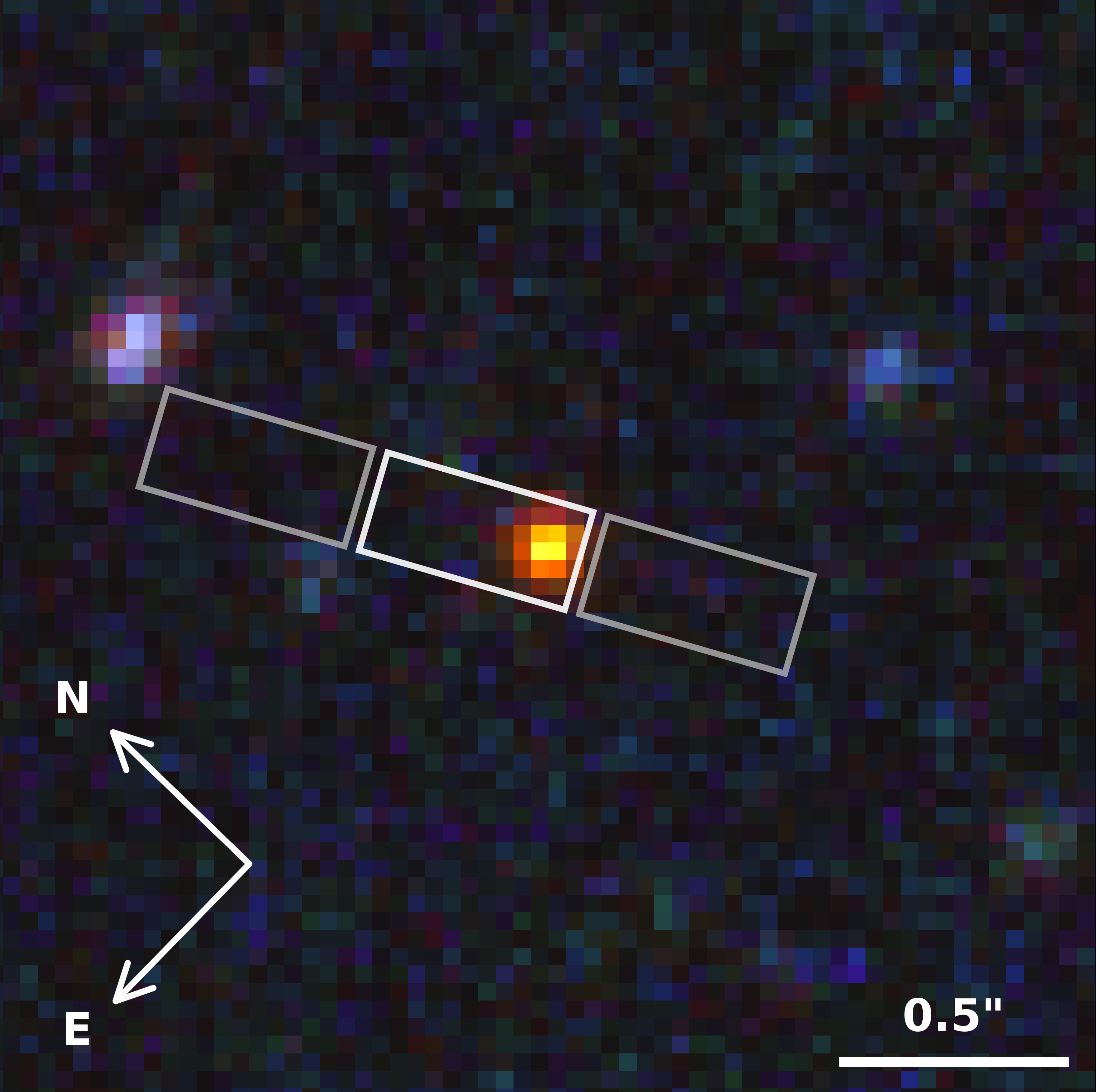}
   \caption[]{RGB NIRCam image of EGS-z11-R0. The gray boxes show the MSA shutters of the NIRSpec/PRISM observation. The stamp has a size of 2.5"$\times$2.5" and it is oriented as indicated by the arrows. The image was created using Trilogy\footnotemark~and GIMP v 3.0.8\footnotemark~by assigning the F090W, F115W filters to the B channel, F150W, F200W, and F277W to the G channel and F356W, F410M, and F444W to the R channel.}
            \label{Fig:cutout_rgb}%
    \end{figure}

The multiwavelength cutouts are shown in Fig. \ref{Fig:cutouts}, with the corresponding flux measurements in all the considered bands reported in Table \ref{tab:egsz11-photometry}. 
The MSA/NIRSpec layout of the PRISM observation is illustrated in the RGB NIRCam cutout of EGS-z11-R0, as depicted in Figure \ref{Fig:cutout_rgb}.

\subsection{Morphology}
We characterize the morphology of EGS-z11-R0 in all JWST/NIRCam detection bands 
(F277W, F356W, F410M, F444W) using \texttt{GALFIT} \citep{2002AJ....124..266P}. 
We note that the analysis was restricted to the NIRCam bands with the highest signal-to-noise ratio$>5$ . Although the source is close to detection in F200W, its significance is formally lower than our threshold (S/N $\simeq$ 4.6). We therefore limited the structural fitting to the higher-S/N images to obtain more robust morphological parameters.

We tested two scenarios --- 
fitting a pure PSF model (generating empirical PSFs via the \texttt{STPSF} package 
in each band) and a PSF plus an extended S\'ersic component. A pure-PSF 
fit yields $\chi^2_{\rm red} = 0.092$--$0.135$ with smooth residuals in 
all bands, while the addition of an extended component does not appreciably 
improve our fits ($\Delta\chi^2 < 0.002$). We therefore conclude that the 
source is consistent with being unresolved in all detection bands. We derive 
an upper limit on its effective radius at $z = 11.452$ by taking half of the 
tabulated STScI empirical FWHM of the detection bands. The F277W band has the smallest FWHM 
($0.092''$), giving a physical size upper limit of $177.8 \, \mathrm{pc}$. 
The F444W band, which has the largest FWHM among the detection bands 
($0.145''$), yields a physical upper limit of $280.1 \, \mathrm{pc}$.

       \begin{figure}
   \centering
\includegraphics[width=1.\linewidth]{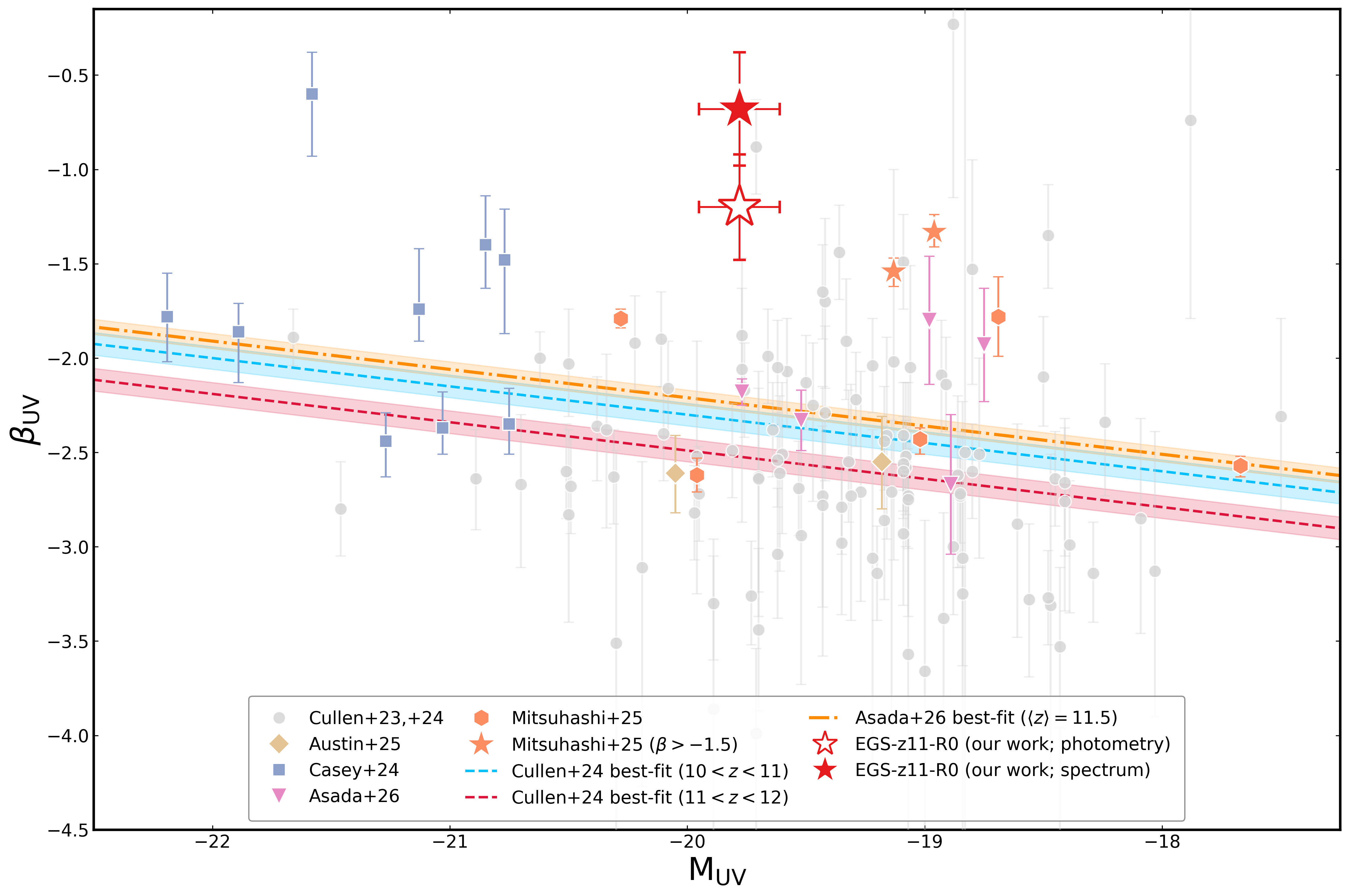}

   \caption{UV continuum slope $\beta_{\rm UV}$ as a function of absolute UV magnitude $M_{\mathrm{UV}}$ 
   at $z \sim 10$-12. Gray circles show literature measurements from \cite{Cullen23,Cullen2024}, while colored symbols denote individual sources from \cite{Casey2024,Austin2025,Asada2026},  and \cite{Mitsuhashi2025}. Star symbols highlight the subset of \cite{Mitsuhashi2025} sources with $\beta_{\rm UV} > -1.5$ and EGS-z11-R0 (this work; filled red star for the $\beta_{\rm UV}$ from spectroscopy and empty red star for the $\beta_{\rm UV}$ from the photometry alone). 
   The blue and red dashed lines show the best-fit $\beta_{\rm UV}$--$M_{\rm UV}$ relations from \citet{Cullen2024} for the $10<z<11$ and $11<z<12$ samples, respectively, while the orange dash-dotted line shows the updated relation from \citet{Asada2026}, derived from a medium-band plus broad-band selected galaxy sample (with shaded regions indicating their associated uncertainties). The latter is systematically redder than the \citet{Cullen2024} relation at similar redshifts, reflecting the improved completeness of medium-band observations for galaxies with red UV continua.
   }
              \label{Fig:uvslope}
    \end{figure}

      \begin{figure*}
   \centering
\includegraphics[width=0.48\linewidth]{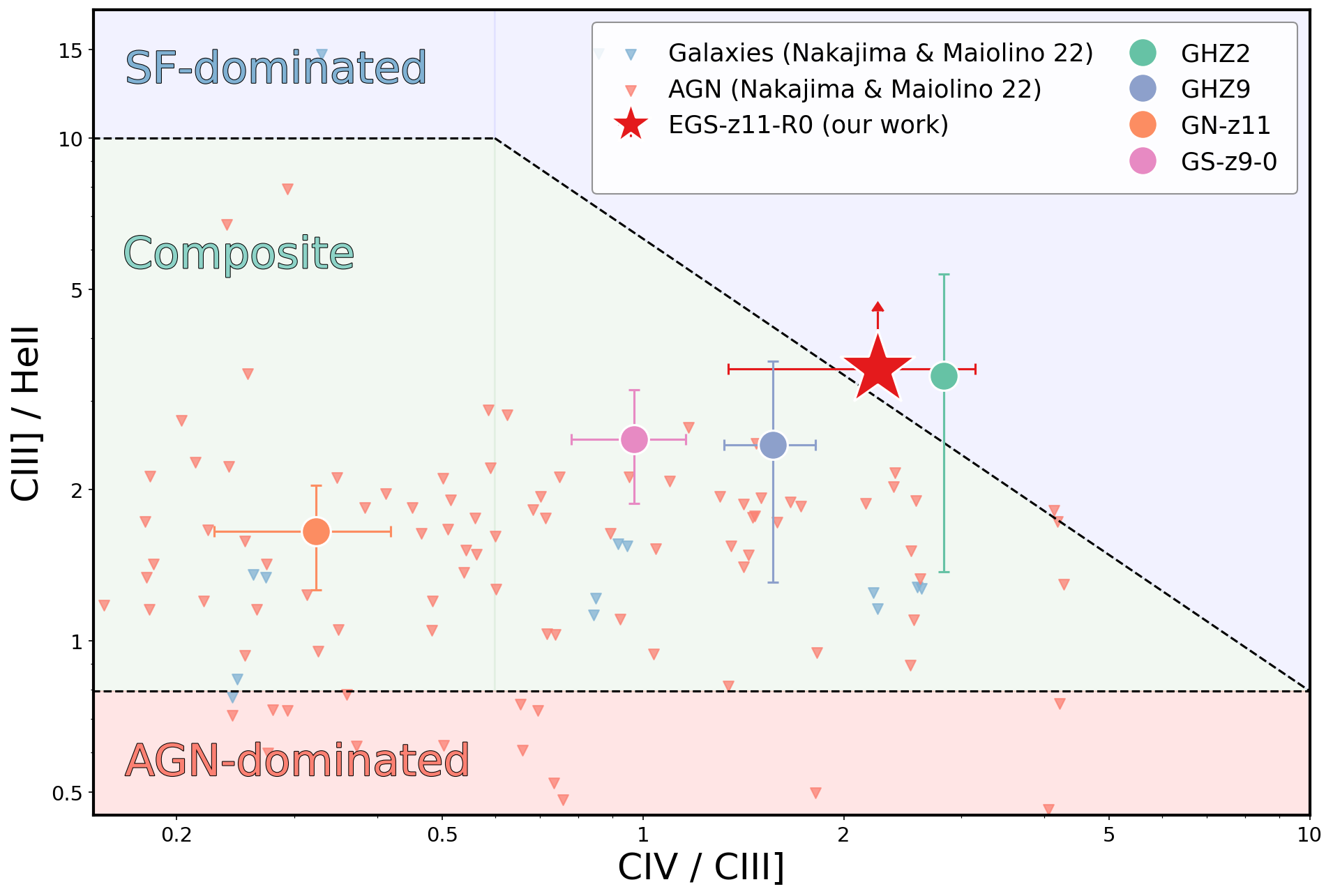}
\includegraphics[width=0.48\linewidth]{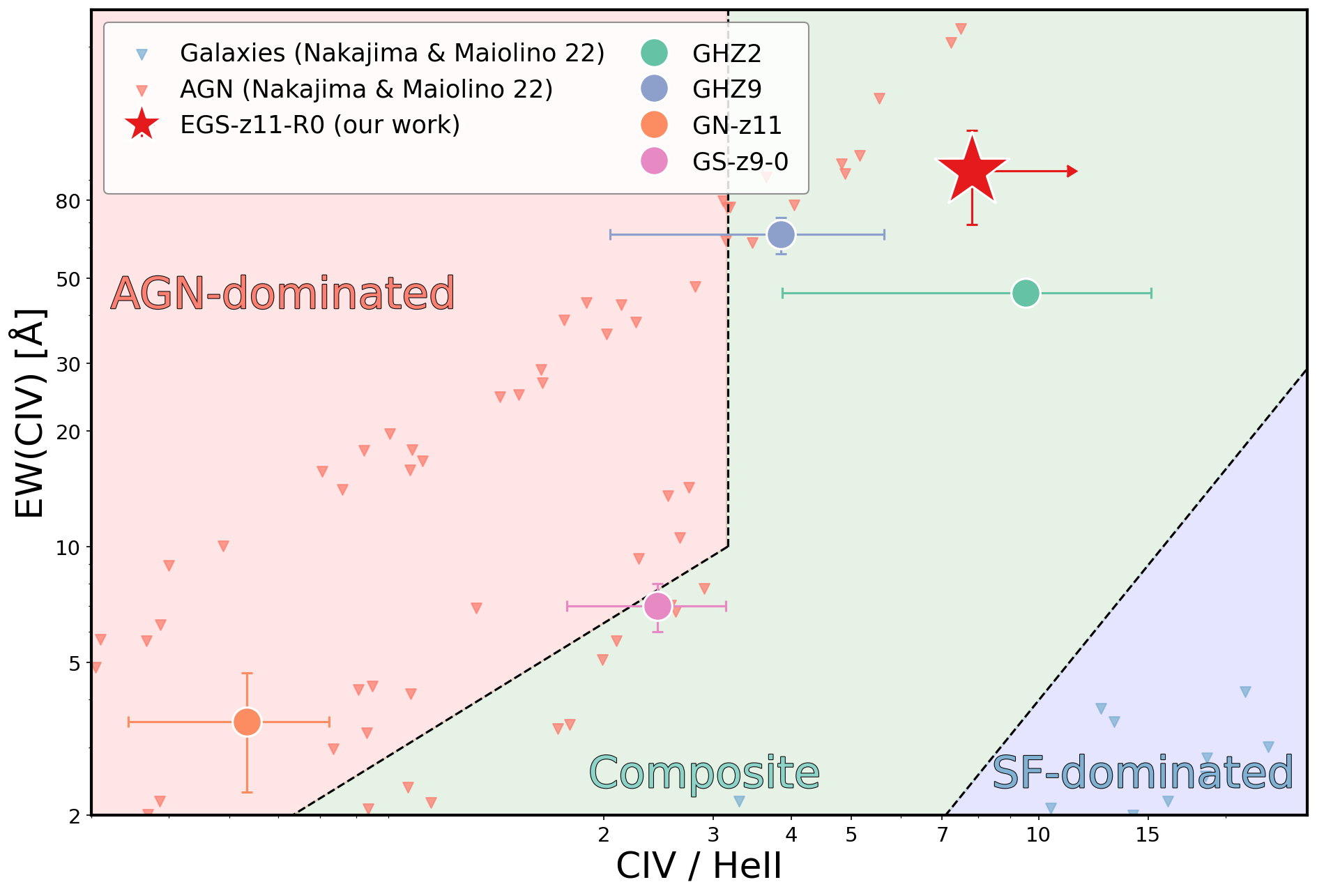}
\includegraphics[width=0.48\linewidth]{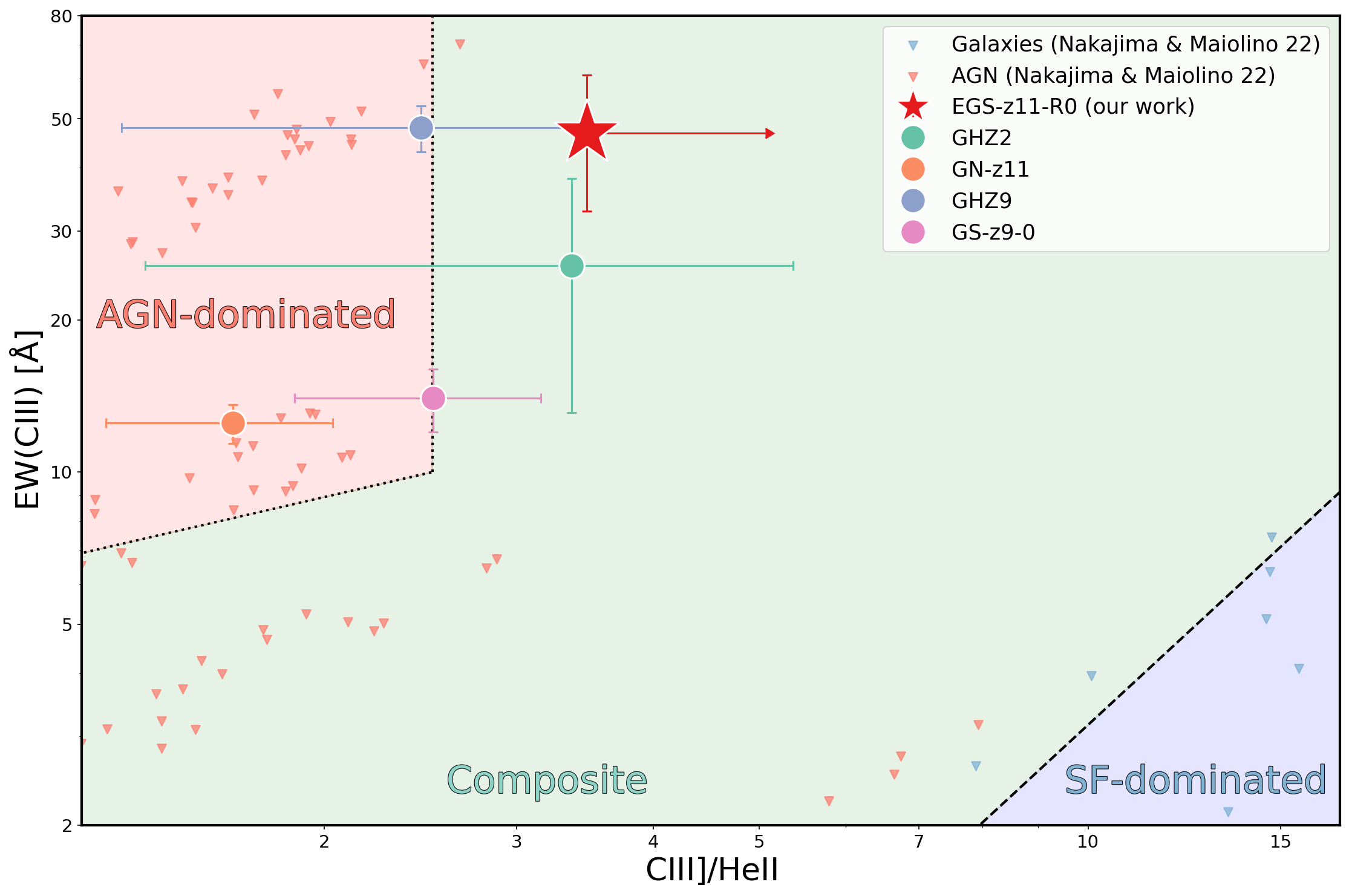}
\includegraphics[width=0.48\linewidth]{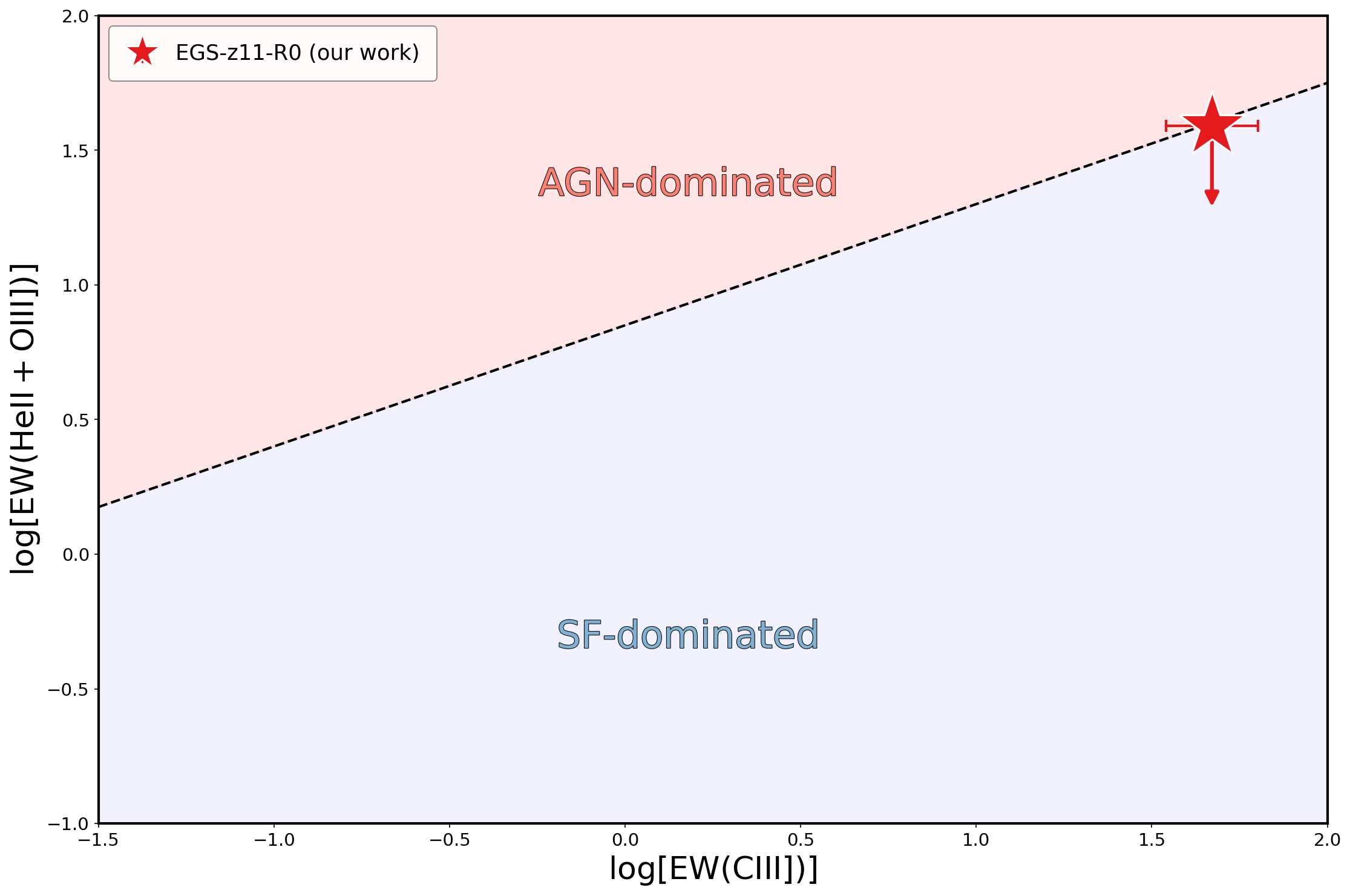}
\caption{
Rest-frame UV diagnostic diagrams for EGS-z11-R0 and comparison with
$z \gtrsim 9$ galaxies. We also display AGN and SFG models from
\citet{Nakajima2022} with $\log U$ and $Z_{\rm gas}$ consistent
within $1\sigma$ with the \textsc{CIGALE} best-fit estimates of
EGS-z11-R0 as red and blue triangles, respectively.
\textit{Top-left panel}: C\,{\sc iv}/C\,{\sc iii}] versus
C\,{\sc iii}]/He\,{\sc ii} line ratios, with dashed curves indicating
the approximate boundaries between AGN-dominated (red shaded area),
composite (green shaded area), and star-formation-dominated
(blue shaded area) ionizing sources.
\textit{Top-right panel}: EW(C\,{\sc iv}) as a function of
C\,{\sc iv}/He\,{\sc ii}.
\textit{Bottom-left panel}: EW(C\,{\sc iii}]) versus
C\,{\sc iii}]/He\,{\sc ii}.
\textit{Bottom-right panel}: the low-resolution equivalent-width
diagnostic proposed by \citet{Aravelo2025},
$\log[\mathrm{EW}(\mathrm{He\,II+O\,III]})]$ versus
$\log[\mathrm{EW}(\mathrm{C\,III]})]$, which explicitly accounts for
the unresolved He\,{\sc ii}+O\,{\sc iii}] blend. The dashed line marks
the AGN/SF demarcation derived by \citet{Aravelo2025}.
The red star marks EGS-z11-R0 (this work), while colored symbols denote
GHZ2, GN-z11, GHZ9, and GS-z9-0. Error bars represent $1\sigma$
uncertainties; the downward arrow in the bottom-right panel indicates
the upper limit on the EW of the unresolved He\,{\sc ii}+O\,{\sc iii}]
blend.
}
\label{diagnostics}
\end{figure*}

         \begin{figure*}
   \centering
\includegraphics[width=1\linewidth]{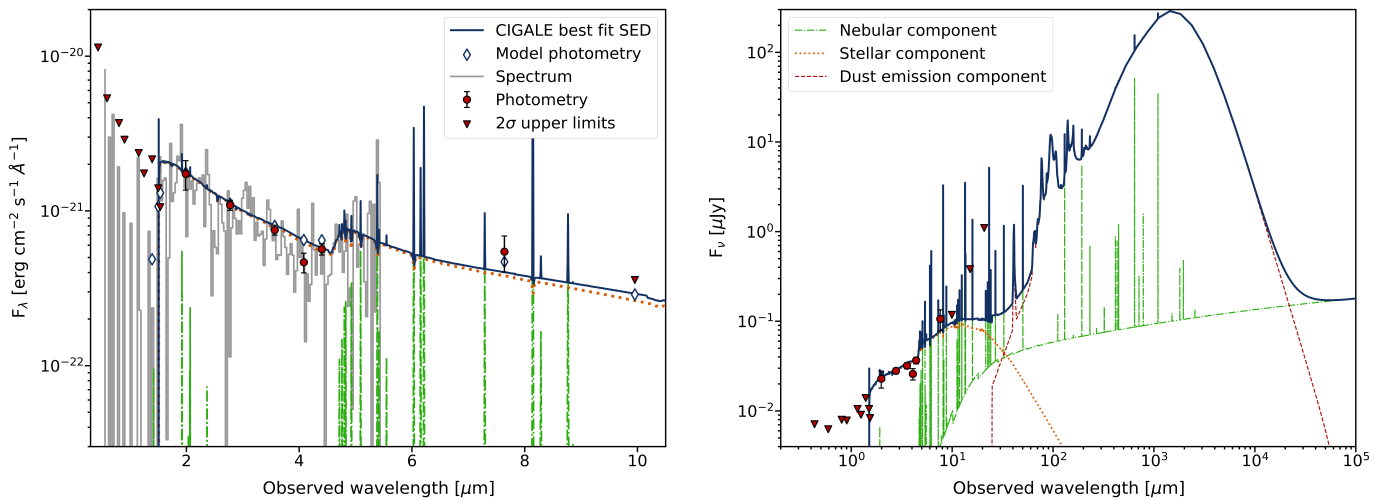}
   \caption{Best fit Spectral Energy Distribution (SED) modeling of EGS-z11-R0 with \textsc{CIGALE}. 
\textbf{Left panel:} Observed \textit{NIRSpec/PRISM} spectrum rebinned with a bin size of $\Delta\lambda = 3\mu$m (gray), overlaid with the best-fit \textsc{CIGALE} model (solid blue line). Blue diamonds indicate the model photometry integrated over the observed bands, red circles show the measured photometric fluxes, and red triangles denote $2\sigma$ upper limits.  The stellar continuum is marked as an orange dotted line, The green dash-dotted curve highlights the contribution from nebular emission lines. 
\textbf{Right panel:} Full best-fit SED extended to far-infrared wavelengths, decomposed into stellar (orange dotted), nebular (green dash-dotted), and dust emission (red dashed) components. The solid blue curve shows the total model flux density.
}
              \label{bestfit}%
    \end{figure*}

          \begin{figure}
   \centering
\includegraphics[width=.9\linewidth]{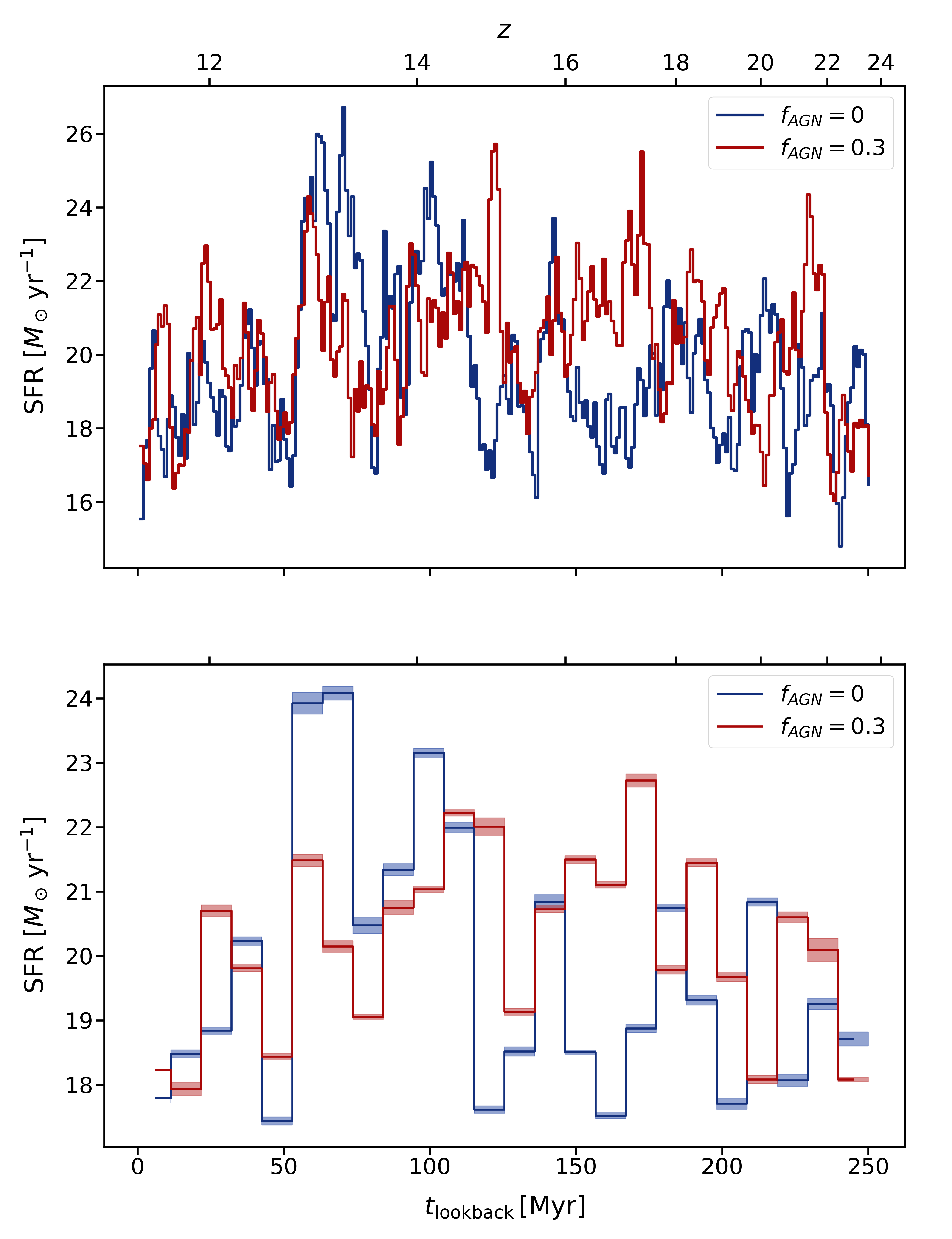}
   \caption{
   Star formation histories (SFHs) derived from the CIGALE modeling of EGS-z11-R0.
The blue curve shows the stochastic SFH corresponding to the best-fit solution
presented in Figure~\ref{bestfit}, where no AGN contribution is required.
The red curve represents the SFH obtained when forcing an AGN fraction of
$f_{\rm AGN}=0.3$ using the \texttt{fritz2006} module (see Table~4).
Top panel: instantaneous stochastic SFHs.
Bottom panel: same SFHs smoothed over a 10 Myr timescale (solid lines) and its associated bin error (shaded regions), computed as the standard error of the mean.
   It is worth noting that in Table \ref{tab:output-cigale-main} the value reported for each physical parameters are referred to the Bayesian analysis of CIGALE, turning into potential offset with the values of the best fit (as this SF history).
   }
              \label{Fig:SFH}%
    \end{figure}

          \begin{figure*}
   \centering
\includegraphics[width=.95\linewidth]{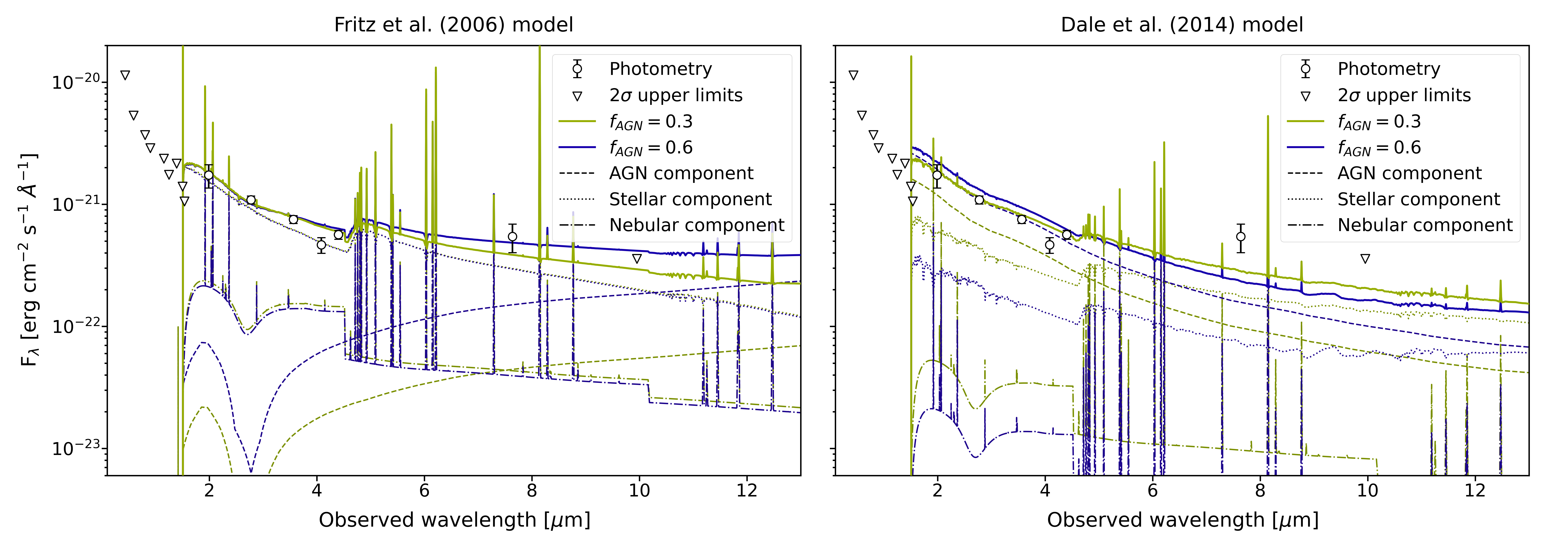}
   \caption{Same as Fig. \ref{bestfit}, but showing CIGALE SED models where the AGN fraction is fixed to different values ($f_{\mathrm{AGN}} = 0.3$ and $f_{\mathrm{AGN}} = 0.6$). 
   The observed photometry (circles) and $2\sigma$ upper limits (triangles) are compared with the total composite models (solid curves), while the individual AGN (dashed), stellar (dotted), and nebular (dash-dotted) components are shown separately to illustrate the impact of varying the AGN contribution on the overall SED shape. Left panel: best-fit SEDs with AGN component modeled using the \texttt{fritz2006} module. Right panel: best-fit SEDs with AGN component modeled using the \texttt{dale2014} module.
   }
              \label{Fig:sed_AGN}%
    \end{figure*}

\renewcommand{\arraystretch}{1.1}
\begin{table}
    \centering
    \caption{EGS-z11-R0 photometry.}
    \begin{tabular}{ccc}
    \hline\hline
        Intrument & Band & Photometry [nJy] \\
    \hline
        HST/ACS & F435W  (CANDELS) & $-1.74 \pm 3.56$ \\
        HST/ACS & F606W (CANDELS)  & $-0.11 \pm 3.12$ \\
        HST/ACS & F775W (CANDELS) & -- \\
        HST/ACS & F814W (CANDELS) & $-4.94 \pm 3.99$ \\
        HST/WFC3 & F105W (CANDELS) & -- \\
        HST/WFC3 & F125W (CANDELS) & $12.13 \pm 4.54$ \\
        HST/WFC3 & F140W (CANDELS) & $13.15 \pm 6.96$ \\
        HST/WFC3 & F160W (CANDELS) & $7.99 \pm 4.17$ \\
        JWST/NIRCam & F090W (UNICORN) & $0.25 \pm 3.91$ \\
        JWST/NIRCam & F115W (CEERS) & $8.76 \pm 5.25$ \\
        JWST/NIRCam & F150W (CEERS) & $11.30\pm 5.25$ \\
        JWST/NIRCam & F200W (CEERS) & $22.86 \pm 4.92$ \\
        JWST/NIRCam & F277W (CEERS) & $27.98 \pm 2.08$ \\
        JWST/NIRCam & F356W (CEERS) & $31.88 \pm 2.36$ \\
        JWST/NIRCam & F410M (CEERS)& $25.86 \pm 3.76$ \\
        JWST/NIRCam & F444W (CEERS)& $36.53 \pm 3.07$ \\
        JWST/MIRI & F770W (MEGA) & $106.12 \pm 27.79$ \\
        JWST/MIRI & F1000W (MEGA)& $55.53 \pm 58.99$ \\
        JWST/MIRI & F1500W (MEGA)& $-335.67 \pm 190.56$ \\
        JWST/MIRI & F2100W (MEGA)& $-756.08 \pm 548.76$ \\
    \hline
    \end{tabular}
\tablefoot{Photometric estimates and 1$\sigma$ uncertainties in the available HST and JWST bands for EGS-z11-R0.}
    \label{tab:egsz11-photometry}
\end{table}

\section{UV slope measurements}
\label{UV_slope}
Figure \ref{Fig:uvslope} places EGS-z11-R0 in the $M_{\rm UV}$--$\beta_{\rm UV}$ 
plane at $10 \lesssim z \lesssim 12$, together with recent 
photometric and spectroscopic measurements from the literature 
\citep{Cullen23,Cullen2024,Austin2025,Casey2024,Asada2026,Mitsuhashi2025}. The dashed relations 
represent the best-fit $\beta_{\rm UV}$--$M_{\rm UV}$ trends derived by 
Cullen et al. (2024) in the redshift intervals $10<z<11$ and 
$11<z<12$. 

Although the majority of galaxies at these epochs exhibit very blue 
rest-frame UV slopes ($\beta_{\rm UV} \lesssim -2$), consistent 
with dust-poor and metal-poor stellar populations, EGS-z11-R0 lies 
significantly above the canonical relations with 
$\beta_{\rm UV} > -1.5$. 
In Sect.~\ref{sec:nirspec} we derived the UV slope $\beta_{\rm UV}$ from the NIRSpec spectrum. As an independent check, we applied the same methodology to the NIRCam photometry. Interpolating the broadband fluxes over the same rest-frame wavelength range yields $\beta_{\rm UV} = -1.20 \pm 0.28$. This value is fully consistent with the spectroscopic estimate, although slightly bluer, and still places the source among galaxies with red UV continua. For clarity, Fig.~\ref{Fig:uvslope} shows both determinations of $\beta_{\rm UV}$, illustrating the uncertainty associated with the slope measurement.

Its location overlaps with the subset 
of ``red'' $z\sim12$ photometric candidates identified by 
\cite{Mitsuhashi2025}, highlighted in Figure \ref{Fig:uvslope} with orange starred 
symbols. These objects deviate markedly from the typical 
$M_{\rm UV}$-$\beta_{\rm UV}$ scaling relations and represent 
a rare population of red UV-continuum sources at cosmic dawn.
As discussed by \cite{Mitsuhashi2025}, such red slopes at 
$z>10$ cannot be easily explained by the age of the stellar population or 
metallicity alone, given the short age of the Universe 
($\sim 350$ Myr at $z\sim12$). Instead, two main physical 
mechanisms have been proposed: (i) significant dust attenuation 
%associated with rapid early dust growth (REF!), 
or (ii) strong nebular continuum emission powered by extremely hot, massive stars 
embedded in dense gas \citep[e.g.][]{Katz2025,Narayanan25b}.
%ADD AGN
%cite Narayanan24-25

In the dust-reddening scenario, values of 
$A_V \sim 0.5$--$1$ mag are required to reproduce 
$\beta_{\rm UV} \gtrsim -1.5$ \citep{Mitsuhashi2025}. 
Such extinction levels are substantially higher than the 
median values inferred for the general $z>10$ population, 
suggesting that these systems may represent a short-lived, 
obscured phase of rapid stellar mass assembly. In this 
framework, red monsters could be the dusty precursors of the 
so-called ``blue monsters'', in which  radiation-driven outflows have 
cleared the line of sight by displacing the dust and gas to distances 
much larger than the effective radius.

Alternatively, a nebular-continuum-dominated spectrum can also 
produce red UV slopes without invoking large dust masses. 
However, reproducing $\beta_{\rm UV} > -1.5$ through nebular 
emission requires extreme physical conditions, including 
high gas densities ($n_{\rm H} \gtrsim 10^4\,{\rm cm^{-3}}$) 
and stellar effective temperatures 
$T_{\rm eff} \gtrsim 5\times10^4$ K, as shown by the Cloudy 
modeling in \cite{Mitsuhashi2025}. These parameters imply 
ionizing photon production efficiencies significantly above 
those of standard stellar populations, potentially pointing 
toward very massive stars or metal-poor starbursts.

We will discuss these possible scenarios in more detail in the next sections.

\addtocounter{footnote}{-1} 
\footnotetext{\url{https://github.com/dancoe/Trilogy}}
\addtocounter{footnote}{1} 
\footnotetext{\url{https://www.gimp.org}}

\section{Emission lines: AGN vs. star formation} 
\label{sec:agnVSsfg}

The detection of multiple rest-frame UV emission lines in the
NIRSpec spectrum of EGS-z11-R0 allows us to investigate the
nature of its ionizing source. 
The diagnostic diagrams presented below require the He\,{\sc ii} flux alone. Throughout this analysis we conservatively assign the full upper limit measured for the unresolved He\,{\sc ii}+O\,{\sc iii}] blend to He\,{\sc ii}. 
This corresponds to the maximum allowed He\,{\sc ii} flux and therefore to the minimum allowed C\,{\sc iv}/He\,{\sc ii} and C\,{\sc iii}]/He\,{\sc ii} ratios. If a non-negligible fraction of the unresolved blend is instead contributed by O\,{\sc iii}], the corresponding upper limit on He\,{\sc ii} becomes lower, shifting EGS-z11-R0 toward the star-forming regions in all three He\,{\sc ii}-based diagnostic diagrams.

In Figure~\ref{diagnostics} we place EGS-z11
(red star) in a set of UV diagnostic diagrams commonly used
to distinguish among AGN-dominated, SF-dominated, and composite ionizing spectra. We compare
line ratios and equivalent widths (EW) with other spectroscopically
confirmed sources at $z \gtrsim 9$ \citep[e.g., GN-z11,
GHZ2, GHZ9, GS-z9-0,][]{Bunker2023B, Maiolino2024, Castellano2024, Curti2024}, following an approach similar to that
adopted by \cite{Napolitano2025b} for GHZ9.

In the top panel of Figure~\ref{diagnostics} we show the C\,{\sc iv}/C\,{\sc iii}]
versus C\,{\sc iii}]/He\,{\sc ii} diagram. These ratios are sensitive
to both the hardness of the ionizing radiation field and the
ionization parameter. The dashed demarcation curves indicate
the approximate separation between the AGN-dominated,
composite, and SF-dominated regimes, calibrated using
photoionization models \citep[e.g.,][]{Feltre2016, Gutkin2016, Hirschmann2019, Nakajima2022}.
EGS-z11-R0 lies in the composite region, consistent within the
uncertainties with both AGN and extreme stellar ionizing
spectra. This degeneracy mirrors what has been reported for
other $z>10$ systems (e.g., GN-z11 and GHZ9), where UV
line-ratio diagnostics alone do not provide a unique classification.

The top and bottom-left panels of Figure \ref{diagnostics} explore the
He\,{\sc ii}-based EW diagnostics, namely EW(C\,{\sc iv}) versus
C\,{\sc iv}/He\,{\sc ii} and EW(C\,{\sc iii}]) versus
C\,{\sc iii}]/He\,{\sc ii}, respectively.

As discussed
by \cite{Napolitano2025b}, the combination of large EWs
($\gtrsim 30$--$50$\,\AA) and elevated metal-to-He\,{\sc ii}
ratios generally requires a hard ionizing spectrum. In EGS-z11,
the observed EWs are significantly larger than typically
expected for standard stellar populations at sub-solar metallicity,
placing the source toward the upper envelope of the
SF-dominated models and overlapping with the composite
region. Pure stellar photoionization models can reproduce
such high EWs only under extreme conditions, including
very low metallicity and high ionization parameter
($\log U \gtrsim -2$), while AGN models naturally predict
strong high-ionization features.

However, unlike the case of GHZ9 \citep{Napolitano2025b},
EGS-z11-R0 does not show unambiguous signatures of a dominant
AGN, such as extremely high-ionization lines (e.g., [He\, {\sc ii}],  
[Ne\,{\sc iii}]) or rest-frame optical ratios firmly located in
the AGN regime. Instead, its position in the diagnostic planes
is consistent with a hard composite spectrum, potentially
arising from a combination of intense star formation and a
moderate AGN contribution.

The bottom-right panel of Figure \ref{diagnostics} presents the low-resolution UV
diagnostic proposed by \citet{Aravelo2025}, which explicitly
accounts for the unresolved He\,{\sc ii}+O\,{\sc iii}] blend and is
based on the equivalent widths of C\,{\sc iii}] and
He\,{\sc ii}+O\,{\sc iii}]. 
EGS-z11-R0 lies very close to the AGN/SF boundary, while
the upper limit on the unresolved blend extends toward the
star-forming region. This result is fully consistent with the other
three diagnostic panels and supports an interpretation of
a composite SF+AGN ionizing spectrum.

Taken together, these diagnostics support our interpretation of EGS-z11-R0 as a composite SF+AGN system, as discussed throughout this work.

The emission-line diagnostics presented in
Figure \ref{diagnostics} suggest that EGS-z11-R0 hosts a hard ionizing radiation
field, but they do not uniquely require an AGN-dominated
scenario. This interpretation is consistent with our SED
modeling (Section \ref{dec:sedfitting}), which favors a composite solution
with a non-negligible AGN fraction while still requiring
significant dust attenuation to reproduce the red UV slope.
In this framework, EGS-z11-R0 may represent a transitional
``red monster'' system in which rapid stellar mass assembly,
dust enrichment, and possible early black-hole growth occur
concurrently.

\section{SED fitting: physical parameters}
\label{dec:sedfitting}
To constrain the physical properties of EGS-z11, we performed
spectral energy distribution (SED) fitting using the available
HST, JWST/NIRCam, and JWST/MIRI photometry listed in
Table~1 and the NIRSpec spectrum. The best-fit solution is
shown in Figure~\ref{bestfit}, where the observed spectrum and
photometric data points are compared to the composite model.

We modeled the SED using \texttt{CIGALE} \citep{cigale-spec}, adopting a flexible
star-formation history (SFH) and including both stellar \citep{bc03} and
nebular emission \citep{inoue11, ferland+98, ferland+13} components. We assumed a Chabrier (2003)
initial mass function and explored sub-solar metallicities,
consistent with the spectroscopic constraints. Dust attenuation
was modeled using a \citet{Calzetti2000}-like law, allowing
for a range of $A_V$ values. In addition, we included an AGN
component parameterized by its fractional contribution to the
total infrared luminosity ($f_{\rm AGN}$). 
For the AGN component, we include both the \texttt{dale2014} \citep{Dale2014} and \texttt{fritz2006} \citep{Fritz06} models. We summarize our SED-fitting set-up and report the prior grids adopted in Table~\ref{tab:priors}.

The left panel of Figure~\ref{bestfit} shows the best-fit model overlayed
on the NIRSpec spectrum. The model successfully reproduces
the red UV continuum slope, the continuum normalization,
and the overall optical-to-mid-IR flux ratios. The right panel
displays the full SED decomposition, separating the stellar,
nebular, and dust components. The nebular contribution
(dash-dotted curve) is non-negligible in the rest-frame UV
and optical, but does not dominate the continuum shape.

The best-fit solution favors a stellar mass of
$\log (M_\star/M_\odot) \sim 8.5$--$9.0$, depending on the
assumed AGN fraction (see also Figure~\ref{Fig:sed_AGN}), placing EGS-z11
among the most massive known galaxies at $z>11$. The
inferred star-formation rate is of the order
${\rm SFR} \sim 10$--$30~M_\odot\,{\rm yr^{-1}}$, consistent
with rapid stellar mass assembly within the first few hundred
Myr of cosmic time. The SFH (Figure \ref{Fig:SFH}) indicates a sustained
or rising star-formation episode rather than a short,
instantaneous burst.

Importantly, the fit requires non-negligible dust attenuation
to reproduce the observed UV slope. The preferred value is
$A_V \sim 0.5$--$1.0$ mag, significantly higher than typical
values measured for the bulk of $z>10$ galaxies. Models with
negligible dust fail to simultaneously match the UV continuum
and the longer-wavelength NIRCam/MIRI photometry. 
In addition, the best-fit solution predicts a dust luminosity of $L_{IR} = (4.80 \pm 1.20) \times10^{11} L_\odot$.
These
results support the interpretation that EGS-z11-R0 is undergoing
an obscured phase of star formation, consistent with the
``red monster'' scenario discussed in previous sections.

\subsection{AGN dependence}
\label{sec:agn}
Figure \ref{Fig:sed_AGN} illustrates the impact of fixing the AGN fraction
($f_{\rm AGN}$) to progressively higher values in our CIGALE
modeling, while Table \ref{tab:output-cigale-main} summarizes the corresponding Bayesian
physical parameters for the different configurations. Although
the open-prior solution presented in Section \ref{dec:sedfitting} does not require
a dominant AGN component, the emission-line diagnostics
(Figure \ref{diagnostics}) allow for a composite stellar+AGN scenario.
We therefore explored forced solutions with $f_{\rm AGN}=0.3$
and $f_{\rm AGN}=0.6$ to assess how increasing AGN
contributions modify the inferred SED and physical parameters.
To this aim, as anticipated at the beginning of this Section, we tested two different AGN prescriptions available in CIGALE: the torus-based models of \cite{Fritz06}, shown in the left panel of Fig. \ref{Fig:sed_AGN}, and the empirical infrared templates of \cite{Dale2014}, shown in the right panel. We note that the AGN fraction has different definitions in the two considered models\footnote{$f_{\rm AGN}$(Dale)=$L_{\rm AGN}/L_{\rm dust}$; $f_{\rm AGN}$(Fritz)=$L_{\rm AGN}/(L_{\rm AGN}+L_{\rm dust})$}.

When forcing $f_{\rm AGN}=0.6$,  in both cases the best-fit is in tension with the observational constraints: the Fritz model exceeds the upper limits in the MIRI/F1000W band, while the Dale model outshines the UV range.
In general, the redistribution of bolometric
luminosity toward the AGN component suppresses the stellar
and dust-reprocessed emission required to match the broadband
photometry.
We also tested the outcome of an extreme case with $f_{\rm AGN}=1$ (AGN only without galaxy). Although this solution could still reproduce the UV, it largely fails in recovering the MIRI F770W, underestimating the optical continuum emission.
In contrast, the $f_{\rm AGN}=0.3$ solution remains broadly
consistent with the observed photometry and spectrum. The models
reproduce the NIRCam  data without violating the
MIRI upper limits. The overall SED shape remains similar to the
open-prior best-fit, indicating that a moderate AGN fraction
does not significantly distort the continuum balance.

The impact of these configurations on the inferred physical
parameters is summarized in Table \ref{tab:output-cigale-main}. 
Increasing the AGN fraction systematically reduces the stellar mass and star-formation
rate, as part of the total luminosity is attributed to
the AGN component rather than to stars. For the
\texttt{dale2014} run, the stellar mass decreases from
$\log(M_\star/M_\odot)=9.64\pm0.13$ (open prior) to
$9.20\pm0.14$ for $f_{\rm AGN}=0.3$, and further down to
$8.64\pm0.32$ for $f_{\rm AGN}=0.6$. This corresponds to
nearly an order-of-magnitude suppression of $M_\star$
in the high-AGN case. A similar trend is observed for the
SFR, which drops from $\sim 42~M_\odot~{\rm yr^{-1}}$
to $\sim 9.6$ and $\sim 3.9~M_\odot~{\rm yr^{-1}}$
for $f_{\rm AGN}=0.3$ and $0.6$, respectively.

A comparable behavior is seen when adopting the
\texttt{fritz2006} AGN module: while $f_{\rm AGN}=0.3$
yields stellar masses $\log(M_\star/M_\odot)\sim9.5$ and
SFRs of $\sim30~M_\odot~{\rm yr^{-1}}$, increasing the
AGN fraction to 0.6 again lowers the stellar mass and SFR.
Importantly, even in the AGN-inclusive runs, substantial
stellar reddening ($E(B-V)_\star \sim 0.25$--0.3) is still
required, indicating that dust attenuation remains necessary
to reproduce the red UV slope.

Overall, these tests demonstrate that while a moderate
AGN fraction ($\sim30\%$, in the dust component) is fully compatible with both the
broadband SED and the composite nature suggested by the
UV line diagnostics, a dominant AGN contribution
($\gtrsim60\%$) is disfavored by the data. The red UV slope
and mid-infrared flux of EGS-z11-R0 are more naturally explained
by a dust-enriched stellar population, possibly accompanied by
a secondary AGN component, rather than by an AGN-dominated
energy budget.

As a complementary test, we repeated the SED fitting using an updated \textsc{CIGALE} implementation that includes AGN-powered nebular emission (Appendix~\ref{app:agnnebular}). In this implementation, $f_{\rm AGN}$ measures the fraction of dust luminosity powered by the AGN and is not directly comparable to the AGN fractions in the standard runs above. The updated fits permit a low AGN-powered dust fraction but do not uniquely constrain the AGN contribution to the emission lines or the observed UV continuum. The inferred instantaneous SFR is also sensitive to the adopted treatment of AGN nebular emission.

\subsection{Comparison to literature high-$z$ red sources}
In a quantitative comparison with other recently reported
$z>10$ systems with red UV slopes, the physical properties of EGS-z11-R0 closely
match the photometric dust-enriched population identified by \cite{Mitsuhashi2025}. Those authors show that reproducing
$\beta_{\rm UV} \gtrsim -1.5$ requires dust attenuation of
$A_V \sim 0.5$-1 mag, stellar masses
$\log (M_\star/M_\odot) \sim 8$--9, and ionization parameters
$\log U \sim -2$ to $-1.5$. Our best-fit solution yields
$A_V \sim 1.0-1.3$, $\log (M_\star/M_\odot) \sim 8.6$-9.6,
and similarly elevated ionization conditions, placing
EGS-z11-R0 fully within the same parameter space. In addition,
the large rest-frame equivalent widths observed for high-ionization UV lines are consistent with the extreme nebular
conditions inferred by Mitsuhashi et al. (2025), although our
SED modeling indicates that dust attenuation remains the
primary driver of the red UV slope.

Compared to GHZ9 at $z=10.145$ studied by \cite{Napolitano2025b}, several similarities and differences
emerge. GHZ9 has a stellar mass of
$M_\star \sim (3$--$7)\times10^8\,M_\odot$ and nebular
reddening $E(B-V) \sim 0.3$ (corresponding to
$A_V \sim 1.2$ for a Calzetti law), comparable to the values
derived here. Its ionization parameter
($\log U \sim -1.9$ to $-1.6$) and extreme UV EWs
(e.g., EW(C\,{\sc iv}) $\sim 65$\,\AA,
EW(C\,{\sc iii}]) $\sim 48$\,\AA)
require a very hard radiation field and are accompanied by
an X-ray detection implying a black-hole mass of order
$10^8\,M_\odot$. In contrast, while EGS-z11-R0 shows similarly
elevated EWs and attenuation  (but larger stellar mass),
we lack independent AGN tracers such as X-ray emission,
and our line-ratio diagnostics remain consistent with a
composite or extreme star-forming solution. 

This comparison suggests that EGS-z11-R0 may
occupy a region of parameter space similar to that of GHZ9 in terms
of stellar mass, ionization state, and dust content, but with a
less dominant or more obscured AGN contribution. In the
framework proposed for ``red monsters'', EGS-z11-R0 could
represent either a dust-enshrouded phase preceding efficient
AGN feedback or a system in which intense star formation
alone is capable of producing the observed hard ionizing
spectrum.

Overall, the best-fit SED solution points toward a compact,
massive, dust-enriched galaxy at $z\sim11.5$, possibly hosting
a moderate AGN contribution during a phase of rapid early
assembly.

\renewcommand{\arraystretch}{1.25}
\begin{table*}
    \centering
    \caption{Physical parameters of EGS-z11-R0 determined by CIGALE Bayesian analysis with different AGN configurations.}
    \begin{tabular}{cccccc}
        \hline\hline
        Parameter & \multicolumn{3}{c}{\texttt{dale2014}} & \multicolumn{2}{c}{\texttt{fritz2006}} \\
        \hline
         & $f_{AGN} \simeq 0$~$^*$ & $f_{AGN} = 0.3$ & $f_{AGN} = 0.6$ & $f_{AGN} = 0.3$ & $f_{AGN} = 0.6$ \\
        \hline
        log$(M_\star/M_\odot)$ &    9.64 $\pm$ 0.13     & 9.20 $\pm$ 0.14   & 8.64 $\pm$ 0.32   & 9.48 $\pm$ 0.18   & 9.29 $\pm$ 0.39 \\
        SFR [$M_\odot$ yr$^{-1}$] & 42.05 $\pm$ 14.52   & 9.64 $\pm$ 1.67   & 3.86 $\pm$ 0.76   & 30.17 $\pm$ 16.07  & 14.64 $\pm$ 5.49 \\
        E(B-V)$_\star$ [mag] &      0.33 $\pm$ 0.03     & 0.29 $\pm$ 0.09   & 0.41 $\pm$ 0.43   & 0.28 $\pm$ 0.08   & 0.25 $\pm$ 0.24 \\
        E(B-V)$_{AGN}$ [mag] &      --                  & --                & --                & 0.32 $\pm$ 0.17   & 0.29 $\pm$ 0.17 \\
        E(B-V)$_{neb}$ [mag] &      0.73 $\pm$ 0.12     & 0.50 $\pm$ 0.29   & 0.65 $\pm$ 0.56   & 0.58 $\pm$ 0.19   & 0.48 $\pm$ 0.31 \\
        \hline
    \end{tabular}
    \tablefoot{$^*$: the physical parameters of this configuration are referred to the run with open priors.}
    \label{tab:output-cigale-main}
\end{table*}

\section{Discussion} 
\subsection{Early dust and the emergence of red monsters at $z > 10$}

The spectroscopic confirmation of EGS-z11-R0 demonstrates that
dust-enriched galaxies already existed at $z \simeq 11.5$, only
$\sim400$ Myr after the Big Bang. Its red UV slope
($\beta_{\rm UV} \sim -1$) and the dust attenuation required by
our SED modeling ($A_V \sim 1$ mag) place the system well
above the canonical $M_{\rm UV}$–$\beta_{\rm UV}$ relation followed by
the majority of $z > 10$ galaxies, which typically exhibit very
blue continua consistent with dust-poor stellar populations.
Together with other recently reported candidates
(e.g., \citealt{Mitsuhashi2025,Donnan25}), EGS-z11-R0 strengthens
the emerging picture that a population of dust-reddened galaxies
may already be present at cosmic dawn.

This result carries important implications for the origin of dust
in the early Universe. Several recent theoretical models predict
that the bulk of dust mass growth in galaxies occurs through
grain growth in the interstellar medium rather than through
direct stellar production. In these models, efficient dust
enrichment requires the ISM to reach sufficiently high
metallicities and densities, implying a delayed rise of the dust
content during the first stages of galaxy evolution
(e.g. \citealt{Narayanan2025}). As a consequence, such
frameworks generally predict that heavily obscured galaxies
should be extremely rare, or absent, at $z \gtrsim 10$.

The discovery of EGS-z11, together with the dusty candidate
reported by \citet{Donnan25}, challenges this expectation.

An alternative interpretation involves the geometry and
dynamical state of the dusty interstellar medium. In the
framework proposed by \citet{Ferrara2024}, early galaxies may
experience a short-lived ``red monster'' phase during which
rapid star formation proceeds within a dust-enshrouded
environment. During this stage, radiation pressure builds up and
eventually drives powerful outflows that expel gas and dust
from the central regions, revealing the young stellar population
and producing the UV-bright ``blue monsters'' commonly
observed at $z > 10$.

Several observational properties of EGS-z11-R0 are consistent with
this scenario. First, the significant dust attenuation inferred
from the SED modeling indicates that the galaxy is observed
during a heavily obscured phase of its evolution. Second, the
star-formation histories derived from the CIGALE modeling
(Fig. \ref{Fig:SFH}) show a rising or approximately constant star-formation
rate rather than a rapidly declining burst. This behavior is
remarkably consistent with the predictions of the
feedback-regulated model of \citet{Ferrara2024}, in which the
dust-obscured phase is characterized by sustained star formation
lasting a substantial fraction of the lifetime of the galaxy. In this
framework, a large fraction of the stellar mass is assembled
during the obscured stage before radiation-driven outflows
disperse the dust and transition the system into the
UV-bright phase.

These results imply that the earliest galaxies may
undergo phases of substantial dust obscuration much earlier than
predicted by models in which dust growth is primarily regulated
by slow ISM processes. If systems such as EGS-z11-R0 represent a
common stage of early galaxy evolution, a non-negligible
fraction of star formation at $z > 10$ may occur in dusty
environments that remain difficult to detect through rest-frame
UV surveys alone. 

\subsection{Possible AGN contribution and implications for early black-hole seeds}

Although the broadband SED and UV line diagnostics are broadly
consistent with a star-formation system, the presence of a
composite stellar+AGN spectrum remains a viable possibility
for EGS-z11-R0 (Sect. \ref{sec:agn}). If part of the observed UV continuum
originates from accretion onto a central black hole, the required
black-hole mass can be estimated from the observed rest-frame
UV flux.
Adopting the observed continuum level
%$F_{\rm UV}\sim 2\times10^{-21}$ erg s$^{-1}$ cm$^{-2}$ \AA$^{-1}$,
and assuming that the AGN radiates at the Eddington limit,
the corresponding black-hole mass required to reproduce the
observed UV emission would be on the order of  
$M_{\rm BH}\simeq3\times10^{5}\,M_\odot$ (in the case of an
assumed AGN contribution to the observed UV continuum (not directly specified by $f_{\rm AGN}$)). 
Such a mass scale is particularly intriguing in the context of
early black-hole seed formation. Black holes with masses of
$\sim10^{5}\,M_\odot$ fall within the range expected for
massive seed channels such as direct-collapse black holes
(DCBHs), as well as primordial black holes (PBHs) that may
have formed in the early Universe. In particular, recent work
by \citet{matteri1} has explored scenarios in which a
population of PBHs with characteristic masses of
$\sim10^{4}$–$10^{5}\,M_\odot$ contributes to the UV luminosity
of super-early galaxies through accretion-powered emission.
In these models, even a small fraction of halos hosting
accreting PBHs can significantly boost the UV luminosity of
high-redshift galaxies and help reproduce the observed
luminosity function at $z\gtrsim10$.

Within this framework, a system such as EGS-z11-R0 could
naturally host an accreting black hole that contributes a fraction
of the observed UV continuum. 

In the context of the radiative-feedback model of
\citet{Ferrara2024}, the possible presence of an accreting black
hole may also be connected to the ``red monster'' phase of
galaxy evolution. 
If a central black hole is already present during this stage,
its accretion luminosity could contribute to the hard ionizing
radiation field and to the observed UV continuum. In this
interpretation, EGS-z11-R0 may represent a system caught during
the early obscured growth phase in which both rapid star
formation and black-hole accretion simultaneously build
 the stellar and black-hole mass of the galaxy.

\section{Summary and Conclusions}

We report the spectroscopic confirmation of EGS-z11-R0 at $z=11.452\pm0.021$ using JWST/NIRSpec observations of the CEERS field. Our main results can be summarized as follows:

\begin{itemize}

\item 
The UV slope of EGS-z11-R0 places it well above the canonical relation $M_{\rm UV}$ - $\beta_{\rm UV}$ followed by the majority galaxies at $z>10$, which typically exhibit very blue and dust-poor continua. The required attenuation ($A_V\sim1$ mag) demonstrates that significant dust reservoirs were already present only $\sim400$ Myr after the Big Bang.

\item 
SED modeling combining spectroscopy with HST, NIRCam, and MIRI photometry yields a stellar mass $\log(M_\star/M_\odot)\sim9.2$–9.6 and a star-formation rate of $\sim10$–40 $M_\odot\,{\rm yr^{-1}}$. The inferred star-formation history is rising or approximately constant, indicating sustained stellar mass growth rather than a short-lived burst.

\item 
The significant dust attenuation, compact morphology, and rising star-formation history are consistent with the radiative-feedback scenario proposed by \cite{Ferrara2024} in which early galaxies undergo a short-lived dust-obscured ``red monster'' phase during which a large fraction of their stellar mass forms before radiation-driven outflows disperse the dust and reveal the UV-bright system.

\item   
Although the data do not require a dominant AGN component, the emission-line diagnostics and SED modeling remain compatible with a composite stellar+AGN spectrum. If part of the UV emission arises from Eddington-limited accretion, the implied black-hole mass would be $M_{\rm BH}\sim10^5\,M_\odot$, a value intriguingly consistent with the expected mass range of massive early black-hole seeds such as direct-collapse black holes or primordial black holes.

\end{itemize}

 Future JWST spectroscopy and deeper mid-infrared, sub-millimeter and radio observations will be crucial to establish the prevalence of these early dust-rich galaxies and to clarify their role in the earliest phases of galaxy and black-hole growth.
   
\begin{acknowledgements}
 We thank the anonymous referee for their careful reading of the manuscript and for their constructive comments and suggestions, which helped to improve the quality and clarity of this work. 
 G.R., P.C., G.G. and A.G. are supported  by the European Union -- NextGeneration EU RFF M4C2 1.1 PRIN 2022 project 2022ZSL4BL INSIGHT. 
This work is also supported by INAF RF2024 Large Grant "UNDUST: UNveiling the Dawn of the Universe with JWST" (P.I. P. Santini).
P.G.P.-G. acknowledges support from grant PID2022-139567NB-I00 funded by Spanish Ministerio de Ciencia, Innovaci\'on y Universidades MCIU/AEI/10.13039/501100011033,
FEDER {\it Una manera de hacer
Europa}. 
The authors thank the PI, Erica Nelson, and the team of the JWST
GO proposal 4106 for making their data public. This work is based on one of the fillers of this program.
We also thank Gabriel Brammer for spotting an error in the GRISM spectrum  that we showed in the first version of this paper.
MC acknowledges INAF GO Grant 2024 ”Revealing the nature of bright galaxies at cosmic dawn with deep JWST spectroscopy”.
This work is based in part on observations made with the NASA/ESA/CSA James Webb Space Telescope. The data were obtained from the Mikulski Archive for Space Telescopes at the Space Telescope Science Institute, which is operated by the Association of Universities for Research in Astronomy, Inc., under NASA contract NAS5-03127 for JWST. These observations are associated with program \#4106.
\end{acknowledgements}

%-------------------------------------------------------------------

\bibliographystyle{aa}
\bibliography{aa.bib}

@ARTICLE{Calzetti2000,
       author = {{Calzetti}, Daniela and {Armus}, Lee and {Bohlin}, Ralph C. and {Kinney}, Anne L. and {Koornneef}, Jan and {Storchi-Bergmann}, Thaisa},
        title = "{The Dust Content and Opacity of Actively Star-forming Galaxies}",
      journal = {\apj},
         year = 2000,
        month = apr,
       volume = {533},
       number = {2},
        pages = {682-695},
          doi = {10.1086/308692},
archivePrefix = {arXiv},
       eprint = {astro-ph/9911459},
 primaryClass = {astro-ph},
       adsurl = {https://ui.adsabs.harvard.edu/abs/2000ApJ...533..682C}
}

@ARTICLE{Santini2026,
       author = {{Santini}, P. and {Castellano}, M. and {Calabr{\`o}}, A. and {Fontana}, A. and {Merlin}, E. and {Bevacqua}, D. and {Bergamini}, P. and {Boquien}, M. and {Cantarella}, S. and {Ciesla}, L. and {Ferrara}, A. and {Finkelstein}, S.~L. and {Fortuni}, F. and {Gandolfi}, G. and {Gasparetto}, T. and {Giallongo}, E. and {Grogin}, N.~A. and {Guida}, S.~T. and {Koekemoer}, A.~M. and {Menci}, N. and {Napolitano}, L. and {Paris}, D. and {Pentericci}, L. and {Perez-Diaz}, B. and {Stoyanova}, B. and {Treu}, T.},
        title = "{The early maturity of high-redshift galaxies: Insights from sSFR, M/L, and SFHs at z {\ensuremath{\sim}} 7─14}",
      journal = {\aap},
         year = 2026,
        month = jun,
       volume = {710},
          eid = {A249},
        pages = {A249},
          doi = {10.1051/0004-6361/202558516},
       adsurl = {https://ui.adsabs.harvard.edu/abs/2026A&A...710A.249S}
}

@ARTICLE{Carniani2025,
       author = {{Carniani}, Stefano and {D'Eugenio}, Francesco and {Ji}, Xihan and {Parlanti}, Eleonora and {Scholtz}, Jan and {Sun}, Fengwu and {Venturi}, Giacomo and {Bakx}, Tom J.~L.~C. and {Curti}, Mirko and {Maiolino}, Roberto and {Tacchella}, Sandro and {Zavala}, Jorge A. and {Hainline}, Kevin and {Witstok}, Joris and {Johnson}, Benjamin D. and {Alberts}, Stacey and {Bunker}, Andrew J. and {Charlot}, St{\'e}phane and {Eisenstein}, Daniel J. and {Helton}, Jakob M. and {Jakobsen}, Peter and {Kumari}, Nimisha and {Robertson}, Brant and {Saxena}, Aayush and {{\"U}bler}, Hannah and {Williams}, Christina C. and {Willmer}, Christopher N.~A. and {Willott}, Chris},
        title = "{The eventful life of a luminous galaxy at z = 14: metal enrichment, feedback, and low gas fraction?}",
      journal = {\aap},
         year = 2025,
        month = apr,
       volume = {696},
          eid = {A87},
        pages = {A87},
          doi = {10.1051/0004-6361/202452451},
archivePrefix = {arXiv},
       eprint = {2409.20533},
 primaryClass = {astro-ph.GA},
       adsurl = {https://ui.adsabs.harvard.edu/abs/2025A&A...696A..87C}
}

@ARTICLE{Castellano2022,
       author = {{Castellano}, Marco and {Fontana}, Adriano and {Treu}, Tommaso and {Santini}, Paola and {Merlin}, Emiliano and {Leethochawalit}, Nicha and {Trenti}, Michele and {Vanzella}, Eros and {Mestric}, Uros and {Bonchi}, Andrea and {Belfiori}, Davide and {Nonino}, Mario and {Paris}, Diego and {Polenta}, Gianluca and {Roberts-Borsani}, Guido and {Boyett}, Kristan and {Brada{\v{c}}}, Maru{\v{s}}a and {Calabr{\`o}}, Antonello and {Glazebrook}, Karl and {Grillo}, Claudio and {Mascia}, Sara and {Mason}, Charlotte and {Mercurio}, Amata and {Morishita}, Takahiro and {Nanayakkara}, Themiya and {Pentericci}, Laura and {Rosati}, Piero and {Vulcani}, Benedetta and {Wang}, Xin and {Yang}, Lilan},
        title = "{Early Results from GLASS-JWST. III. Galaxy Candidates at z  9-15}",
      journal = {\apjl},
         year = 2022,
        month = oct,
       volume = {938},
       number = {2},
          eid = {L15},
        pages = {L15},
          doi = {10.3847/2041-8213/ac94d0},
archivePrefix = {arXiv},
       eprint = {2207.09436},
 primaryClass = {astro-ph.GA},
       adsurl = {https://ui.adsabs.harvard.edu/abs/2022ApJ...938L..15C}
}

@ARTICLE{Napolitano2024,
       author = {{Napolitano}, L. and {Pentericci}, L. and {Santini}, P. and {Calabr{\`o}}, A. and {Mascia}, S. and {Llerena}, M. and {Castellano}, M. and {Dickinson}, M. and {Finkelstein}, S.~L. and {Amor{\'\i}n}, R. and {Arrabal Haro}, P. and {Bagley}, M. and {Bhatawdekar}, R. and {Cleri}, N.~J. and {Davis}, K. and {Gardner}, J.~P. and {Gawiser}, E. and {Giavalisco}, M. and {Hathi}, N. and {Holwerda}, B.~W. and {Hu}, W. and {Jung}, I. and {Kartaltepe}, J.~S. and {Koekemoer}, A.~M. and {Larson}, R.~L. and {Merlin}, E. and {Mobasher}, B. and {Papovich}, C. and {Park}, H. and {Pirzkal}, N. and {Trump}, J.~R. and {Wilkins}, S.~M. and {Yung}, L.~Y.~A.},
        title = "{Peering into cosmic reionization: Ly{\ensuremath{\alpha}} visibility evolution from galaxies at z = 4.5-8.5 with JWST}",
      journal = {\aap},
         year = 2024,
        month = aug,
       volume = {688},
          eid = {A106},
        pages = {A106},
          doi = {10.1051/0004-6361/202449644},
archivePrefix = {arXiv},
       eprint = {2402.11220},
 primaryClass = {astro-ph.GA},
       adsurl = {https://ui.adsabs.harvard.edu/abs/2024A&A...688A.106N}
}

@ARTICLE{PerezMontero2023,
       author = {{P{\'e}rez-Montero}, E. and {Amor{\'\i}n}, R. and {P{\'e}rez-D{\'\i}az}, B. and {V{\'\i}lchez}, J.~M. and {Garc{\'\i}a-Benito}, R.},
        title = "{Assessing model-based carbon and oxygen abundance derivation from ultraviolet emission lines in AGNs}",
      journal = {\mnras},
         year = 2023,
        month = may,
       volume = {521},
       number = {1},
        pages = {1556-1569},
          doi = {10.1093/mnras/stad621},
archivePrefix = {arXiv},
       eprint = {2302.12560},
 primaryClass = {astro-ph.GA},
       adsurl = {https://ui.adsabs.harvard.edu/abs/2023MNRAS.521.1556P}
}

@ARTICLE{Feltre2012,
       author = {{Feltre}, A. and {Hatziminaoglou}, E. and {Fritz}, J. and {Franceschini}, A.},
        title = "{Smooth and clumpy dust distributions in AGN: a direct comparison of two commonly explored infrared emission models}",
      journal = {\mnras},
         year = 2012,
        month = oct,
       volume = {426},
       number = {1},
        pages = {120-127},
          doi = {10.1111/j.1365-2966.2012.21695.x},
archivePrefix = {arXiv},
       eprint = {1207.2668},
 primaryClass = {astro-ph.CO},
       adsurl = {https://ui.adsabs.harvard.edu/abs/2012MNRAS.426..120F}
}

@ARTICLE{Berg2018,
       author = {{Berg}, Danielle A. and {Erb}, Dawn K. and {Auger}, Matthew W. and {Pettini}, Max and {Brammer}, Gabriel B.},
        title = "{A Window on the Earliest Star Formation: Extreme Photoionization Conditions of a High-ionization, Low-metallicity Lensed Galaxy at z {\ensuremath{\sim}} 2*}",
      journal = {\apj},
         year = 2018,
        month = jun,
       volume = {859},
       number = {2},
          eid = {164},
        pages = {164},
          doi = {10.3847/1538-4357/aab7fa},
archivePrefix = {arXiv},
       eprint = {1803.02340},
 primaryClass = {astro-ph.GA},
       adsurl = {https://ui.adsabs.harvard.edu/abs/2018ApJ...859..164B}
}

@ARTICLE{Nanayakkara2019,
       author = {{Nanayakkara}, Themiya and {Brinchmann}, Jarle and {Boogaard}, Leindert and {Bouwens}, Rychard and {Cantalupo}, Sebastiano and {Feltre}, Anna and {Kollatschny}, Wolfram and {Marino}, Raffaella Anna and {Maseda}, Michael and {Matthee}, Jorryt and {Paalvast}, Mieke and {Richard}, Johan and {Verhamme}, Anne},
        title = "{Exploring He II {\ensuremath{\lambda}}1640 emission line properties at z {\ensuremath{\sim}}2-4}",
      journal = {\aap},
         year = 2019,
        month = apr,
       volume = {624},
          eid = {A89},
        pages = {A89},
          doi = {10.1051/0004-6361/201834565},
archivePrefix = {arXiv},
       eprint = {1902.05960},
 primaryClass = {astro-ph.GA},
       adsurl = {https://ui.adsabs.harvard.edu/abs/2019A&A...624A..89N}
}

@ARTICLE{Weida2024,
       author = {{Hu}, Weida and {Papovich}, Casey and {Dickinson}, Mark and {Kennicutt}, Robert and {Shen}, Lu and {Amor{\'\i}n}, Ricardo O. and {Arrabal Haro}, Pablo and {Bagley}, Micaela B. and {Bhatawdekar}, Rachana and {Cleri}, Nikko J. and {Cole}, Justin W. and {Dekel}, Avishai and {de la Vega}, Alexander and {Finkelstein}, Steven L. and {Grogin}, Norman A. and {Hathi}, Nimish P. and {Hirschmann}, Michaela and {Holwerda}, Benne W. and {Hutchison}, Taylor A. and {Jung}, Intae and {Koekemoer}, Anton M. and {Kartaltepe}, Jeyhan S. and {Lucas}, Ray A. and {Llerena}, Mario and {Mascia}, S. and {Mobasher}, Bahram and {Napolitano}, L. and {Newman}, Jeffrey A. and {Pentericci}, Laura and {P{\'e}rez-Gonz{\'a}lez}, Pablo G. and {Trump}, Jonathan R. and {Wilkins}, Stephen M. and {Yung}, L.~Y. Aaron},
        title = "{Characterizing the Average Interstellar Medium Conditions of Galaxies at z {\ensuremath{\sim}} 5.6─9 with Ultraviolet and Optical Nebular Lines}",
      journal = {\apj},
         year = 2024,
        month = aug,
       volume = {971},
       number = {1},
          eid = {21},
        pages = {21},
          doi = {10.3847/1538-4357/ad5015},
archivePrefix = {arXiv},
       eprint = {2401.12402},
 primaryClass = {astro-ph.GA},
       adsurl = {https://ui.adsabs.harvard.edu/abs/2024ApJ...971...21H}
}

@ARTICLE{Aravelo2025,
       author = {{Arevalo-Gonzalez}, Flor and {Braun}, Titanilla and {Trussler}, James and {Conselice}, Christopher J. and {Harvey}, Thomas and {Adams}, Nathan and {Austin}, Duncan and {Li}, Qiong and {Juod{\v{z}}balis}, Ignas and {Nakajima}, Kimihiko},
        title = "{New methods of identifying AGN in the early Universe using spectroscopy and photometry in the JWST era}",
      journal = {\mnras},
         year = 2025,
        month = dec,
       volume = {544},
       number = {3},
        pages = {2737-2757},
          doi = {10.1093/mnras/staf1836},
archivePrefix = {arXiv},
       eprint = {2501.09585},
 primaryClass = {astro-ph.GA},
       adsurl = {https://ui.adsabs.harvard.edu/abs/2025MNRAS.544.2737A}
}

@ARTICLE{Dekel2023,
       author = {{Dekel}, Avishai and {Sarkar}, Kartick C. and {Birnboim}, Yuval and {Mandelker}, Nir and {Li}, Zhaozhou},
        title = "{Efficient formation of massive galaxies at cosmic dawn by feedback-free starbursts}",
      journal = {\mnras},
         year = 2023,
        month = aug,
       volume = {523},
       number = {3},
        pages = {3201-3218},
          doi = {10.1093/mnras/stad1557},
archivePrefix = {arXiv},
       eprint = {2303.04827},
 primaryClass = {astro-ph.GA},
       adsurl = {https://ui.adsabs.harvard.edu/abs/2023MNRAS.523.3201D}
}

@ARTICLE{Somerville2025,
       author = {{Somerville}, Rachel S. and {Yung}, L.~Y. Aaron and {Lancaster}, Lachlan and {Menon}, Shyam and {Sommovigo}, Laura and {Finkelstein}, Steven L.},
        title = "{Density-modulated star formation efficiency: implications for the observed abundance of ultraviolet luminous galaxies at z > 10}",
      journal = {\mnras},
         year = 2025,
        month = dec,
       volume = {544},
       number = {4},
        pages = {3774-3798},
          doi = {10.1093/mnras/staf1824},
archivePrefix = {arXiv},
       eprint = {2505.05442},
 primaryClass = {astro-ph.GA},
       adsurl = {https://ui.adsabs.harvard.edu/abs/2025MNRAS.544.3774S}
}

@ARTICLE{Chavez2026,
       author = {{Marques-Chaves}, R. and {{\'A}lvarez-M{\'a}rquez}, J. and {Colina}, L. and {Kendrew}, S. and {Abdurro'uf} and {Blanco-Prieto}, C. and {Boogaard}, L.~A. and {Castellano}, M. and {Caputi}, K.~I. and {Crespo-G{\'o}mez}, A. and {Fontana}, A. and {Fudamoto}, Y. and {Fujimoto}, S. and {Garc{\'\i}a-Mar{\'\i}n}, M. and {Harikane}, Y. and {Harish}, S. and {Hashimoto}, T. and {Hsiao}, T. and {Iani}, E. and {Inoue}, A.~K. and {Langeroodi}, D. and {Lin}, R. and {Melinder}, J. and {Napolitano}, L. and {{\"O}stlin}, G. and {P{\'e}rez-Gonz{\'a}lez}, P.~G. and {Prieto-Jim{\'e}nez}, C. and {Rinaldi}, P. and {Rodr{\'\i}guez Del Pino}, B. and {Santini}, P. and {Sugahara}, Y. and {Varo-O'Ferrall}, A. and {Wright}, G. and {Zavala}, J.},
        title = "{PRISMS: U37126, a very blue ISM-naked starburst at z = 10.255 with a nearly 100\% Lyman continuum escape fraction}",
      journal = {\aap},
         year = 2026,
        month = jul,
       volume = {711},
          eid = {A301},
        pages = {A301},
          doi = {10.1051/0004-6361/202659281},
archivePrefix = {arXiv},
       eprint = {2602.02322},
 primaryClass = {astro-ph.GA},
       adsurl = {https://ui.adsabs.harvard.edu/abs/2026A&A...711A.301M}
}

@ARTICLE{Chabrier,
       author = {{Chabrier}, Gilles},
        title = "{Galactic Stellar and Substellar Initial Mass Function}",
      journal = {\pasp},
         year = 2003,
        month = jul,
       volume = {115},
       number = {809},
        pages = {763-795},
          doi = {10.1086/376392},
archivePrefix = {arXiv},
       eprint = {astro-ph/0304382},
 primaryClass = {astro-ph},
       adsurl = {https://ui.adsabs.harvard.edu/abs/2003PASP..115..763C}
}

@ARTICLE{2020A&A...641A...6P,
       author = {{Planck Collaboration} and {Aghanim}, N. and {Akrami}, Y. and {Ashdown}, M. and {Aumont}, J. and {Baccigalupi}, C. and {Ballardini}, M. and {Banday}, A.~J. and {Barreiro}, R.~B. and {Bartolo}, N. and {Basak}, S. and {Battye}, R. and {Benabed}, K. and {Bernard}, J.-P. and {Bersanelli}, M. and {Bielewicz}, P. and {Bock}, J.~J. and {Bond}, J.~R. and {Borrill}, J. and {Bouchet}, F.~R. and {Boulanger}, F. and {Bucher}, M. and {Burigana}, C. and {Butler}, R.~C. and {Calabrese}, E. and {Cardoso}, J.-F. and {Carron}, J. and {Challinor}, A. and {Chiang}, H.~C. and {Chluba}, J. and {Colombo}, L.~P.~L. and {Combet}, C. and {Contreras}, D. and {Crill}, B.~P. and {Cuttaia}, F. and {de Bernardis}, P. and {de Zotti}, G. and {Delabrouille}, J. and {Delouis}, J.-M. and {Di Valentino}, E. and {Diego}, J.~M. and {Dor{\'e}}, O. and {Douspis}, M. and {Ducout}, A. and {Dupac}, X. and {Dusini}, S. and {Efstathiou}, G. and {Elsner}, F. and {En{\ss}lin}, T.~A. and {Eriksen}, H.~K. and {Fantaye}, Y. and {Farhang}, M. and {Fergusson}, J. and {Fernandez-Cobos}, R. and {Finelli}, F. and {Forastieri}, F. and {Frailis}, M. and {Fraisse}, A.~A. and {Franceschi}, E. and {Frolov}, A. and {Galeotta}, S. and {Galli}, S. and {Ganga}, K. and {G{\'e}nova-Santos}, R.~T. and {Gerbino}, M. and {Ghosh}, T. and {Gonz{\'a}lez-Nuevo}, J. and {G{\'o}rski}, K.~M. and {Gratton}, S. and {Gruppuso}, A. and {Gudmundsson}, J.~E. and {Hamann}, J. and {Handley}, W. and {Hansen}, F.~K. and {Herranz}, D. and {Hildebrandt}, S.~R. and {Hivon}, E. and {Huang}, Z. and {Jaffe}, A.~H. and {Jones}, W.~C. and {Karakci}, A. and {Keih{\"a}nen}, E. and {Keskitalo}, R. and {Kiiveri}, K. and {Kim}, J. and {Kisner}, T.~S. and {Knox}, L. and {Krachmalnicoff}, N. and {Kunz}, M. and {Kurki-Suonio}, H. and {Lagache}, G. and {Lamarre}, J.-M. and {Lasenby}, A. and {Lattanzi}, M. and {Lawrence}, C.~R. and {Le Jeune}, M. and {Lemos}, P. and {Lesgourgues}, J. and {Levrier}, F. and {Lewis}, A. and {Liguori}, M. and {Lilje}, P.~B. and {Lilley}, M. and {Lindholm}, V. and {L{\'o}pez-Caniego}, M. and {Lubin}, P.~M. and {Ma}, Y.-Z. and {Mac{\'\i}as-P{\'e}rez}, J.~F. and {Maggio}, G. and {Maino}, D. and {Mandolesi}, N. and {Mangilli}, A. and {Marcos-Caballero}, A. and {Maris}, M. and {Martin}, P.~G. and {Martinelli}, M. and {Mart{\'\i}nez-Gonz{\'a}lez}, E. and {Matarrese}, S. and {Mauri}, N. and {McEwen}, J.~D. and {Meinhold}, P.~R. and {Melchiorri}, A. and {Mennella}, A. and {Migliaccio}, M. and {Millea}, M. and {Mitra}, S. and {Miville-Desch{\^e}nes}, M.-A. and {Molinari}, D. and {Montier}, L. and {Morgante}, G. and {Moss}, A. and {Natoli}, P. and {N{\o}rgaard-Nielsen}, H.~U. and {Pagano}, L. and {Paoletti}, D. and {Partridge}, B. and {Patanchon}, G. and {Peiris}, H.~V. and {Perrotta}, F. and {Pettorino}, V. and {Piacentini}, F. and {Polastri}, L. and {Polenta}, G. and {Puget}, J.-L. and {Rachen}, J.~P. and {Reinecke}, M. and {Remazeilles}, M. and {Renzi}, A. and {Rocha}, G. and {Rosset}, C. and {Roudier}, G. and {Rubi{\~n}o-Mart{\'\i}n}, J.~A. and {Ruiz-Granados}, B. and {Salvati}, L. and {Sandri}, M. and {Savelainen}, M. and {Scott}, D. and {Shellard}, E.~P.~S. and {Sirignano}, C. and {Sirri}, G. and {Spencer}, L.~D. and {Sunyaev}, R. and {Suur-Uski}, A.-S. and {Tauber}, J.~A. and {Tavagnacco}, D. and {Tenti}, M. and {Toffolatti}, L. and {Tomasi}, M. and {Trombetti}, T. and {Valenziano}, L. and {Valiviita}, J. and {Van Tent}, B. and {Vibert}, L. and {Vielva}, P. and {Villa}, F. and {Vittorio}, N. and {Wandelt}, B.~D. and {Wehus}, I.~K. and {White}, M. and {White}, S.~D.~M. and {Zacchei}, A. and {Zonca}, A.},
        title = "{Planck 2018 results. VI. Cosmological parameters}",
      journal = {\aap},
         year = 2020,
        month = sep,
       volume = {641},
          eid = {A6},
        pages = {A6},
          doi = {10.1051/0004-6361/201833910},
archivePrefix = {arXiv},
       eprint = {1807.06209},
 primaryClass = {astro-ph.CO},
       adsurl = {https://ui.adsabs.harvard.edu/abs/2020A&A...641A...6P}
}

@ARTICLE{Kokorev2025,
       author = {{Kokorev}, Vasily and {Ch{\'a}vez Ortiz}, {\'O}scar A. and {Taylor}, Anthony J. and {Finkelstein}, Steven L. and {Arrabal Haro}, Pablo and {Dickinson}, Mark and {Chisholm}, John and {Fujimoto}, Seiji and {noz}, Julian B. Mu and {Endsley}, Ryan and {Hu}, Weida and {Napolitano}, Lorenzo and {Wilkins}, Stephen M. and {Akins}, Hollis B. and {Amori{\'\i}n}, Ricardo and {Casey}, Caitlin M. and {Cheng}, Yingjie and {Cleri}, Nikko J. and {Cole}, Justin and {Cullen}, Fergus and {Daddi}, Emanuele and {Davis}, Kelcey and {Donnan}, Callum T. and {Dunlop}, James S. and {Fern{\'a}ndez}, Vital and {Giavalisco}, Mauro and {Grogin}, Norman A. and {Hathi}, Nimish and {Hirschmann}, Michaela and {Kartaltepe}, Jeyhan S. and {Koekemoer}, Anton M. and {Leung}, Ho-Hin and {Lucas}, Ray A. and {McLeod}, Derek and {Papovich}, Casey and {Pentericci}, Laura and {P{\'e}rez-Gonz{\'a}lez}, Pablo G. and {Somerville}, Rachel S. and {Wang}, Xin and {Yung}, L.~Y. Aaron and {Zavala}, Jorge A.},
        title = "{CAPERS Observations of Two UV-bright Galaxies at z > 10. More Evidence for Bursting Star Formation in the Early Universe}",
      journal = {\apjl},
         year = 2025,
        month = jul,
       volume = {988},
       number = {1},
          eid = {L10},
        pages = {L10},
          doi = {10.3847/2041-8213/ade8f5},
archivePrefix = {arXiv},
       eprint = {2504.12504},
 primaryClass = {astro-ph.GA},
       adsurl = {https://ui.adsabs.harvard.edu/abs/2025ApJ...988L..10K}
}

@ARTICLE{Morales2024,
       author = {{Morales}, Alexa M. and {Finkelstein}, Steven L. and {Leung}, Gene C.~K. and {Bagley}, Micaela B. and {Cleri}, Nikko J. and {Dave}, Romeel and {Dickinson}, Mark and {Ferguson}, Henry C. and {Hathi}, Nimish P. and {Jones}, Ewan and {Koekemoer}, Anton M. and {Papovich}, Casey and {P{\'e}rez-Gonz{\'a}lez}, Pablo G. and {Pirzkal}, Nor and {Smith}, Britton and {Wilkins}, Stephen M. and {Yung}, L.~Y. Aaron},
        title = "{Rest-frame UV Colors for Faint Galaxies at z {\ensuremath{\sim}} 9{\textendash}16 with the JWST NGDEEP Survey}",
      journal = {\apjl},
         year = 2024,
        month = apr,
       volume = {964},
       number = {2},
          eid = {L24},
        pages = {L24},
          doi = {10.3847/2041-8213/ad2de4},
archivePrefix = {arXiv},
       eprint = {2311.04294},
 primaryClass = {astro-ph.GA},
       adsurl = {https://ui.adsabs.harvard.edu/abs/2024ApJ...964L..24M}
}

@ARTICLE{2023ApJ...951L...1P,
       author = {{P{\'e}rez-Gonz{\'a}lez}, Pablo G. and {Costantin}, Luca and {Langeroodi}, Danial and {Rinaldi}, Pierluigi and {Annunziatella}, Marianna and {Ilbert}, Olivier and {Colina}, Luis and {N{\o}rgaard-Nielsen}, Hans Ulrik and {Greve}, Thomas R. and {{\"O}stlin}, G{\"o}ran and {Wright}, Gillian and {Alonso-Herrero}, Almudena and {{\'A}lvarez-M{\'a}rquez}, Javier and {Caputi}, Karina I. and {Eckart}, Andreas and {Le F{\`e}vre}, Olivier and {Labiano}, {\'A}lvaro and {Garc{\'\i}a-Mar{\'\i}n}, Macarena and {Hjorth}, Jens and {Kendrew}, Sarah and {Pye}, John P. and {Tikkanen}, Tuomo and {van der Werf}, Paul and {Walter}, Fabian and {Ward}, Martin and {Bik}, Arjan and {Boogaard}, Leindert and {Bosman}, Sarah E.~I. and {G{\'o}mez}, Alejandro Crespo and {Gillman}, Steven and {Iani}, Edoardo and {Jermann}, Iris and {Melinder}, Jens and {Meyer}, Romain A. and {Moutard}, Thibaud and {van Dishoek}, Ewine and {Henning}, Thomas and {Lagage}, Pierre-Olivier and {Guedel}, Manuel and {Peissker}, Florian and {Ray}, Tom and {Vandenbussche}, Bart and {Garc{\'\i}a-Argum{\'a}nez}, {\'A}ngela and {Mar{\'\i}a M{\'e}rida}, Rosa},
        title = "{Life beyond 30: Probing the -20 < M $_{UV}$ < -17 Luminosity Function at 8 < z < 13 with the NIRCam Parallel Field of the MIRI Deep Survey}",
      journal = {\apjl},
         year = 2023,
        month = jul,
       volume = {951},
       number = {1},
          eid = {L1},
        pages = {L1},
          doi = {10.3847/2041-8213/acd9d0},
archivePrefix = {arXiv},
       eprint = {2302.02429},
 primaryClass = {astro-ph.GA},
       adsurl = {https://ui.adsabs.harvard.edu/abs/2023ApJ...951L...1P}
}

@ARTICLE{Asada2026,
       author = {{Asada}, Yoshihisa and {Willott}, Chris J. and {Muzzin}, Adam and {Brada{\v{c}}}, Maru{\v{s}}a and {Brammer}, Gabriel and {Desprez}, Guillaume and {Iyer}, Kartheik G. and {Marchesini}, Danilo and {Martis}, Nicholas S. and {Noirot}, Ga{\"e}l and {Sarrouh}, Ghassan T.~E. and {Sawicki}, Marcin and {Withers}, Sunna and {Fujimoto}, Seiji and {Felicioni}, Giordano and {Goovaerts}, Ilias and {Jude{\v{z}}}, Jon and {Jagga}, Naadiyah and {Merchant}, Maya and {M{\'e}rida}, Rosa M. and {Robbins}, Luke},
        title = "{Earliest Galaxy Evolution in the CANUCS+Technicolor Fields: Galaxy Properties at z {\ensuremath{\sim}} 10─16 Seen with the Full NIRCam Medium- and Broadband Filters}",
      journal = {\apj},
         year = 2026,
        month = jan,
       volume = {996},
       number = {2},
          eid = {115},
        pages = {115},
          doi = {10.3847/1538-4357/ae1f8d},
archivePrefix = {arXiv},
       eprint = {2507.03124},
 primaryClass = {astro-ph.GA},
       adsurl = {https://ui.adsabs.harvard.edu/abs/2026ApJ...996..115A}
}

@ARTICLE{Casey2024,
       author = {{Casey}, Caitlin M. and {Akins}, Hollis B. and {Shuntov}, Marko and {Ilbert}, Olivier and {Paquereau}, Louise and {Franco}, Maximilien and {Hayward}, Christopher C. and {Finkelstein}, Steven L. and {Boylan-Kolchin}, Michael and {Robertson}, Brant E. and {Allen}, Natalie and {Brinch}, Malte and {Cooper}, Olivia R. and {Ding}, Xuheng and {Drakos}, Nicole E. and {Faisst}, Andreas L. and {Fujimoto}, Seiji and {Gillman}, Steven and {Harish}, Santosh and {Hirschmann}, Michaela and {Jin}, Shuowen and {Kartaltepe}, Jeyhan S. and {Koekemoer}, Anton M. and {Kokorev}, Vasily and {Liu}, Daizhong and {Long}, Arianna S. and {Magdis}, Georgios and {Maraston}, Claudia and {Martin}, Crystal L. and {McCracken}, Henry Joy and {McKinney}, Jed and {Mobasher}, Bahram and {Rhodes}, Jason and {Rich}, R. Michael and {Sanders}, David B. and {Silverman}, John D. and {Toft}, Sune and {Vijayan}, Aswin P. and {Weaver}, John R. and {Wilkins}, Stephen M. and {Yang}, Lilan and {Zavala}, Jorge A.},
        title = "{COSMOS-Web: Intrinsically Luminous z {\ensuremath{\gtrsim}} 10 Galaxy Candidates Test Early Stellar Mass Assembly}",
      journal = {\apj},
         year = 2024,
        month = apr,
       volume = {965},
       number = {1},
          eid = {98},
        pages = {98},
          doi = {10.3847/1538-4357/ad2075},
archivePrefix = {arXiv},
       eprint = {2308.10932},
 primaryClass = {astro-ph.GA},
       adsurl = {https://ui.adsabs.harvard.edu/abs/2024ApJ...965...98C}
}

@ARTICLE{Austin2025,
       author = {{Austin}, Duncan and {Conselice}, Christopher J. and {Adams}, Nathan J. and {Harvey}, Thomas and {Duan}, Qiao and {Trussler}, James and {Li}, Qiong and {Juod{\v{z}}balis}, Ignas and {Ormerod}, Katherine and {Ferreira}, Leonardo and {Westcott}, Lewi and {Harris}, Honor and {Wilkins}, Stephen M. and {Bhatawdekar}, Rachana and {Caruana}, Joseph and {Coe}, Dan and {Cohen}, Seth H. and {Driver}, Simon P. and {D'Silva}, Jordan C.~J. and {Frye}, Brenda and {Furtak}, Lukas J. and {Grogin}, Norman A. and {Hathi}, Nimish P. and {Holwerda}, Benne W. and {Jansen}, Rolf A. and {Koekemoer}, Anton M. and {Marshall}, Madeline A. and {Nonino}, Mario and {Ortiz}, III, Rafael and {Pirzkal}, Nor and {Robotham}, Aaron and {Ryan}, Jr., Russell E. and {Summers}, Jake and {Willmer}, Christopher N.~A. and {Windhorst}, Rogier A. and {Yan}, Haojing and {Zackrisson}, Erik},
        title = "{EPOCHS. III. Unbiased UV Continuum Slopes at 6.5 < z < 13 from Combined PEARLS GTO and Public JWST/NIRCam Imaging}",
      journal = {\apj},
         year = 2025,
        month = dec,
       volume = {995},
       number = {1},
          eid = {43},
        pages = {43},
          doi = {10.3847/1538-4357/ae07db},
archivePrefix = {arXiv},
       eprint = {2404.10751},
 primaryClass = {astro-ph.GA},
       adsurl = {https://ui.adsabs.harvard.edu/abs/2025ApJ...995...43A}
}

@ARTICLE{Cullen23,
       author = {{Cullen}, Fergus and {McLure}, R.~J. and {McLeod}, D.~J. and {Dunlop}, J.~S. and {Donnan}, C.~T. and {Carnall}, A.~C. and {Bowler}, R.~A.~A. and {Begley}, R. and {Hamadouche}, M.~L. and {Stanton}, T.~M.},
        title = "{The ultraviolet continuum slopes ({\ensuremath{\beta}}) of galaxies at z ≃ 8-16 from JWST and ground-based near-infrared imaging}",
      journal = {\mnras},
         year = 2023,
        month = mar,
       volume = {520},
       number = {1},
        pages = {14-23},
          doi = {10.1093/mnras/stad073},
archivePrefix = {arXiv},
       eprint = {2208.04914},
 primaryClass = {astro-ph.GA},
       adsurl = {https://ui.adsabs.harvard.edu/abs/2023MNRAS.520...14C}
}

@ARTICLE{Cullen2024,
       author = {{Cullen}, F. and {McLeod}, D.~J. and {McLure}, R.~J. and {Dunlop}, J.~S. and {Donnan}, C.~T. and {Carnall}, A.~C. and {Keating}, L.~C. and {Magee}, D. and {Arellano-Cordova}, K.~Z. and {Bowler}, R.~A.~A. and {Begley}, R. and {Flury}, S.~R. and {Hamadouche}, M.~L. and {Stanton}, T.~M.},
        title = "{The ultraviolet continuum slopes of high-redshift galaxies: evidence for the emergence of dust-free stellar populations at z > 10}",
      journal = {\mnras},
         year = 2024,
        month = jun,
       volume = {531},
       number = {1},
        pages = {997-1020},
          doi = {10.1093/mnras/stae1211},
archivePrefix = {arXiv},
       eprint = {2311.06209},
 primaryClass = {astro-ph.GA},
       adsurl = {https://ui.adsabs.harvard.edu/abs/2024MNRAS.531..997C}
}

@ARTICLE{PPG,
       author = {{P{\'e}rez-Gonz{\'a}lez}, Pablo G. and {{\"O}stlin}, G{\"o}ran and {Costantin}, Luca and {Melinder}, Jens and {Finkelstein}, Steven L. and {Somerville}, Rachel S. and {Annunziatella}, Marianna and {{\'A}lvarez-M{\'a}rquez}, Javier and {Colina}, Luis and {Dekel}, Avishai and {Ferguson}, Henry C. and {Li}, Zhaozhou and {Yung}, L.~Y. Aaron and {Bagley}, Micaela B. and {Boogaard}, Leindert A. and {Burgarella}, Denis and {Calabr{\`o}}, Antonello and {Caputi}, Karina I. and {Cheng}, Yingjie and {Dickinson}, Mark and {Eckart}, Andreas and {Giavalisco}, Mauro and {Gillman}, Steven and {Greve}, Thomas R. and {Hamed}, Mahmoud and {Hathi}, Nimish P. and {Hjorth}, Jens and {Huertas-Company}, Marc and {Kartaltepe}, Jeyhan S. and {Koekemoer}, Anton M. and {Kokorev}, Vasily and {Labiano}, {\'A}lvaro and {Langeroodi}, Danial and {Leung}, Gene C.~K. and {Natarajan}, Priyamvada and {Papovich}, Casey and {Peissker}, Florian and {Pentericci}, Laura and {Pirzkal}, Nor and {Rinaldi}, Pierluigi and {van der Werf}, Paul and {Walter}, Fabian},
        title = "{The Rise of the Galactic Empire: Ultraviolet Luminosity Functions at z {\ensuremath{\sim}} 17 and z {\ensuremath{\sim}} 25 Estimated with the MIDIS+NGDEEP Ultra-deep JWST/NIRCam Data Set}",
      journal = {\apj},
         year = 2025,
        month = oct,
       volume = {991},
       number = {2},
          eid = {179},
        pages = {179},
          doi = {10.3847/1538-4357/adf8c9},
archivePrefix = {arXiv},
       eprint = {2503.15594},
 primaryClass = {astro-ph.GA},
       adsurl = {https://ui.adsabs.harvard.edu/abs/2025ApJ...991..179P}
}

@ARTICLE{Gandolfi25,
       author = {{Gandolfi}, G. and {Rodighiero}, G. and {Bisigello}, L. and {Grazian}, A. and {Finkelstein}, S.~L. and {Dickinson}, M. and {Castellano}, M. and {Merlin}, E. and {Calabr{\`o}}, A. and {Papovich}, C. and {Bianchetti}, A. and {Ba{\~n}ados}, E. and {Benotto}, P. and {Catone}, M. and {Buitrago}, F. and {Daddi}, E. and {Girardi}, G. and {Giulietti}, M. and {Hirschmann}, M. and {Holwerda}, B.~W. and {Arrabal Haro}, P. and {Lapi}, A. and {Lucas}, R.~A. and {Lyu}, Y. and {Massardi}, M. and {Pacucci}, F. and {P{\'e}rez-Gonz{\'a}lez}, P.~G. and {Ronconi}, T. and {Tarrasse}, M. and {Wilkins}, S. and {Vulcani}, B. and {Yung}, L.~Y.~A. and {Zavala}, J.~A. and {Backhaus}, B. and {Bagley}, M. and {Buat}, V. and {Burgarella}, D. and {Kartaltepe}, J. and {Khusanova}, Y. and {Kirkpatrick}, A. and {Kocevski}, D. and {Koekemoer}, A.~M. and {Lambrides}, E. and {Pirzkal}, N. and {Yang}, G.},
        title = "{Ultrahigh-redshift or closer-by, dust-obscured galaxies?: Deciphering the nature of faint, previously missed F200W dropouts in CEERS}",
      journal = {\aap},
         year = 2026,
        month = apr,
       volume = {708},
          eid = {A195},
        pages = {A195},
          doi = {10.1051/0004-6361/202554009},
archivePrefix = {arXiv},
       eprint = {2502.02637},
 primaryClass = {astro-ph.GA},
       adsurl = {https://ui.adsabs.harvard.edu/abs/2026A&A...708A.195G}
}

@ARTICLE{Naidu2022,
       author = {{Naidu}, Rohan P. and {Oesch}, Pascal A. and {van Dokkum}, Pieter and {Nelson}, Erica J. and {Suess}, Katherine A. and {Brammer}, Gabriel and {Whitaker}, Katherine E. and {Illingworth}, Garth and {Bouwens}, Rychard and {Tacchella}, Sandro and {Matthee}, Jorryt and {Allen}, Natalie and {Bezanson}, Rachel and {Conroy}, Charlie and {Labbe}, Ivo and {Leja}, Joel and {Leonova}, Ecaterina and {Magee}, Dan and {Price}, Sedona H. and {Setton}, David J. and {Strait}, Victoria and {Stefanon}, Mauro and {Toft}, Sune and {Weaver}, John R. and {Weibel}, Andrea},
        title = "{Two Remarkably Luminous Galaxy Candidates at z {\ensuremath{\approx}} 10-12 Revealed by JWST}",
      journal = {\apjl},
         year = 2022,
        month = nov,
       volume = {940},
       number = {1},
          eid = {L14},
        pages = {L14},
          doi = {10.3847/2041-8213/ac9b22},
archivePrefix = {arXiv},
       eprint = {2207.09434},
 primaryClass = {astro-ph.GA},
       adsurl = {https://ui.adsabs.harvard.edu/abs/2022ApJ...940L..14N}
}

@ARTICLE{Finkelstein2,
       author = {{Finkelstein}, Steven L. and {Leung}, Gene C.~K. and {Bagley}, Micaela B. and {Dickinson}, Mark and {Ferguson}, Henry C. and {Papovich}, Casey and {Akins}, Hollis B. and {Arrabal Haro}, Pablo and {Dav{\'e}}, Romeel and {Dekel}, Avishai and {Kartaltepe}, Jeyhan S. and {Kocevski}, Dale D. and {Koekemoer}, Anton M. and {Pirzkal}, Nor and {Somerville}, Rachel S. and {Yung}, L.~Y. Aaron and {Amor{\'\i}n}, Ricardo O. and {Backhaus}, Bren E. and {Behroozi}, Peter and {Bisigello}, Laura and {Bromm}, Volker and {Casey}, Caitlin M. and {Ch{\'a}vez Ortiz}, {\'O}scar A. and {Cheng}, Yingjie and {Chworowsky}, Katherine and {Cleri}, Nikko J. and {Cooper}, M.~C. and {Davis}, Kelcey and {de la Vega}, Alexander and {Elbaz}, David and {Franco}, Maximilien and {Fontana}, Adriano and {Fujimoto}, Seiji and {Giavalisco}, Mauro and {Grogin}, Norman A. and {Holwerda}, Benne W. and {Huertas-Company}, Marc and {Hirschmann}, Michaela and {Iyer}, Kartheik G. and {Jogee}, Shardha and {Jung}, Intae and {Larson}, Rebecca L. and {Lucas}, Ray A. and {Mobasher}, Bahram and {Morales}, Alexa M. and {Morley}, Caroline V. and {Mukherjee}, Sagnick and {P{\'e}rez-Gonz{\'a}lez}, Pablo G. and {Ravindranath}, Swara and {Rodighiero}, Giulia and {Rowland}, Melanie J. and {Tacchella}, Sandro and {Taylor}, Anthony J. and {Trump}, Jonathan R. and {Wilkins}, Stephen M.},
        title = "{The Complete CEERS Early Universe Galaxy Sample: A Surprisingly Slow Evolution of the Space Density of Bright Galaxies at z {\ensuremath{\sim}} 8.5{\textendash}14.5}",
      journal = {\apjl},
         year = 2024,
        month = jul,
       volume = {969},
       number = {1},
          eid = {L2},
        pages = {L2},
          doi = {10.3847/2041-8213/ad4495},
archivePrefix = {arXiv},
       eprint = {2311.04279},
 primaryClass = {astro-ph.GA},
       adsurl = {https://ui.adsabs.harvard.edu/abs/2024ApJ...969L...2F}
}

@ARTICLE{Castellano2023,
       author = {{Castellano}, Marco and {Fontana}, Adriano and {Treu}, Tommaso and {Merlin}, Emiliano and {Santini}, Paola and {Bergamini}, Pietro and {Grillo}, Claudio and {Rosati}, Piero and {Acebron}, Ana and {Leethochawalit}, Nicha and {Paris}, Diego and {Bonchi}, Andrea and {Belfiori}, Davide and {Calabr{\`o}}, Antonello and {Correnti}, Matteo and {Nonino}, Mario and {Polenta}, Gianluca and {Trenti}, Michele and {Boyett}, Kristan and {Brammer}, G. and {Broadhurst}, Tom and {Caminha}, Gabriel B. and {Chen}, Wenlei and {Filippenko}, Alexei V. and {Fortuni}, Flaminia and {Glazebrook}, Karl and {Mascia}, Sara and {Mason}, Charlotte A. and {Menci}, Nicola and {Meneghetti}, Massimo and {Mercurio}, Amata and {Metha}, Benjamin and {Morishita}, Takahiro and {Nanayakkara}, Themiya and {Pentericci}, Laura and {Roberts-Borsani}, Guido and {Roy}, Namrata and {Vanzella}, Eros and {Vulcani}, Benedetta and {Yang}, Lilan and {Wang}, Xin},
        title = "{Early Results from GLASS-JWST. XIX. A High Density of Bright Galaxies at z {\ensuremath{\approx}} 10 in the A2744 Region}",
      journal = {\apjl},
         year = 2023,
        month = may,
       volume = {948},
       number = {2},
          eid = {L14},
        pages = {L14},
          doi = {10.3847/2041-8213/accea5},
archivePrefix = {arXiv},
       eprint = {2212.06666},
 primaryClass = {astro-ph.GA},
       adsurl = {https://ui.adsabs.harvard.edu/abs/2023ApJ...948L..14C}
}

@ARTICLE{Harikane2022,
       author = {{Harikane}, Yuichi and {Inoue}, Akio K. and {Mawatari}, Ken and {Hashimoto}, Takuya and {Yamanaka}, Satoshi and {Fudamoto}, Yoshinobu and {Matsuo}, Hiroshi and {Tamura}, Yoichi and {Dayal}, Pratika and {Yung}, L.~Y. Aaron and {Hutter}, Anne and {Pacucci}, Fabio and {Sugahara}, Yuma and {Koekemoer}, Anton M.},
        title = "{A Search for H-Dropout Lyman Break Galaxies at z 12-16}",
      journal = {\apj},
         year = 2022,
        month = apr,
       volume = {929},
       number = {1},
          eid = {1},
        pages = {1},
          doi = {10.3847/1538-4357/ac53a9},
archivePrefix = {arXiv},
       eprint = {2112.09141},
 primaryClass = {astro-ph.GA},
       adsurl = {https://ui.adsabs.harvard.edu/abs/2022ApJ...929....1H}
}

@ARTICLE{Austin,
       author = {{Austin}, Duncan and {Adams}, Nathan and {Conselice}, Christopher J. and {Harvey}, Thomas and {Ormerod}, Katherine and {Trussler}, James and {Li}, Qiong and {Ferreira}, Leonardo and {Dayal}, Pratika and {Juod{\v{z}}balis}, Ignas},
        title = "{A Large Population of Faint 8 < z < 16 Galaxies Found in the First JWST NIRCam Observations of the NGDEEP Survey}",
      journal = {\apjl},
         year = 2023,
        month = jul,
       volume = {952},
       number = {1},
          eid = {L7},
        pages = {L7},
          doi = {10.3847/2041-8213/ace18d},
archivePrefix = {arXiv},
       eprint = {2302.04270},
 primaryClass = {astro-ph.GA},
       adsurl = {https://ui.adsabs.harvard.edu/abs/2023ApJ...952L...7A}
}

@ARTICLE{Bagley23,
       author = {{Bagley}, Micaela B. and {Finkelstein}, Steven L. and {Koekemoer}, Anton M. and {Ferguson}, Henry C. and {Arrabal Haro}, Pablo and {Dickinson}, Mark and {Kartaltepe}, Jeyhan S. and {Papovich}, Casey and {P{\'e}rez-Gonz{\'a}lez}, Pablo G. and {Pirzkal}, Nor and {Somerville}, Rachel S. and {Willmer}, Christopher N.~A. and {Yang}, Guang and {Yung}, L.~Y. Aaron and {Fontana}, Adriano and {Grazian}, Andrea and {Grogin}, Norman A. and {Hirschmann}, Michaela and {Kewley}, Lisa J. and {Kirkpatrick}, Allison and {Kocevski}, Dale D. and {Lotz}, Jennifer M. and {Medrano}, Aubrey and {Morales}, Alexa M. and {Pentericci}, Laura and {Ravindranath}, Swara and {Trump}, Jonathan R. and {Wilkins}, Stephen M. and {Calabr{\`o}}, Antonello and {Cooper}, M.~C. and {Costantin}, Luca and {de la Vega}, Alexander and {Hilbert}, Bryan and {Hutchison}, Taylor A. and {Larson}, Rebecca L. and {Lucas}, Ray A. and {McGrath}, Elizabeth J. and {Ryan}, Russell and {Wang}, Xin and {Wuyts}, Stijn},
        title = "{CEERS Epoch 1 NIRCam Imaging: Reduction Methods and Simulations Enabling Early JWST Science Results}",
      journal = {\apjl},
         year = 2023,
        month = mar,
       volume = {946},
       number = {1},
          eid = {L12},
        pages = {L12},
          doi = {10.3847/2041-8213/acbb08},
archivePrefix = {arXiv},
       eprint = {2211.02495},
 primaryClass = {astro-ph.IM},
       adsurl = {https://ui.adsabs.harvard.edu/abs/2023ApJ...946L..12B}
}

@ARTICLE{matteri1,
       author = {{Matteri}, Antonio and {Pallottini}, Andrea and {Ferrara}, Andrea},
        title = "{Can primordial black holes explain the overabundance of bright super-early galaxies?}",
      journal = {\aap},
         year = 2025,
        month = may,
       volume = {697},
          eid = {A65},
        pages = {A65},
          doi = {10.1051/0004-6361/202553701},
archivePrefix = {arXiv},
       eprint = {2503.01968},
 primaryClass = {astro-ph.GA},
       adsurl = {https://ui.adsabs.harvard.edu/abs/2025A&A...697A..65M}
}

@ARTICLE{Finkelstein2025,
       author = {{Finkelstein}, Steven L. and {Bagley}, Micaela B. and {Arrabal Haro}, Pablo and {Dickinson}, Mark and {Ferguson}, Henry C. and {Kartaltepe}, Jeyhan S. and {Kocevski}, Dale D. and {Koekemoer}, Anton M. and {Lotz}, Jennifer M. and {Papovich}, Casey and {P{\'e}rez-Gonz{\'a}lez}, Pablo G. and {Pirzkal}, Nor and {Somerville}, Rachel S. and {Trump}, Jonathan R. and {Yang}, Guang and {Yung}, L.~Y. Aaron and {Fontana}, Adriano and {Grazian}, Andrea and {Grogin}, Norman A. and {Kewley}, Lisa J. and {Kirkpatrick}, Allison and {Larson}, Rebecca L. and {Pentericci}, Laura and {Ravindranath}, Swara and {Wilkins}, Stephen M. and {Almaini}, Omar and {Amor{\'\i}n}, Ricardo O. and {Barro}, Guillermo and {Bhatawdekar}, Rachana and {Bisigello}, Laura and {Brooks}, Madisyn and {Buat}, V{\'e}ronique and {Buitrago}, Fernando and {Burgarella}, Denis and {Calabr{\`o}}, Antonello and {Castellano}, Marco and {Cheng}, Yingjie and {Cleri}, Nikko J. and {Cole}, Justin W. and {Cooper}, M.~C. and {Cooper}, Olivia R. and {Costantin}, Luca and {Cox}, Isa G. and {Croton}, Darren and {Daddi}, Emanuele and {Davis}, Kelcey and {Dekel}, Avishai and {Elbaz}, David and {Fern{\'a}ndez}, Vital and {Fujimoto}, Seiji and {Gandolfi}, Giovanni and {Gardner}, Jonathan P. and {Gawiser}, Eric and {Giavalisco}, Mauro and {G{\'o}mez-Guijarro}, Carlos and {Guo}, Yuchen and {Gupta}, Ansh R. and {Hathi}, Nimish P. and {Harish}, Santosh and {Henry}, Aur{\'e}lien and {Hirschmann}, Michaela and {Hu}, Weida and {Hutchison}, Taylor A. and {Iyer}, Kartheik G. and {Jaskot}, Anne E. and {Jha}, Saurabh W. and {Jung}, Intae and {Kassin}, Susan A. and {Kokorev}, Vasily and {Kurczynski}, Peter and {Leung}, Gene C.~K. and {Llerena}, Mario and {Long}, Arianna S. and {Lucas}, Ray A. and {Lu}, Shiying and {McGrath}, Elizabeth J. and {McIntosh}, Daniel H. and {Merlin}, Emiliano and {Mobasher}, Bahram and {Morales}, Alexa M. and {Napolitano}, Lorenzo and {Pacucci}, Fabio and {Pandya}, Viraj and {Rafelski}, Marc and {Rodighiero}, Giulia and {Rose}, Caitlin and {Santini}, Paola and {Seill{\'e}}, Lise-Marie and {Simons}, Raymond C. and {Shen}, Lu and {Straughn}, Amber N. and {Tacchella}, Sandro and {Taylor}, Anthony J. and {Vanderhoof}, Brittany N. and {Vega-Ferrero}, Jes{\'u}s and {Weiner}, Benjamin J. and {Willmer}, Christopher N.~A. and {Zhu}, Peixin and {Bell}, Eric F. and {Wuyts}, Stijn and {Holwerda}, Benne W. and {Wang}, Xin and {Wang}, Weichen and {Zavala}, Jorge A. and {CEERS Collaboration}},
        title = "{The Cosmic Evolution Early Release Science Survey (CEERS)}",
      journal = {\apjl},
         year = 2025,
        month = apr,
       volume = {983},
       number = {1},
          eid = {L4},
        pages = {L4},
          doi = {10.3847/2041-8213/adbbd3},
archivePrefix = {arXiv},
       eprint = {2501.04085},
 primaryClass = {astro-ph.GA},
       adsurl = {https://ui.adsabs.harvard.edu/abs/2025ApJ...983L...4F}
}

@ARTICLE{Ferrara2024,
       author = {{Ferrara}, A.},
        title = "{Super-early JWST galaxies, outflows, and Ly{\ensuremath{\alpha}} visibility in the Epoch of Reionization}",
      journal = {\aap},
         year = 2024,
        month = apr,
       volume = {684},
          eid = {A207},
        pages = {A207},
          doi = {10.1051/0004-6361/202348321},
archivePrefix = {arXiv},
       eprint = {2310.12197},
 primaryClass = {astro-ph.GA},
       adsurl = {https://ui.adsabs.harvard.edu/abs/2024A&A...684A.207F}
}

@ARTICLE{Narayanan25b,
       author = {{Narayanan}, Desika and {Stark}, Daniel P. and {Finkelstein}, Steven L. and {Torrey}, Paul and {Li}, Qi and {Cullen}, Fergus and {Topping}, Micheal W. and {Marinacci}, Federico and {Sales}, Laura V. and {Shen}, Xuejian and {Vogelsberger}, Mark},
        title = "{The Ultraviolet Slopes of Early Universe Galaxies: The Impact of Bursty Star Formation, Dust, and Nebular Continuum Emission}",
      journal = {\apj},
         year = 2025,
        month = mar,
       volume = {982},
       number = {1},
          eid = {7},
        pages = {7},
          doi = {10.3847/1538-4357/adb41c},
archivePrefix = {arXiv},
       eprint = {2408.13312},
 primaryClass = {astro-ph.GA},
       adsurl = {https://ui.adsabs.harvard.edu/abs/2025ApJ...982....7N}
}

@ARTICLE{Narayanan2025,
       author = {{Narayanan}, Desika and {Torrey}, Paul and {Stark}, Daniel and {Chisholm}, John and {Finkelstein}, Steven and {Garcia}, Alex and {Kelley-Derzon}, Jessica and {Marinacci}, Federico and {Sales}, Laura and {Savitch}, Ethan and {Vogelsberger}, Mark and {Zimmerman}, Dhruv},
        title = "{The Growth of Dust in Galaxies in the First Billion Years with Applications to Blue Monsters}",
      journal = {The Open Journal of Astrophysics},
         year = 2026,
        month = apr,
       volume = {9},
        pages = {59986},
          doi = {10.33232/001c.159986},
archivePrefix = {arXiv},
       eprint = {2509.18266},
 primaryClass = {astro-ph.GA},
       adsurl = {https://ui.adsabs.harvard.edu/abs/2026OJAp....959986N}
}

@ARTICLE{Katz2025,
       author = {{Katz}, Harley and {Cameron}, Alex J. and {Saxena}, Aayush and {Barrufet}, Laia and {Choustikov}, Nichloas and {Cleri}, Nikko J. and {de Graff}, Anna and {Ellis}, Richard S. and {Fosbury}, Robert A.~E. and {Heintz}, Kasper E. and {Maseda}, Michael and {Matthee}, Jorryt and {McConachie}, Ian and {Oesch}, Pascal A.},
        title = "{21 Balmer Jump Street: The Nebular Continuum at High Redshift and Implications for the Bright Galaxy Problem, UV Continuum Slopes, and Early Stellar Populations}",
      journal = {The Open Journal of Astrophysics},
         year = 2025,
        month = jul,
       volume = {8},
          eid = {104},
        pages = {104},
          doi = {10.33232/001c.142570},
archivePrefix = {arXiv},
       eprint = {2408.03189},
 primaryClass = {astro-ph.GA},
       adsurl = {https://ui.adsabs.harvard.edu/abs/2025OJAp....8E.104K}
}

@ARTICLE{Hainline2011,
       author = {{Hainline}, Kevin N. and {Shapley}, Alice E. and {Greene}, Jenny E. and {Steidel}, Charles C.},
        title = "{The Rest-frame Ultraviolet Spectra of UV-selected Active Galactic Nuclei at z \raisebox{-0.5ex}\textasciitilde 2-3}",
      journal = {\apj},
         year = 2011,
        month = may,
       volume = {733},
       number = {1},
          eid = {31},
        pages = {31},
          doi = {10.1088/0004-637X/733/1/31},
archivePrefix = {arXiv},
       eprint = {1012.0075},
 primaryClass = {astro-ph.CO},
       adsurl = {https://ui.adsabs.harvard.edu/abs/2011ApJ...733...31H}
}

@ARTICLE{Hainline2012,
       author = {{Hainline}, Kevin N. and {Shapley}, Alice E. and {Greene}, Jenny E. and {Steidel}, Charles C. and {Reddy}, Naveen A. and {Erb}, Dawn K.},
        title = "{Stellar Populations of Ultraviolet-selected Active Galactic Nuclei Host Galaxies at z \raisebox{-0.5ex}\textasciitilde 2-3}",
      journal = {\apj},
         year = 2012,
        month = nov,
       volume = {760},
       number = {1},
          eid = {74},
        pages = {74},
          doi = {10.1088/0004-637X/760/1/74},
archivePrefix = {arXiv},
       eprint = {1206.3308},
 primaryClass = {astro-ph.CO},
       adsurl = {https://ui.adsabs.harvard.edu/abs/2012ApJ...760...74H}
}

@ARTICLE{Taylor2025,
       author = {{Taylor}, Anthony J. and {Kokorev}, Vasily and {Kocevski}, Dale D. and {Akins}, Hollis B. and {Cullen}, Fergus and {Dickinson}, Mark and {Finkelstein}, Steven L. and {Arrabal Haro}, Pablo and {Bromm}, Volker and {Giavalisco}, Mauro and {Inayoshi}, Kohei and {Juneau}, St{\'e}phanie and {Leung}, Gene C.~K. and {P{\'e}rez-Gonz{\'a}lez}, Pablo G. and {Somerville}, Rachel S. and {Trump}, Jonathan R. and {Amor{\'\i}n}, Ricardo O. and {Barro}, Guillermo and {Burgarella}, Denis and {Brooks}, Madisyn and {Carnall}, Adam C. and {Casey}, Caitlin M. and {Cheng}, Yingjie and {Chisholm}, John and {Chworowsky}, Katherine and {Davis}, Kelcey and {Donnan}, Callum T. and {Dunlop}, James S. and {Ellis}, Richard S. and {Fern{\'a}ndez}, Vital and {Fujimoto}, Seiji and {Grogin}, Norman A. and {Gupta}, Ansh R. and {Hathi}, Nimish P. and {Jung}, Intae and {Hirschmann}, Michaela and {Kartaltepe}, Jeyhan S. and {Koekemoer}, Anton M. and {Larson}, Rebecca L. and {Leung}, Ho-Hin and {Llerena}, Mario and {Lucas}, Ray A. and {McLeod}, Derek J. and {McLure}, Ross and {Napolitano}, Lorenzo and {Papovich}, Casey and {Stanton}, Thomas M. and {Tripodi}, Roberta and {Wang}, Xin and {Wilkins}, Stephen M. and {Yung}, L.~Y. Aaron and {Zavala}, Jorge A.},
        title = "{CAPERS-LRD-z9: A Gas-enshrouded Little Red Dot Hosting a Broad-line Active Galactic Nucleus at z = 9.288}",
      journal = {\apjl},
         year = 2025,
        month = aug,
       volume = {989},
       number = {1},
          eid = {L7},
        pages = {L7},
          doi = {10.3847/2041-8213/ade789},
archivePrefix = {arXiv},
       eprint = {2505.04609},
 primaryClass = {astro-ph.GA},
       adsurl = {https://ui.adsabs.harvard.edu/abs/2025ApJ...989L...7T}
}

@ARTICLE{Tang25,
       author = {{Tang}, Mengtao and {Stark}, Daniel P. and {Mason}, Charlotte A. and {Gelli}, Viola and {Chen}, Zuyi and {Topping}, Michael W.},
        title = "{The JWST Spectroscopic Properties of Galaxies at z = 9{\ensuremath{-}}14}",
      journal = {\apj},
         year = 2026,
        month = apr,
       volume = {1001},
       number = {1},
          eid = {38},
        pages = {38},
          doi = {10.3847/1538-4357/ae4edb},
archivePrefix = {arXiv},
       eprint = {2507.08245},
 primaryClass = {astro-ph.GA},
       adsurl = {https://ui.adsabs.harvard.edu/abs/2026ApJ..1001...38T}
}

@ARTICLE{Rodighiero2023,
       author = {{Rodighiero}, Giulia and {Bisigello}, Laura and {Iani}, Edoardo and {Marasco}, Antonino and {Grazian}, Andrea and {Sinigaglia}, Francesco and {Cassata}, Paolo and {Gruppioni}, Carlotta},
        title = "{JWST unveils heavily obscured (active and passive) sources up to z   13}",
      journal = {\mnras},
         year = 2023,
        month = jan,
       volume = {518},
       number = {1},
        pages = {L19-L24},
          doi = {10.1093/mnrasl/slac115},
archivePrefix = {arXiv},
       eprint = {2208.02825},
 primaryClass = {astro-ph.GA},
       adsurl = {https://ui.adsabs.harvard.edu/abs/2023MNRAS.518L..19R}
}

@ARTICLE{Heintz2024dja,
       author = {{Heintz}, Kasper E. and {Watson}, Darach and {Brammer}, Gabriel and {Vejlgaard}, Simone and {Hutter}, Anne and {Strait}, Victoria B. and {Matthee}, Jorryt and {Oesch}, Pascal A. and {Jakobsson}, P{\'a}ll and {Tanvir}, Nial R. and {Laursen}, Peter and {Naidu}, Rohan P. and {Mason}, Charlotte A. and {Killi}, Meghana and {Jung}, Intae and {Hsiao}, Tiger Yu-Yang and {Abdurro'uf} and {Coe}, Dan and {Arrabal Haro}, Pablo and {Finkelstein}, Steven L. and {Toft}, Sune},
        title = "{Strong damped Lyman-{\ensuremath{\alpha}} absorption in young star-forming galaxies at redshifts 9 to 11}",
      journal = {Science},
         year = 2024,
        month = may,
       volume = {384},
       number = {6698},
        pages = {890-894},
          doi = {10.1126/science.adj0343},
archivePrefix = {arXiv},
       eprint = {2306.00647},
 primaryClass = {astro-ph.GA},
       adsurl = {https://ui.adsabs.harvard.edu/abs/2024Sci...384..890H}
}

@ARTICLE{deGraaff2025,
       author = {{de Graaff}, Anna and {Brammer}, Gabriel and {Weibel}, Andrea and {Lewis}, Zach and {Maseda}, Michael V. and {Oesch}, Pascal A. and {Bezanson}, Rachel and {Boogaard}, Leindert A. and {Cleri}, Nikko J. and {Cooper}, Olivia R. and {Gottumukkala}, Rashmi and {Greene}, Jenny E. and {Hirschmann}, Michaela and {Hviding}, Raphael E. and {Katz}, Harley and {Labb{\'e}}, Ivo and {Leja}, Joel and {Matthee}, Jorryt and {McConachie}, Ian and {Miller}, Tim B. and {Naidu}, Rohan P. and {Price}, Sedona H. and {Rix}, Hans-Walter and {Setton}, David J. and {Suess}, Katherine A. and {Wang}, Bingjie and {Whitaker}, Katherine E. and {Williams}, Christina C.},
        title = "{RUBIES: A complete census of the bright and red distant Universe with JWST/NIRSpec}",
      journal = {\aap},
         year = 2025,
        month = may,
       volume = {697},
          eid = {A189},
        pages = {A189},
          doi = {10.1051/0004-6361/202452186},
archivePrefix = {arXiv},
       eprint = {2409.05948},
 primaryClass = {astro-ph.GA},
       adsurl = {https://ui.adsabs.harvard.edu/abs/2025A&A...697A.189D}
}

@ARTICLE{Napolitano2025b,
       author = {{Napolitano}, Lorenzo and {Castellano}, Marco and {Pentericci}, Laura and {Vignali}, Cristian and {Gilli}, Roberto and {Fontana}, Adriano and {Santini}, Paola and {Treu}, Tommaso and {Calabr{\`o}}, Antonello and {Llerena}, Mario and {Piconcelli}, Enrico and {Zappacosta}, Luca and {Mascia}, Sara and {Tripodi}, Roberta and {Arrabal Haro}, Pablo and {Bergamini}, Pietro and {Bakx}, Tom J.~L.~C. and {Dickinson}, Mark and {Glazebrook}, Karl and {Henry}, Alaina and {Leethochawalit}, Nicha and {Mazzolari}, Giovanni and {Merlin}, Emiliano and {Morishita}, Takahiro and {Nanayakkara}, Themiya and {Paris}, Diego and {Puccetti}, Simonetta and {Roberts-Borsani}, Guido and {Rojas Ruiz}, Sofia and {Rosati}, Piero and {Vanzella}, Eros and {Vito}, Fabio and {Vulcani}, Benedetta and {Wang}, Xin and {Yoon}, Ilsang and {Zavala}, Jorge A.},
        title = "{The Dual Nature of GHZ9: Coexisting Active Galactic Nuclei and Star Formation Activity in a Remote X-Ray Source at z = 10.145}",
      journal = {\apj},
         year = 2025,
        month = aug,
       volume = {989},
       number = {1},
          eid = {75},
        pages = {75},
          doi = {10.3847/1538-4357/ade706},
       adsurl = {https://ui.adsabs.harvard.edu/abs/2025ApJ...989...75N}
}

@ARTICLE{Feltre2016,
       author = {{Feltre}, A. and {Charlot}, S. and {Gutkin}, J.},
        title = "{Nuclear activity versus star formation: emission-line diagnostics at ultraviolet and optical wavelengths}",
      journal = {\mnras},
         year = 2016,
        month = mar,
       volume = {456},
       number = {3},
        pages = {3354-3374},
          doi = {10.1093/mnras/stv2794},
archivePrefix = {arXiv},
       eprint = {1511.08217},
 primaryClass = {astro-ph.GA},
       adsurl = {https://ui.adsabs.harvard.edu/abs/2016MNRAS.456.3354F}
}

@ARTICLE{Gutkin2016,
       author = {{Gutkin}, Julia and {Charlot}, St{\'e}phane and {Bruzual}, Gustavo},
        title = "{Modelling the nebular emission from primeval to present-day star-forming galaxies}",
      journal = {\mnras},
         year = 2016,
        month = oct,
       volume = {462},
       number = {2},
        pages = {1757-1774},
          doi = {10.1093/mnras/stw1716},
archivePrefix = {arXiv},
       eprint = {1607.06086},
 primaryClass = {astro-ph.GA},
       adsurl = {https://ui.adsabs.harvard.edu/abs/2016MNRAS.462.1757G}
}

@ARTICLE{Nakajima2022,
       author = {{Nakajima}, K. and {Maiolino}, R.},
        title = "{Diagnostics for PopIII galaxies and direct collapse black holes in the early universe}",
      journal = {\mnras},
         year = 2022,
        month = jul,
       volume = {513},
       number = {4},
        pages = {5134-5147},
          doi = {10.1093/mnras/stac1242},
archivePrefix = {arXiv},
       eprint = {2204.11870},
 primaryClass = {astro-ph.GA},
       adsurl = {https://ui.adsabs.harvard.edu/abs/2022MNRAS.513.5134N}
}

@ARTICLE{Hirschmann2019,
       author = {{Hirschmann}, Michaela and {Charlot}, Stephane and {Feltre}, Anna and {Naab}, Thorsten and {Somerville}, Rachel S. and {Choi}, Ena},
        title = "{Synthetic nebular emission from massive galaxies - II. Ultraviolet-line diagnostics of dominant ionizing sources}",
      journal = {\mnras},
         year = 2019,
        month = jul,
       volume = {487},
       number = {1},
        pages = {333-353},
          doi = {10.1093/mnras/stz1256},
archivePrefix = {arXiv},
       eprint = {1811.07909},
 primaryClass = {astro-ph.GA},
       adsurl = {https://ui.adsabs.harvard.edu/abs/2019MNRAS.487..333H}
}

@ARTICLE{Bunker2023B,
       author = {{Bunker}, Andrew J. and {Saxena}, Aayush and {Cameron}, Alex J. and {Willott}, Chris J. and {Curtis-Lake}, Emma and {Jakobsen}, Peter and {Carniani}, Stefano and {Smit}, Renske and {Maiolino}, Roberto and {Witstok}, Joris and {Curti}, Mirko and {D'Eugenio}, Francesco and {Jones}, Gareth C. and {Ferruit}, Pierre and {Arribas}, Santiago and {Charlot}, Stephane and {Chevallard}, Jacopo and {Giardino}, Giovanna and {de Graaff}, Anna and {Looser}, Tobias J. and {L{\"u}tzgendorf}, Nora and {Maseda}, Michael V. and {Rawle}, Tim and {Rix}, Hans-Walter and {Del Pino}, Bruno Rodr{\'\i}guez and {Alberts}, Stacey and {Egami}, Eiichi and {Eisenstein}, Daniel J. and {Endsley}, Ryan and {Hainline}, Kevin and {Hausen}, Ryan and {Johnson}, Benjamin D. and {Rieke}, George and {Rieke}, Marcia and {Robertson}, Brant E. and {Shivaei}, Irene and {Stark}, Daniel P. and {Sun}, Fengwu and {Tacchella}, Sandro and {Tang}, Mengtao and {Williams}, Christina C. and {Willmer}, Christopher N.~A. and {Baker}, William M. and {Baum}, Stefi and {Bhatawdekar}, Rachana and {Bowler}, Rebecca and {Boyett}, Kristan and {Chen}, Zuyi and {Circosta}, Chiara and {Helton}, Jakob M. and {Ji}, Zhiyuan and {Kumari}, Nimisha and {Lyu}, Jianwei and {Nelson}, Erica and {Parlanti}, Eleonora and {Perna}, Michele and {Sandles}, Lester and {Scholtz}, Jan and {Suess}, Katherine A. and {Topping}, Michael W. and {{\"U}bler}, Hannah and {Wallace}, Imaan E.~B. and {Whitler}, Lily},
        title = "{JADES NIRSpec Spectroscopy of GN-z11: Lyman-{\ensuremath{\alpha}} emission and possible enhanced nitrogen abundance in a z = 10.60 luminous galaxy}",
      journal = {\aap},
         year = 2023,
        month = sep,
       volume = {677},
          eid = {A88},
        pages = {A88},
          doi = {10.1051/0004-6361/202346159},
archivePrefix = {arXiv},
       eprint = {2302.07256},
 primaryClass = {astro-ph.GA},
       adsurl = {https://ui.adsabs.harvard.edu/abs/2023A&A...677A..88B}
}

@ARTICLE{Maiolino2024,
       author = {{Maiolino}, Roberto and {Scholtz}, Jan and {Witstok}, Joris and {Carniani}, Stefano and {D'Eugenio}, Francesco and {de Graaff}, Anna and {{\"U}bler}, Hannah and {Tacchella}, Sandro and {Curtis-Lake}, Emma and {Arribas}, Santiago and {Bunker}, Andrew and {Charlot}, St{\'e}phane and {Chevallard}, Jacopo and {Curti}, Mirko and {Looser}, Tobias J. and {Maseda}, Michael V. and {Rawle}, Timothy D. and {Rodr{\'\i}guez del Pino}, Bruno and {Willott}, Chris J. and {Egami}, Eiichi and {Eisenstein}, Daniel J. and {Hainline}, Kevin N. and {Robertson}, Brant and {Williams}, Christina C. and {Willmer}, Christopher N.~A. and {Baker}, William M. and {Boyett}, Kristan and {DeCoursey}, Christa and {Fabian}, Andrew C. and {Helton}, Jakob M. and {Ji}, Zhiyuan and {Jones}, Gareth C. and {Kumari}, Nimisha and {Laporte}, Nicolas and {Nelson}, Erica J. and {Perna}, Michele and {Sandles}, Lester and {Shivaei}, Irene and {Sun}, Fengwu},
        title = "{A small and vigorous black hole in the early Universe}",
      journal = {\nat},
         year = 2024,
        month = mar,
       volume = {627},
       number = {8002},
        pages = {59-63},
          doi = {10.1038/s41586-024-07052-5},
archivePrefix = {arXiv},
       eprint = {2305.12492},
 primaryClass = {astro-ph.GA},
       adsurl = {https://ui.adsabs.harvard.edu/abs/2024Natur.627...59M}
}

@ARTICLE{Castellano2024,
       author = {{Castellano}, Marco and {Napolitano}, Lorenzo and {Fontana}, Adriano and {Roberts-Borsani}, Guido and {Treu}, Tommaso and {Vanzella}, Eros and {Zavala}, Jorge A. and {Arrabal Haro}, Pablo and {Calabr{\`o}}, Antonello and {Llerena}, Mario and {Mascia}, Sara and {Merlin}, Emiliano and {Paris}, Diego and {Pentericci}, Laura and {Santini}, Paola and {Bakx}, Tom J.~L.~C. and {Bergamini}, Pietro and {Cupani}, Guido and {Dickinson}, Mark and {Filippenko}, Alexei V. and {Glazebrook}, Karl and {Grillo}, Claudio and {Kelly}, Patrick L. and {Malkan}, Matthew A. and {Mason}, Charlotte A. and {Morishita}, Takahiro and {Nanayakkara}, Themiya and {Rosati}, Piero and {Sani}, Eleonora and {Wang}, Xin and {Yoon}, Ilsang},
        title = "{JWST NIRSpec Spectroscopy of the Remarkable Bright Galaxy GHZ2/GLASS-z12 at Redshift 12.34}",
      journal = {\apj},
         year = 2024,
        month = sep,
       volume = {972},
       number = {2},
          eid = {143},
        pages = {143},
          doi = {10.3847/1538-4357/ad5f88},
archivePrefix = {arXiv},
       eprint = {2403.10238},
 primaryClass = {astro-ph.GA},
       adsurl = {https://ui.adsabs.harvard.edu/abs/2024ApJ...972..143C}
}

@ARTICLE{Curti2024,
       author = {{Curti}, Mirko and {Witstok}, Joris and {Jakobsen}, Peter and {Kobayashi}, Chiaki and {Curtis-Lake}, Emma and {Hainline}, Kevin and {Ji}, Xihan and {D'Eugenio}, Francesco and {Chevallard}, Jacopo and {Maiolino}, Roberto and {Scholtz}, Jan and {Carniani}, Stefano and {Arribas}, Santiago and {Baker}, William M. and {Bhatawdekar}, Rachana and {Boyett}, Kristan and {Bunker}, Andrew J. and {Cameron}, Alex and {Cargile}, Phillip A. and {Charlot}, St{\'e}phane and {Eisenstein}, Daniel J. and {Ji}, Zhiyuan and {Johnson}, Benjamin D. and {Kumari}, Nimisha and {Maseda}, Michael V. and {Robertson}, Brant and {Silcock}, Maddie S. and {Tacchella}, Sandro and {{\"U}bler}, Hannah and {Venturi}, Giacomo and {Williams}, Christina C. and {Willmer}, Christopher N.~A. and {Willott}, Chris},
        title = "{JADES: The star formation and chemical enrichment history of a luminous galaxy at z {\ensuremath{\sim}} 9.43 probed by ultra-deep JWST/NIRSpec spectroscopy}",
      journal = {\aap},
         year = 2025,
        month = may,
       volume = {697},
          eid = {A89},
        pages = {A89},
          doi = {10.1051/0004-6361/202451410},
       adsurl = {https://ui.adsabs.harvard.edu/abs/2025A&A...697A..89C}
}

@ARTICLE{Dale2014,
       author = {{Dale}, Daniel A. and {Helou}, George and {Magdis}, Georgios E. and {Armus}, Lee and {D{\'\i}az-Santos}, Tanio and {Shi}, Yong},
        title = "{A Two-parameter Model for the Infrared/Submillimeter/Radio Spectral Energy Distributions of Galaxies and Active Galactic Nuclei}",
      journal = {\apj},
         year = 2014,
        month = mar,
       volume = {784},
       number = {1},
          eid = {83},
        pages = {83},
          doi = {10.1088/0004-637X/784/1/83},
archivePrefix = {arXiv},
       eprint = {1402.1495},
 primaryClass = {astro-ph.GA},
       adsurl = {https://ui.adsabs.harvard.edu/abs/2014ApJ...784...83D}
}

@ARTICLE{Napolitano2025a,
       author = {{Napolitano}, L. and {Castellano}, M. and {Pentericci}, L. and {Arrabal Haro}, P. and {Fontana}, A. and {Treu}, T. and {Bergamini}, P. and {Calabr{\`o}}, A. and {Mascia}, S. and {Morishita}, T. and {Roberts-Borsani}, G. and {Santini}, P. and {Vanzella}, E. and {Vulcani}, B. and {Zakharova}, D. and {Bakx}, T. and {Dickinson}, M. and {Grillo}, C. and {Leethochawalit}, N. and {Llerena}, M. and {Merlin}, E. and {Paris}, D. and {Rojas-Ruiz}, S. and {Rosati}, P. and {Wang}, X. and {Yoon}, I. and {Zavala}, J.},
        title = "{Seven wonders of Cosmic Dawn: JWST confirms a high abundance of galaxies and AGN at z ≃ 9{\textendash}11 in the GLASS field}",
      journal = {\aap},
         year = 2025,
        month = jan,
       volume = {693},
          eid = {A50},
        pages = {A50},
          doi = {10.1051/0004-6361/202452090},
archivePrefix = {arXiv},
       eprint = {2410.10967},
 primaryClass = {astro-ph.GA},
       adsurl = {https://ui.adsabs.harvard.edu/abs/2025A&A...693A..50N}
}

@ARTICLE{Napolitano2026,
       author = {{Napolitano}, L. and {Pentericci}, L. and {Dickinson}, M. and {Arrabal Haro}, P. and {Taylor}, A.~J. and {Calabr{\`o}}, A. and {Bhagwat}, A. and {Santini}, P. and {Arevalo-Gonzalez}, F. and {Begley}, R. and {Castellano}, M. and {Ciardi}, B. and {Donnan}, C.~T. and {Dottorini}, D. and {Dunlop}, J.~S. and {Finkelstein}, S.~L. and {Fontana}, A. and {Giavalisco}, M. and {Hirschmann}, M. and {Jung}, I. and {Koekemoer}, A.~M. and {Kokorev}, V. and {Llerena}, M. and {Lucas}, R.~A. and {Mascia}, S. and {Merlin}, E. and {P{\'e}rez-Gonz{\'a}lez}, P.~G. and {Stanton}, T.~M. and {Tripodi}, R. and {Wang}, X. and {Weiner}, B.~J.},
        title = "{Ly{\ensuremath{\alpha}} visibility from z = 4.5 to 11 in the UDS field: Evidence for a high neutral hydrogen fraction and small ionized bubbles at z {\ensuremath{\sim}} 7}",
      journal = {\aap},
         year = 2026,
        month = apr,
       volume = {708},
          eid = {A102},
        pages = {A102},
          doi = {10.1051/0004-6361/202556912},
archivePrefix = {arXiv},
       eprint = {2508.14171},
 primaryClass = {astro-ph.GA},
       adsurl = {https://ui.adsabs.harvard.edu/abs/2026A&A...708A.102N}
}

@ARTICLE{Foreman_Mackey2013,
       author = {{Foreman-Mackey}, Daniel and {Hogg}, David W. and {Lang}, Dustin and {Goodman}, Jonathan},
        title = "{emcee: The MCMC Hammer}",
      journal = {\pasp},
         year = 2013,
        month = mar,
       volume = {125},
       number = {925},
        pages = {306},
          doi = {10.1086/670067},
archivePrefix = {arXiv},
       eprint = {1202.3665},
 primaryClass = {astro-ph.IM},
       adsurl = {https://ui.adsabs.harvard.edu/abs/2013PASP..125..306F}
}

@ARTICLE{Astropy2013,
       author = {{Astropy Collaboration} and {Robitaille}, Thomas P. and {Tollerud}, Erik J. and {Greenfield}, Perry and {Droettboom}, Michael and {Bray}, Erik and {Aldcroft}, Tom and {Davis}, Matt and {Ginsburg}, Adam and {Price-Whelan}, Adrian M. and {Kerzendorf}, Wolfgang E. and {Conley}, Alexander and {Crighton}, Neil and {Barbary}, Kyle and {Muna}, Demitri and {Ferguson}, Henry and {Grollier}, Fr{\'e}d{\'e}ric and {Parikh}, Madhura M. and {Nair}, Prasanth H. and {Unther}, Hans M. and {Deil}, Christoph and {Woillez}, Julien and {Conseil}, Simon and {Kramer}, Roban and {Turner}, James E.~H. and {Singer}, Leo and {Fox}, Ryan and {Weaver}, Benjamin A. and {Zabalza}, Victor and {Edwards}, Zachary I. and {Azalee Bostroem}, K. and {Burke}, D.~J. and {Casey}, Andrew R. and {Crawford}, Steven M. and {Dencheva}, Nadia and {Ely}, Justin and {Jenness}, Tim and {Labrie}, Kathleen and {Lim}, Pey Lian and {Pierfederici}, Francesco and {Pontzen}, Andrew and {Ptak}, Andy and {Refsdal}, Brian and {Servillat}, Mathieu and {Streicher}, Ole},
        title = "{Astropy: A community Python package for astronomy}",
      journal = {\aap},
         year = 2013,
        month = oct,
       volume = {558},
          eid = {A33},
        pages = {A33},
          doi = {10.1051/0004-6361/201322068},
archivePrefix = {arXiv},
       eprint = {1307.6212},
 primaryClass = {astro-ph.IM},
       adsurl = {https://ui.adsabs.harvard.edu/abs/2013A&A...558A..33A}
}

@ARTICLE{Mitsuhashi2025,
       author = {{Mitsuhashi}, Ikki and {Suess}, Katherine A. and {Leja}, Joel and {Dayal}, Pratika and {Feldmann}, Robert and {Fujimoto}, Seiji and {Katz}, Harley and {Nanayakkara}, Themiya and {Narayanan}, Desika and {Price}, Sedona H. and {Weaver}, John R. and {Williams}, Christina C. and {Labbe}, Ivo and {Bezanson}, Rachel and {Atek}, Hakim and {Brammer}, Gabriel and {Cutler}, Sam E. and {Furtak}, Lukas J. and {Pan}, Richard and {Wang}, Bingjie and {Whitaker}, Katherine E. and {The Uncover/Megascience Team}},
        title = "{Discovery of Red Galaxy Candidates at z {\ensuremath{\sim}} 12: Early Dust Growth or Significant Nebular Emission with High-temperature Stars?}",
      journal = {\apj},
         year = 2026,
        month = jul,
       volume = {1005},
       number = {1},
          eid = {108},
        pages = {108},
          doi = {10.3847/1538-4357/ae7102},
archivePrefix = {arXiv},
       eprint = {2510.13240},
 primaryClass = {astro-ph.GA},
       adsurl = {https://ui.adsabs.harvard.edu/abs/2026ApJ..1005..108M}
}

@ARTICLE{Donnan25,
       author = {{Donnan}, Callum T. and {Dickinson}, Mark and {Taylor}, Anthony J. and {Arrabal Haro}, Pablo and {Finkelstein}, Steven L. and {Stanton}, Thomas M. and {Jung}, Intae and {Papovich}, Casey and {Akins}, Hollis B. and {Koekemoer}, Anton M. and {McLeod}, Derek J. and {Napolitano}, Lorenzo and {Amor{\'\i}n}, Ricardo O. and {Begley}, Ryan and {Burgarella}, Denis and {Carnall}, Adam C. and {Casey}, Caitlin M. and {Calabr{\`o}}, Antonello and {Cullen}, Fergus and {Dunlop}, James S. and {Ellis}, Richard S. and {Fern{\'a}ndez}, Vital and {Giavalisco}, Mauro and {Hirschmann}, Michaela and {Hu}, Weida and {Illingworth}, Garth and {Kartaltepe}, Jeyhan S. and {Kocevski}, Dale D. and {Kokorev}, Vasily and {Leung}, Ho-Hin and {Lucas}, Ray A. and {Morales}, Alexa M. and {McLure}, Ross and {Pentericci}, Laura and {P{\'e}rez-Gonz{\'a}lez}, Pablo G. and {Somerville}, Rachel S. and {Stevenson}, Struan and {Trump}, Jonathan R. and {Yung}, L.~Y. Aaron and {Zavala}, Jorge A.},
        title = "{Very Bright, Very Blue, and Very Red: JWST CAPERS Analysis of Highly Luminous Galaxies with Extreme Ultraviolet Slopes at z = 10}",
      journal = {\apj},
         year = 2025,
        month = nov,
       volume = {993},
       number = {2},
          eid = {224},
        pages = {224},
          doi = {10.3847/1538-4357/ae0a1f},
archivePrefix = {arXiv},
       eprint = {2507.10518},
 primaryClass = {astro-ph.GA},
       adsurl = {https://ui.adsabs.harvard.edu/abs/2025ApJ...993..224D}
}

@ARTICLE{2002AJ....124..266P,
       author = {{Peng}, Chien Y. and {Ho}, Luis C. and {Impey}, Chris D. and {Rix}, Hans-Walter},
        title = "{Detailed Structural Decomposition of Galaxy Images}",
      journal = {\aj},
         year = 2002,
        month = jul,
       volume = {124},
       number = {1},
        pages = {266-293},
          doi = {10.1086/340952},
archivePrefix = {arXiv},
       eprint = {astro-ph/0204182},
 primaryClass = {astro-ph},
       adsurl = {https://ui.adsabs.harvard.edu/abs/2002AJ....124..266P}
}

@ARTICLE{Backhaus2025,
       author = {{Backhaus}, Bren E. and {Kirkpatrick}, Allison and {Yang}, Guang and {Troiani}, Gregory and {Hamblin}, Kurt and {Kartaltepe}, Jeyhan S. and {Kocevski}, Dale D. and {Koekemoer}, Anton M. and {Lambrides}, Erini and {Papovich}, Casey and {Ronayne}, Kaila},
        title = "{MEGA Mass Assembly with JWST: The MIRI EGS Galaxy and Active Galactic Nucleus Survey}",
      journal = {\aj},
         year = 2025,
        month = dec,
       volume = {170},
       number = {6},
          eid = {300},
        pages = {300},
          doi = {10.3847/1538-3881/ae0cc4},
archivePrefix = {arXiv},
       eprint = {2503.19078},
 primaryClass = {astro-ph.GA},
       adsurl = {https://ui.adsabs.harvard.edu/abs/2025AJ....170..300B}
}

@ARTICLE{Cox+25,
       author = {{Cox}, Isa G. and {Kartaltepe}, Jeyhan S. and {Bagley}, Micaela B. and {Finkelstein}, Steven L. and {Rose}, Caitlin and {Khostovan}, Ali Ahmad and {Chworowsky}, Katherine and {Ilbert}, Olivier and {Koekemoer}, Anton M. and {Ferguson}, Henry C. and {Arrabal Haro}, Pablo and {Backhaus}, Bren E. and {Dickinson}, Mark and {Fontana}, Adriano and {Guo}, Yuchen and {Grazian}, Andrea and {Grogin}, Norman A. and {Harish}, Santosh and {Hathi}, Nimish P. and {Holwerda}, Benne W. and {Iyer}, Kartheik G. and {Kewley}, Lisa J. and {Kirkpatrick}, Allison and {Kocevski}, Dale D. and {Larson}, Rebecca L. and {Lotz}, Jennifer M. and {Lucas}, Ray A. and {Pacucci}, Fabio and {Papovich}, Casey and {Pentericci}, Laura and {P{\'e}rez-Gonz{\'a}lez}, Pablo G. and {Pirzkal}, Nor and {Ravindranath}, Swara and {Somerville}, Rachel S. and {Trump}, Jonathan R. and {Wilkins}, Stephen M. and {Yang}, Guang and {Yung}, L.~Y. Aaron},
        title = "{The CEERS Photometric and Physical Parameter Catalog}",
      journal = {\apj},
         year = 2026,
        month = jul,
       volume = {1005},
       number = {1},
          eid = {62},
        pages = {62},
          doi = {10.3847/1538-4357/ae6f13},
archivePrefix = {arXiv},
       eprint = {2510.08743},
 primaryClass = {astro-ph.GA},
       adsurl = {https://ui.adsabs.harvard.edu/abs/2026ApJ..1005...62C}
}

@ARTICLE{Merlin+24,
       author = {{Merlin}, E. and {Santini}, P. and {Paris}, D. and {Castellano}, M. and {Fontana}, A. and {Treu}, T. and {Finkelstein}, S.~L. and {Dunlop}, J.~S. and {Arrabal Haro}, P. and {Bagley}, M. and {Boyett}, K. and {Calabr{\`o}}, A. and {Correnti}, M. and {Davis}, K. and {Dickinson}, M. and {Donnan}, C.~T. and {Ferguson}, H.~C. and {Fortuni}, F. and {Giavalisco}, M. and {Glazebrook}, K. and {Grazian}, A. and {Grogin}, N.~A. and {Hathi}, N. and {Hirschmann}, M. and {Kartaltepe}, J.~S. and {Kewley}, L.~J. and {Kirkpatrick}, A. and {Kocevski}, D.~D. and {Koekemoer}, A.~M. and {Leung}, G. and {Lotz}, J.~M. and {Lucas}, R.~A. and {Magee}, D.~K. and {Marchesini}, D. and {Mascia}, S. and {McLeod}, D.~J. and {McLure}, R.~J. and {Nanayakkara}, T. and {Napolitano}, L. and {Nonino}, M. and {Papovich}, C. and {Pentericci}, L. and {P{\'e}rez-Gonz{\'a}lez}, P.~G. and {Pirzkal}, N. and {Ravindranath}, S. and {Roberts-Borsani}, G. and {Somerville}, R.~S. and {Trenti}, M. and {Trump}, J.~R. and {Vulcani}, B. and {Wang}, X. and {Watson}, P.~J. and {Wilkins}, S.~M. and {Yang}, G. and {Yung}, L.~Y.~A.},
        title = "{ASTRODEEP-JWST: NIRCam-HST multi-band photometry and redshifts for half a million sources in six extragalactic deep fields}",
      journal = {\aap},
         year = 2024,
        month = nov,
       volume = {691},
          eid = {A240},
        pages = {A240},
          doi = {10.1051/0004-6361/202451409},
archivePrefix = {arXiv},
       eprint = {2409.00169},
 primaryClass = {astro-ph.GA},
       adsurl = {https://ui.adsabs.harvard.edu/abs/2024A&A...691A.240M}
}

@ARTICLE{Merlin16,
       author = {{Merlin}, E. and {Bourne}, N. and {Castellano}, M. and {Ferguson}, H.~C. and {Wang}, T. and {Derriere}, S. and {Dunlop}, J.~S. and {Elbaz}, D. and {Fontana}, A.},
        title = "{T-PHOT version 2.0: Improved algorithms for background subtraction, local convolution, kernel registration, and new options}",
      journal = {\aap},
         year = 2016,
        month = nov,
       volume = {595},
          eid = {A97},
        pages = {A97},
          doi = {10.1051/0004-6361/201628751},
archivePrefix = {arXiv},
       eprint = {1609.00146},
 primaryClass = {astro-ph.IM},
       adsurl = {https://ui.adsabs.harvard.edu/abs/2016A&A...595A..97M}
}

@INPROCEEDINGS{Perrin14,
       author = {{Perrin}, Marshall D. and {Sivaramakrishnan}, Anand and {Lajoie}, Charles-Philippe and {Elliott}, Erin and {Pueyo}, Laurent and {Ravindranath}, Swara and {Albert}, Lo{\"\i}c.},
        title = "{Updated point spread function simulations for JWST with WebbPSF}",
        booktitle = {Space Telescopes and Instrumentation 2014: Optical, Infrared, and Millimeter Wave},
         year = 2014,
       editor = {{Oschmann}, Jr., Jacobus M. and {Clampin}, Mark and {Fazio}, Giovanni G. and {MacEwen}, Howard A.},
       series = {Society of Photo-Optical Instrumentation Engineers (SPIE) Conference Series},
       volume = {9143},
        month = aug,
          eid = {91433X},
        pages = {91433X},
          doi = {10.1117/12.2056689},
       adsurl = {https://ui.adsabs.harvard.edu/abs/2014SPIE.9143E..3XP}
}

@ARTICLE{Fritz06,
       author = {{Fritz}, J. and {Franceschini}, A. and {Hatziminaoglou}, E.},
        title = "{Revisiting the infrared spectra of active galactic nuclei with a new torus emission model}",
      journal = {\mnras},
         year = 2006,
        month = mar,
       volume = {366},
       number = {3},
        pages = {767-786},
          doi = {10.1111/j.1365-2966.2006.09866.x},
archivePrefix = {arXiv},
       eprint = {astro-ph/0511428},
 primaryClass = {astro-ph},
       adsurl = {https://ui.adsabs.harvard.edu/abs/2006MNRAS.366..767F}
}

@ARTICLE{bc03,
       author = {{Bruzual}, G. and {Charlot}, S.},
        title = "{Stellar population synthesis at the resolution of 2003}",
      journal = {\mnras},
         year = 2003,
        month = oct,
       volume = {344},
       number = {4},
        pages = {1000-1028},
          doi = {10.1046/j.1365-8711.2003.06897.x},
archivePrefix = {arXiv},
       eprint = {astro-ph/0309134},
 primaryClass = {astro-ph},
       adsurl = {https://ui.adsabs.harvard.edu/abs/2003MNRAS.344.1000B}
}

\begin{appendix}
\appendix

\section{SED fitting configuration}
In table \ref{tab:priors} we report the priors and the grids used in the CIGALE runs adopted in the paper.

\renewcommand{\arraystretch}{1.2}
\begin{table*}
    \centering
    \caption{Prior grids for free parameters used in our CIGALE SED-fitting runs.}
    \begin{tabular}{ccc}
    \hline\hline
    CIGALE parameter & Grid values & Description \\
    \hline
    \multicolumn{3}{c}{Stochastic SFH [\texttt{sfh\_stochasticity\_physgaussproc}]} \\
    \hline
        Age$_\text{main}$ & 10, 100, 250, 500 & Age of the main population in Myr \\
        Basline code & 1 & Baseline SFH (1: constant) \\
        $\sigma_\text{reg}$ & 0.70 & Amplitude for regulator PSD (nat-log) \\
        $\sigma_\text{dyn}$ & 0.34 & Amplitude for GMC/dynamical PSD  (nat-log)\\
        $\tau_\text{inflow}$ & 100, 120, 140 & Inflow correlation timescale in Myr \\
        $\tau_\text{eq}$ & 15 & Equilibrium/depletion timescale in Myr \\
        $\tau_\text{GMC}$ & 5 & GMC/dynamical timescale in Myr \\
        $T_\text{burst, short}$ & 10, 50 & Short-timescale averaging window for burstiness metrics in Myr\\
        $T_\text{burst, long}$ & 100, 500 & Long-timescale averaging window for burstiness metrics in Myr\\
        $T_\text{burst, cut}$ & 30, 50 & Cutoff timescale for high-frequency power fraction in Myr\\
    \hline
    \multicolumn{3}{c}{SSP [\texttt{bc03}]} \\
    \hline
        $Z$ & 0.0001, 0.02, 0.05 & Stellar metallicity \\
    \hline
    \multicolumn{3}{c}{Nebular [\texttt{nebular}]} \\
    \hline
        log($U$) & -4.0, -3.0, -2.0, -1.0 & Ionisation parameter \\
        $Z_\text{gas}$ & 0.0001, 0.02, 0.051 & Gas metallicity \\
        $N_e$ & 100, 1000 & Electron density \\
    \hline
    \multicolumn{3}{c}{Dust attenuation [\texttt{dustatt\_modified\_starburst}]} \\
    \hline
        E(B-V)$_\text{lines}$ & 0.01, 0.1, 0.25, 0.5, 0.75, 1, 1.5, 2, 2.5, 3 & Color excess of the nebular lines light \\
        E(B-V)$_\text{factor}$ & 0.44, 1 & Reduction factor to compute the stellar continuum attenuation \\
        $R_V$ & 4.05 &  $A_V$ / E(B-V) (Calzetti law) \\
    \hline
    \multicolumn{3}{c}{Dust emission [\texttt{dale2014}]} \\
    \hline
        $f_{AGN}$ & 0.0, 0.1, 0.5, 0.9 & AGN fraction \\
        $\alpha$ & 2, 3 & Alpha slope \\
    \hline
    \multicolumn{3}{c}{Dust emission [\texttt{dl2014}]\,$^\dagger$} \\
    \hline
        $\alpha$ & 2, 3 & Powerlaw slope $dU/dM \propto U^\alpha$ \\
    \hline
    \multicolumn{3}{c}{AGN [\texttt{fritz2006}]\,$^\dagger$} \\
    \hline
        $\psi$ & 30.1, 70.1 & Angle between equatorial axis and line of sight \\
        $f_{AGN}$ & 0.3, 0.6 & AGN fraction \\
        $\lambda_{f_{AGN}}$ & 0/0 & Wavelength range in microns where to compute \\
         &  & the AGN fraction (0/0: total dust luminosity) \\
        E(B-V) & 0.03, 0.1, 0.3, 0.5 & E(B-V) for the extinction in the polar direction in magnitudes \\
    \hline
    \end{tabular}
    \label{tab:priors}
    \tablefoot{$\dagger$: These modules were used in place of \texttt{dale2014}.}
\end{table*}

\section{SED fitting including AGN nebular emission}
\label{app:agnnebular}

The SED modeling presented in Section~\ref{dec:sedfitting} uses the standard \textsc{CIGALE} AGN prescriptions, in which the AGN component contributes continuum emission but the nebular emission is associated with stellar photoionization. The JWST/MIRI F770W excess may include H$\alpha$ emission at the redshift of EGS-z11-R0. To assess the sensitivity of the inferred properties to AGN-powered line emission, we performed additional fits with a custom implementation of the \textsc{SKIRTOR} AGN module in \textsc{CIGALE}, including nebular emission powered by the AGN ionizing radiation (P\'{e}rez-D\'{\i}az \& Merlin, in preparation).

\begin{figure}
\centering
\includegraphics[width=0.5\textwidth]{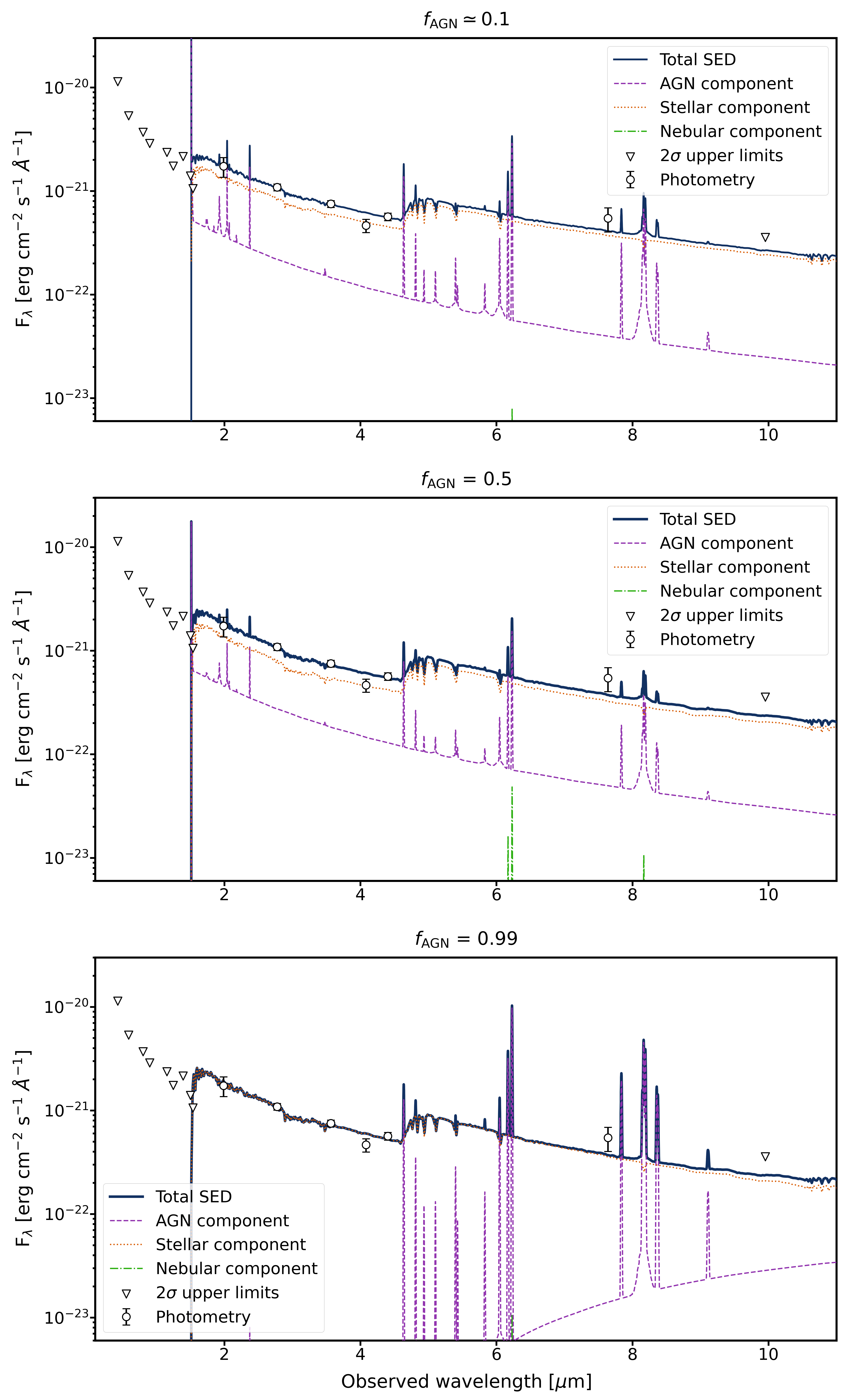}
\caption{Representative SED fits with the updated \textsc{CIGALE} implementation including AGN-powered nebular emission (P\'{e}rez-D\'{\i}az \& Merlin, in preparation). The upper and lower panels illustrate a solution with a modest AGN contribution and an extreme AGN-powered dust-luminosity fraction, respectively. In this implementation $f_{\rm AGN}$ refers to the AGN fraction of the total dust luminosity, not to the observed UV-continuum fraction. The attenuation parameters are discussed using the Bayesian estimates reported in Table~\ref{tab:output-cigale-nebular}.}
\label{fig:appendix_cigale}
\end{figure}

The parameter $f_{\rm AGN}$ is the fraction of the total dust luminosity powered by the AGN, $f_{\rm AGN}=L_{\rm dust,AGN}/(L_{\rm dust,AGN}+L_{\rm dust,stars})$. It is therefore neither a bolometric AGN fraction nor a measure of the AGN contribution to the observed UV continuum. We explored a run with $f_{\rm AGN}$ free and two runs with $f_{\rm AGN}$ fixed to 0.5 and 0.99. Table~\ref{tab:output-cigale-nebular} reports Bayesian parameter estimates for these three configurations, while Figure~\ref{fig:appendix_cigale} illustrates representative best-fitting SEDs.

With $f_{\rm AGN}$ free, the Bayesian estimate is $f_{\rm AGN}=0.12\pm0.13$, indicating that AGN-dominated dust emission is not required within this model family. The corresponding Bayesian attenuation parameters are $E(B-V)_\star=0.161\pm0.064$ mag, $E(B-V)_{\rm AGN}=0.45\pm0.68$ mag, and $E(B-V)_{\rm neb}=0.37\pm0.15$ mag. For the run fixed at $f_{\rm AGN}=0.5$, these become $E(B-V)_\star=0.107\pm0.085$ mag, $E(B-V)_{\rm AGN}=0.99\pm0.81$ mag, and $E(B-V)_{\rm neb}=0.24\pm0.19$ mag. In the extreme $f_{\rm AGN}=0.99$ configuration, the corresponding values are $E(B-V)_\star=0.10\pm0.21$ mag, $E(B-V)_{\rm AGN}=1.35\pm0.85$ mag, and $E(B-V)_{\rm neb}=0.23\pm0.47$ mag. Although the AGN attenuation tends to increase in the forced high-$f_{\rm AGN}$ runs, the broad Bayesian uncertainties preclude a firm conclusion about the relative attenuation of the stellar and AGN components or the physical plausibility of a particular geometry.

The updated models can enhance the predicted H$\alpha$ contribution to F770W, although the available broad-band measurements do not isolate the line flux or establish how much of it is powered by the AGN. The Bayesian instantaneous SFR estimates in Table~\ref{tab:output-cigale-nebular} are much lower than the SFR values reported for the standard models in Section~\ref{dec:sedfitting}. These quantities depend on the adopted model and SFR timescale and should not be compared without this distinction; in particular, an SFR averaged over 100 Myr is not equivalent to the instantaneous SFR.

It is worth noting that the extreme ($f_{\rm AGN}=0.99$) solution may appear counterintuitive in Fig. \ref{fig:appendix_cigale}, as the UV–optical SED is not dominated by the AGN component. This is because ($f_{\rm AGN}$) refers to the fraction of the total dust luminosity powered by the AGN, rather than its contribution to the observed UV–optical emission. In this configuration, the AGN-powered emission is predominantly reprocessed by dust and emerges at far-infrared wavelengths, outside the spectral range displayed in the figure. Thus, the apparently stellar-dominated UV–optical SED is not inconsistent with the imposed high AGN fraction.

In conclusion, the updated fits support a stellar-dominated interpretation of the observed UV-optical continuum, with dust reddening remaining a viable explanation for its red slope, but they do not uniquely constrain the contribution of AGN photoionization to the emission lines. Further spectroscopy and longer-wavelength data are required to distinguish the alternative ionizing sources and dust geometries.

\renewcommand{\arraystretch}{1.25}
\begin{table*}[htp]
    \centering
    \caption{Bayesian physical parameters of EGS-z11-R0 from the updated CIGALE model including AGN-powered nebular emission. Here $f_{\rm AGN}$ is the fraction of total dust luminosity powered by the AGN; the SFR is instantaneous.}
    \begin{tabular}{cccc}
        \hline\hline
         Parameter & $f_{\rm AGN}$ free$^*$ & $f_{AGN} = 0.5$ & $f_{AGN} = 0.99$ \\
        \hline
        log$(M_\star/M_\odot)$ & 9.39 $\pm$ 0.12 & 9.34 $\pm$ 0.12 & 9.19 $\pm$ 0.27 \\
        SFR$_{\rm inst}$ [$M_\odot$ yr$^{-1}$] & 0.91 $\pm$ 2.62 & 0.95 $\pm$ 3.32 & 0.0013 $\pm$ 0.033 \\
        E(B-V)$_\star$ [mag] & 0.1614 $\pm$ 0.064 & 0.107 $\pm$ 0.085 & 0.10 $\pm$ 0.21 \\
        E(B-V)$_{AGN}$ [mag] & 0.45 $\pm$ 0.68 & 0.99 $\pm$ 0.81 & 1.35 $\pm$ 0.85 \\
        E(B-V)$_{neb}$ [mag] & 0.37 $\pm$ 0.15 & 0.24 $\pm$ 0.19 & 0.23 $\pm$ 0.47 \\
        $f_\mathrm{AGN}$ & 0.12 $\pm$ 0.13 & -- & -- \\
        \hline
    \end{tabular}
    \tablefoot{$^*$ The first column corresponds to the run with $f_{\rm AGN}$ free; its Bayesian estimate is $0.12\pm0.13$. Other columns have fixed $f_{\rm AGN}$. All tabulated parameter estimates and the attenuation values quoted in the text are Bayesian; the plotted SED curves represent individual best-fitting models. The SFR is instantaneous, not averaged over 100 Myr.}
    \label{tab:output-cigale-nebular}
\end{table*}

\end{appendix}

\end{document}